\documentclass[fleqn,usenatbib]{mnras}

\usepackage{newtxtext,newtxmath}
\usepackage[T1]{fontenc}

\DeclareRobustCommand{\VAN}[3]{#2}
\let\VANthebibliography\thebibliography
\def\thebibliography{\DeclareRobustCommand{\VAN}[3]{##3}\VANthebibliography}

\usepackage{graphicx}	% Including figure files
\usepackage{amsmath}	% Advanced maths commands
\usepackage{graphicx}
\usepackage{subcaption}
\usepackage{comment}
\usepackage{pdflscape}	% Landscape pages
\usepackage{physics}

\title[Physical properties of star-forming clumps]{Physical properties of star-forming clumps in nearby galaxies from CLAUDS and HSC-SSP}

\author[J. J. Popp et al.]{
Jürgen J. Popp,$^{1}$\thanks{E-mail: jurgen.popp@open.ac.uk}
Hugh Dickinson,$^{1}$
Stephen Serjeant,$^{1}$
Lucy F. Fortson$^{2}$
and Vihang Mehta$^{3}$
\\
$^{1}$School of Physical Sciences, The Open University, Milton Keynes, MK7 6AA, UK\\
$^{2}$School of Physics and Astronomy, University of Minnesota, 116 Church Street SE, Minneapolis, MN 55455, USA\\
$^{3}$IPAC, Mail Code 314-6, California Institute of Technology, 1200 E. California Blvd., Pasadena, CA, 91125, USA\\
}

\date{Accepted XXX. Received YYY; in original form ZZZ}

\pubyear{\the\year{}}

\begin{document}
\label{firstpage}
\pagerange{\pageref{firstpage}--\pageref{lastpage}}
\maketitle

\begin{abstract}
Giant Star-forming Clumps (GSFCs) are kpc-scale regions of enhanced star-formation with stellar masses of $10^6$ to $10^{10}\,M_\odot$ that are commonly observed in high-redshift ($z \gtrsim 1$) galaxies but their formation and role in galaxy evolution remain unclear. We applied a Faster R-CNN object detection framework (FRCNN) to identify star-forming clumps in astrophysical imaging data from a mass-complete sample of $\sim$210,000 low-redshift galaxies ($z\lesssim0.32$), located in the HSC-SSP Wide and Deep fields observed by the Hyper Suprime-Cam Subaru Strategic Survey (HSC-SSP) and CFHT Large Area U-band Deep Survey (CLAUDS). We present a population of $\sim$710,000 detected clumps for which we measured the (\textit{u})\textit{grizy}-photometry and derived the physical properties (stellar mass, age, dust extinction and star-formation rate) for each clump by fitting the photometric data to stellar population models. We estimate that the clump Stellar Mass Function (cSMF) for young clumps ($<100\,\mathrm{Myr}$) follows a power law with an exponent of $\alpha=1.98\pm0.02$ down to a completeness limit of $\sim10^{8}\,M_\odot$ which is consistent with an in situ formation through violent disk instabilities (VDI). The clumps show elevated star-formation rates and negative gradients in their ages and stellar masses as a function of galactocentric distance. These gradients are steeper than those measured for the intra-clump regions of the host galaxies and are robust against uncertainties introduced by different modelling assumptions. For nearby galaxies we argue that the observed low-redshift clumps are likely analogues of high-redshift GSFCs and that the inward migration of star-forming clumps directly contributes to the central bulge formation of the host galaxy. 
\end{abstract}


%%%%%%%%%%%%%%%%%%%%%%%%%%%%%%%%%%%%%%%%%%%%%%%%%%

%%%%%%%%%%%%%%%%% BODY OF PAPER %%%%%%%%%%%%%%%%%%

\section{Introduction}\label{sec:introduction}
Early Hubble Space Telescope (HST) deep field observations of high-redshift ($z>1$) star-forming galaxies (SFGs) revealed galaxy morphologies which differ from the smooth disk morphologies that are mainly seen from their low-redshift counterparts. The irregular and chaotic morphologies at higher redshifts \citep{Cowie1995,Bergh1996,Conselice2004,Elmegreen2005a,Elmegreen2007a,Elmegreen2009,FoersterSchreiber2009,FoersterSchreiber2011} are dominated by several giant star-forming knots or `clumps' (GSFC, or clumps for short) which are much more luminous and larger in extent than typical \textsc{Hii} regions of local galaxies. These GSFCs appear as bright, kpc-scale substructures in restframe ultraviolet (UV) and optical images, suggesting that these are regions of enhanced star-formation within the galaxy which are tightly connected to the evolution of the host galaxy \citep[e.g.][]{Guo2012,Guo2015,Guo2018,Wuyts2012,Wuyts2013,Shibuya2016,Soto2017}. 

Since first being observed, star-forming clumps have been intensively studied through observations and simulations. Understanding the formation and evolution of clumps is important to understanding their role in galaxy evolution, such as whether they contribute directly to the build-up of their host galaxy features or whether they are transient events that only contribute to the total stellar mass formed in a galaxy. The current limits in the spatial resolution of observational data and in galaxy simulations still leave room for different theoretical explanations of how clumps form and evolve in SFGs throughout cosmic history.

Two principal modes of clump formation have been proposed. The first is the formation by gravitational or violent disk instabilities \citep[VDI,][]{Dekel2009,Dekel2013a} in a gas-rich disk \citep{Elmegreen2005,Bournaud2007,Dekel2009a,Genzel2011,Bournaud2013,Mandelker2014,Romeo2014,Guo2012,Guo2015,Fisher2017,HuertasCompany2020,Claeyssens2025}. The constant supply of accreted cold gas from the intergalactic medium (IGM) provides a high gas-fraction in the disks so that the instabilities are sustained in high-redshift galaxies \citep[e.g.][]{Tacconi2020}. This can then lead to the fragmentation of the gaseous disk and the formation of clumps (in situ clumps). The second mode of formation is believed to be triggered by galaxy-galaxy interactions or minor mergers \citep{Conselice2009,Hopkins2013,Wuyts2014,Ribeiro2017,Mandelker2016,Zanella2019}. Gravitational interactions between galaxies compress the gas present in the host galaxy and initiate strong star-forming activity and/or remnants of accreted smaller galaxies can themselves survive as clumps or clump-like objects (ex situ clumps). Most of the recent studies suggest that the majority of observed GSFCs form in situ as a result of the VDI process and only a minor fraction originates from mergers.

The contribution of clumps to the evolution of their host galaxy is strongly dependent on the longevity of the clumps. Simulation studies show that short-lived clumps ($<100\,\rm{Myr}$) are quickly disrupted by outflows and tidal interactions due to their high specific star-formation rates (sSFR) and the resulting strong stellar feedback \citep{Murray2009,Hopkins2012,Hopkins2014,Buck2017,Oklopcic2016}. The disrupted clumps contribute newly formed stars to the formation of thick disks but have only a very weak effect on the stellar structure of the host galaxy, otherwise \citep{Genzel2008,Newman2012a}.

Other simulations indicate that clumps survive at least a few orbital timescales and dynamical friction would lead them to migrate towards the galactic centre where they contribute to the growth of the galactic bulge \citep[e.g.][]{Bournaud2007,Elmegreen2008,Ceverino2010,Bournaud2013,Mandelker2014}. This scenario finds support in observations of radial colour gradients of clumps that show redder colours closer to the galactic centre and bluer colours farther away \citep[e.g.][]{FoersterSchreiber2011a,Tadaki2014,Shibuya2014,Soto2017,Guo2018}. The colour gradients are also expected to be reflected by decreasing stellar ages with increasing galactocentric distance of the clumps with age differences between the outer and inner clumps predicted to be of a few hundred Myr \citep{Dekel2009,Ceverino2010,Dekel2022}. 

While migrating inwards, long-lived clumps are also expected to increase in stellar mass as more and more matter is accreted from their surrounding environment. Observations of radial stellar mass gradients of clumps also support the inward migration scenario \citep[e.g.][]{Guo2018,Zanella2019,Ambachew2022,Dekel2022,Kalita2025} as well as predictions from theoretical models and simulations \citep[e.g.][]{Mandelker2014,Mandelker2016,Dekel2022}. Furthermore, the distribution of the clump stellar masses is important to understand the formation and evolution of the star-forming clumps. If clumps were formed in situ through gravitational or VDI processes, the clump stellar mass function (cSMF) would have a similar power-law slope as stellar mass functions (SMFs) of stellar clusters with slope values of $\alpha \simeq -2.0$ at its massive end \citep{DessaugesZavadsky2018}.

The accurate measurement of the physical properties of clumps is challenging and depends on the spatial resolution and sensitivity of the observations as clumps are only marginally resolved and may even be unresolved \citep[e.g.][]{DessaugesZavadsky2017,DessaugesZavadsky2018,HuertasCompany2020}. Reported stellar masses of high-redshift GSFCs are estimated to be in the range of $10^6$-$10^{10}\,M_\odot$ \citep[e.g.][]{FoersterSchreiber2011a,Zanella2019,HuertasCompany2020,Ambachew2022,Kalita2025} but \citet{HuertasCompany2020} argue that those clump stellar masses are likely overestimated due to sample incompleteness, background light contamination and limits in spatial resolution. Instead, corrected stellar masses of $<10^9\,M_\odot$ for the majority of clumps are more in agreement with observations of clumps in gravitationally lensed galaxies \citep{Livermore2012,Cava2017,Mestric2022,Claeyssens2023,Messa2024,Claeyssens2025}.

The physical sizes of clumps can also be overestimated due to the observational limits \citep[e.g.][]{Fisher2017,Tamburello2017,Meng2020}. While observations of GSFCs in high-redshift galaxies have reported physical clump sizes of $\sim 1$ kpc \citep[e.g.][]{Elmegreen2007a,FoersterSchreiber2011a,Guo2018,Zanella2019}, other works have suggested that the observed kpc-scale structures are possibly blended smaller clumps or sub-clumps that are undetected due to the limited spatial resolution \citep[e.g.][]{Tamburello2015,DessaugesZavadsky2017,Behrendt2016,Meng2020,Faure2021,Mestric2022}. Individual clumps might form a larger clump complex that appears as a kpc-scale GSFC in unresolved observations. This is supported by observations from lensed galaxies that have reported clumps with physical sizes ranging from $\sim 10$ pc to a few hundred pc \citep[e.g.][]{Livermore2012,Adamo2013,Cava2017,Zick2020,Vanzella2022,Messa2022,Mestric2022,Claeyssens2023,Claeyssens2025}. Depending on the lensing magnification, some of the observed objects have been resolved into multiple compact systems that form together a clump complex or a clustered stellar region \citep{Claeyssens2023}.

% Specifically, recent JWST observations of the Sparkler Galaxy, a SFG at redshift $z=1.378$ \citep{Mowla2022} that is gravitationally lensed by the SMACS J0723.3-7323 galaxy cluster at $z=0.388$, found clumps with sizes between $\sim10$ and $\sim50$ pc \citep{Claeyssens2023,Adamo2023,Tomasetti2025,Giunchi2025}. The authors estimated clump stellar masses between $10^{6.3}$ and $10^{7.4}\,M_\odot$ with more than half of the detected clumps older than 1 Gyr. This suggests that the clumps are likely to survive long enough to migrate inwards rather than being quickly dissolved \citep{Giunchi2025}. 

So far, however, most of the studies that measured the physical properties or explored the substructure of clumps have analysed intermediate- and high-redshift galaxies ($z > 0.5$) or small samples of gravitationally lensed high-redshift galaxies. Only a few studies have observed clumps in low-redshift or nearby galaxies \citep[e.g.][]{Overzier2009,Fisher2014,Messa2019,Mehta2021,Lenkic2021,EspinosaPonce2022,Adams2022} and robust measurements of the physical properties for large samples of clumps are still missing. This is partly due to the availability of high-resolution imaging data from HST and the James Webb Space Telescope (JWST) for high-redshift galaxies but also due to the scarcity of low-redshift galaxies that host possible analogues of the observed clump complexes at higher redshift and that are challenging to find in only small samples of observed galaxies.

We perform several steps to address the challenge of finding star-forming clumps in low-redshift galaxies and to quantify the distributions of their physical properties. To identify possible star-forming clumps in a large sample of low-redshift galaxies observed by the CFHT Large Area U-band Deep Survey \citep[CLAUDS,][]{Sawicki2019} and the Hyper Suprime-Cam Subaru Strategic Program \citep[HSC-SSP,][]{Aihara2017}, we make use of an advanced Deep Learning (DL)-based object detection model \citep[][hereafter Paper I]{Popp2026b} that was the first of its kind to use the \textsc{Zoobot} foundation model \citep{Walmsley2023} for downstream tasks like object detection. Based on these clump detections, we measure the physical properties of the star-forming clumps in low-redshift galaxies and compare the results with those resulting from high-redshift studies. Specifically, we analyse the clump stellar mass distribution and the variations of the clump properties as a function of the clump's distance to the galactic centre to test current theories on clump formation and evolution. Our catalogue of $\sim$710,000 star-forming clumps, including the measured photometry and estimated physical properties, is made public \citep{Popp2026} and described in \citet[][hereafter Paper II]{Popp2026d}.

This paper is organised as follows. Section \ref{sec:data} describes the galaxy sample used in this work. In Section \ref{sec:clump_det_phot_sed} we briefly summarise our method to identify star-forming clumps using the DL-based object detection model, the applied photometry method to measure the fluxes of the detected clumps and the inference of the physical properties of the clumps with SED fitting. We present the measured and inferred physical properties of our clump sample in Section \ref{sec:results}, specifically the cSMF and the observed stellar mass and age gradients. We discuss our results in Section \ref{sec:discussion} and conclude with a brief summary in Section \ref{sec:conclusion}.

Throughout this paper, we express all magnitudes in the AB system \citep{Oke1983}. For simplicity, we use the terms `low-redshift' for a redshift range of $z\leq 0.5$ and `high-redshift' for a range of $z>0.5$. Logarithmic quantities are either referenced to a base of $10$ using the notation $\log$ or to a base of $e$ using the notation $\ln$. In this work we adopt the Planck 2015 \citep{Ade2016} cosmological parameters with $(\Omega_m, \Omega_{\Lambda}, h) = (0.31, 0.69, 0.68)$.

% ================================================
% = Galaxy sample
% ================================================
\section{Galaxy sample}\label{sec:data}
For our analysis we combined imaging data from CLAUDS \citep{Sawicki2019} and the HSC-SSP Wide survey \citep{Aihara2017}. The HSC-SSP Wide survey is a multiband (five \textit{grizy} broadband and additional narrowband filters) imaging survey that covers $1400\,\mathrm{deg}^2$ of the sky. CLAUDS provides \textit{u}-band imaging data for those areas where it overlaps with the HSC-SSP survey for the XMM-LSS, E-COSMOS, ELAIS-N1 and DEEP2-3 fields. The \textit{u}-band imaging data was acquired using two different filters. The older $u^\star$ filter was mainly used for the XMM-LSS field and later replaced by the $u$ filter which was used for the E-COSMOS, ELAIS-N1 and DEEP2-3 fields. The two different \textit{u}-band filters cover slightly different wavelength ranges and are treated differently throughout image processing and photometry related tasks (Tab. \ref{tab:hsc_data_survey_overview}).

\begin{table}
	\centering
	\caption[Filter bands, depths and wavelength ranges used by the CLAUDS and HSC-SSP Wide survey.]{Filter bands, depths and wavelength ranges used by the CLAUDS \citep{Sawicki2019} and HSC-SSP Wide survey \citep{Aihara2017}. The CLAUDS \textit{u}-band filter combines the CFHT MegaCam filters $u$ and $u^\star$. The $u$-filter is used for the E-COSMOS, ELAIS-N1 and DEEP2-3 fields while the older $u^\star$-filter only for the XMM-LSS field. The limiting magnitude is shown as the $5\sigma$ point source depth and seeing as the median seeing FWHM with the seeing range of our galaxy sample in brackets.}
    \label{tab:hsc_data_survey_overview}
	\footnotesize
    \begin{tabular}{llrrr}
		\hline
		\multicolumn{1}{c}{Survey} & \multicolumn{1}{c}{Filter-} & \multicolumn{1}{c}{Seeing} & \multicolumn{1}{c}{Lim. mag.} & \multicolumn{1}{c}{Wavelength} \\ 
         & \multicolumn{1}{c}{band} & \multicolumn{1}{c}{[arcsec]} & \multicolumn{1}{c}{$[m_{\mathrm{AB}}]$} & \multicolumn{1}{c}{range, [nm]} \\
		\hline
		CLAUDS   & \textit{u}/$u^\star$ & 0.92        & 27.1 & 310-397 / \\
                 &            & (0.69-1.15) &      & 336-412 \\
        HSC Wide & \textit{g} & 0.79        & 26.5 & 400-500 \\
                 &            & (0.47-1.35) &      & \\
        HSC Wide & \textit{r} & 0.74        & 26.1 & 550-695 \\
                 &            & (0.41-1.33) &      & \\
        HSC Wide & \textit{i} & 0.61        & 25.9 & 695-845 \\
                 &            & (0.35-1.01) &      & \\
        HSC Wide & \textit{z} & 0.68        & 25.1 & 845-930 \\
                 &            & (0.41-1.21) &      & \\
        HSC Wide & \textit{y} & 0.67        & 24.4 & 930-1,070 \\
                 &            & (0.40-1.76) &      & \\
		\hline
	\end{tabular}
\end{table}

The selection of our target galaxies is described in detail in Paper I. Briefly, we started with a preselection of target galaxies from the SDSS Data Release 18 \citep{Almeida2023} catalogue. By doing so we obtain an initial sample of galaxies with robust measurements of the Petrosian radius that we used to define the cutout size to ensure a comparable visual size of the target galaxies in each cutout. As a result of the preselection, our sample is magnitude-limited by the median $5\sigma$ depth for SDSS photometric observations in the $r_{\mathrm{SDSS}}$-band magnitude of $r_{\mathrm{SDSS}} \leq 22.7\,m_{\mathrm{AB}}$ and limited to a maximum redshift of $z \leq 0.5$.

We further required the galaxies to have a minimum extent defined by a SDSS \textit{r}-band 90\% Petrosian radius of $\geq3\,\mathrm{arcsec}$ so that morphological features are resolvable. This preselection was then crossmatched (within 1.0 arcsec) with sources detected in the \textit{i}-band from the latest Public Data Release 3 \citep[PDR3,][]{Aihara2022} of the HSC-SSP Wide survey, which resulted in 710,271 HSC-SSP galaxies with clean five filter band photometry (\textit{grizy}, Table \ref{tab:hsc_data_preprocess}). We refer to this set of galaxies with only 5-filter band photometry as the HSC-SSP only galaxy sample.

We then crossmatched the HSC-SSP set with the CLAUDS data for the XMM-LSS, E-COSMOS and DEEP2-3 fields. The ELAIS-N1 field is covered by the HSC-SSP Deep and Ultra-Deep surveys but not by the HSC-SSP Wide survey (using the same crossmatching distance of 1.0 arcsec) so that this set consists of 14,231 galaxies with six filter band photometry data (\textit{ugrizy}, Table \ref{tab:hsc_data_preprocess}). We refer to this smaller subset of galaxies with \textit{ugrizy} photometry as the (combined) CLAUDS/HSC-SSP galaxy sample.

We generated image cutouts for each galaxy and filter band as separate files containing the sky-subtracted and calibrated science image and the corresponding variance map (see Paper II).

Apart from the imaging data, both surveys provide catalogues with multiple photometry measurements in the different filter bands and inferred physical properties of the detected source objects, which are described in \citet{Desprez2023} for the CLAUDS catalogue and in \citet{Aihara2022} for the HSC-SSP PDR3 catalogue. For consistency, we use and report the published source properties from the HSC-SSP PDR3 catalogue as they are provided for all objects in our selected sample. All object identifiers used in this work refer to the object IDs in the HSC-SSP PDR3 catalogue.

To determine the 90\% mass-completeness limit for our sample of galaxies over the observed redshift range, we followed the method described by \citet{Pozzetti2010}. For every galaxy, a stellar mass limit $M_{\star,\mathrm{lim}}$ was calculated that represents the mass a galaxy would have if its apparent magnitude was equal to the SDSS \textit{i}-band detection limit at $i_{\mathrm{lim}} = 22.2\,m_{\mathrm{AB}}$. With the observed stellar mass $M_\star$ of the galaxy, $M_{\star,\mathrm{lim}}$ can be calculated as:
\begin{equation}
    \log(M_{\star,\mathrm{lim}}) = \log(M_\star) + 0.4 (i-i_{\mathrm{lim}}),
\end{equation}
where $i$ is the SDSS \textit{i}-band magnitude of the galaxy. Then, for each redshift bin between $0.0 \leq z \leq 0.5$ using a bin width of $\Delta z = 0.01$, we selected the faintest 20\% of the galaxies and determined the 90th percentile of $M_{\star,\mathrm{lim}}$ for each subset of galaxies.

\begin{figure}
    \centering
    \includegraphics[width=0.4\textwidth]{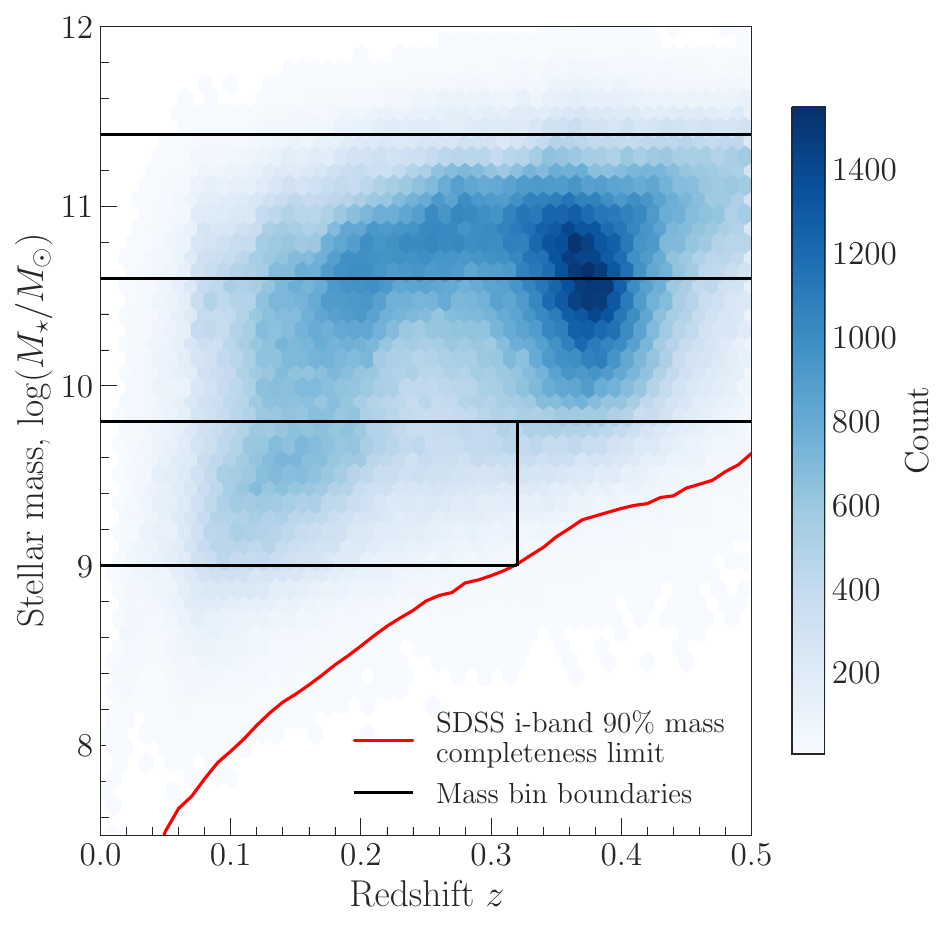}
    \caption[Galaxy stellar mass as a function of redshift for the sample of CLAUDS and HSC-SSP galaxies.]{Galaxy stellar mass as a function of redshift for the sample of CLAUDS and HSC-SSP galaxies. The red solid line indicates the 90\% mass-completeness limit as a function of redshift. The black outlines mark the mass-complete sample for the mass bins $9.0 \leq \log(M_\star/M_\odot) < 9.8$ (redshift $z<0.32$), $9.8 \leq \log(M_\star/M_\odot) < 10.6$ and $10.6 \leq \log(M_\star/M_\odot) < 11.4$.}
    \label{fig:hsc_data_galaxy_completeness}
\end{figure}

Our galaxy sample is mass-complete for galaxies with stellar mass $\log(M_\star/M_\odot) \geq 9.0$ and redshift $z \leq 0.32$ (Fig. \ref{fig:hsc_data_galaxy_completeness}). The mass-complete sample used in this paper includes 7,778 galaxies with six filter band (\textit{ugrizy}) imaging data and 389,500 galaxies with only five filter band (\textit{grizy}) imaging data (Table \ref{tab:hsc_data_preprocess}).

\begin{table}
	\centering
	\caption[Number of galaxies contained in the sample.]{Number of galaxies contained in the sample. The mass-complete sample consists of all galaxies with redshift $z \leq 0.32$ and stellar mass $\log(M_\star/M_\odot) \geq 9.0$. The stellar mass bins for galaxies with redshift $z \leq 0.32$ are: $9.0 \leq \log(M_\star/M_\odot) < 9.8$ (low), $9.8 \leq \log(M_\star/M_\odot) < 10.6$ (medium) and $10.6 \leq \log(M_\star/M_\odot) < 11.4$ (high).}
    \label{tab:hsc_data_preprocess}
	\footnotesize
        \begin{tabular}{lrrr}
		\hline
		Selection & \multicolumn{1}{c}{CLAUDS/} & \multicolumn{1}{c}{HSC} & \multicolumn{1}{c}{Total} \\
         & \multicolumn{1}{c}{HSC} & \multicolumn{1}{c}{only} & \\
        \hline
        Galaxies     & 14,231 &   696,040 &    710,271 \\ 
        \textit{thereof:} mass-complete        &  7,778 &   389,500 &    397,278 \\
        \textit{thereof:} low-mass galaxies    &  2,073 &    97,074 &     99,147 \\
        \textit{thereof:} medium-mass galaxies &  3,002 &   151,856 &    154,858 \\
        \textit{thereof:} high-mass galaxies   &  2,703 &   140,570 &    143,273 \\
		\hline
	\end{tabular}
\end{table}

% ================================================
% = Clump detection, photometry and SED fitting
% ================================================
\section{Clump detection, photometry and SED fitting}\label{sec:clump_det_phot_sed}

% -----------------------------------------------
% == Clump identification
\subsection{Clump identification}\label{sec:clump_det}
We identified potential star-forming clumps or clump candidates in our galaxy sample using the DL-based object detection models that are described in detail in Paper I and which we also used in Paper II. The models are built on the Faster R-CNN object detection framework \citep[FRCNN,][]{Ren2015} and use the \textsc{Zoobot} foundation deep learning model \citep{Walmsley2023} in its latest version v2.0 as their feature extracting backbone. We applied the 6-channel object detection model that accepts \textit{ugrizy} filter band data as input on our CLAUDS/HSC-SSP galaxy sample and the 5-channel model that uses only imaging data from the five \textit{grizy} filter bands on the HSC-SSP only galaxy sample. 

% \begin{figure*}
%     \centering
%     \includegraphics[width=1.0\textwidth]{figures/example_peak_detection.pdf}
%     \caption[Galaxy examples showing the postprocessed FRCNN model detections with extracted flux peaks.]{Three galaxy examples showing the postprocessed FRCNN model detections with extracted flux peaks. The model detections are shown as boxes where the colour indicates the object class. Flux peaks are marked with red crosses. The galaxies are shown with their \textit{u}-band images.}
%     \label{fig:hsc_det_over_peaks_example}
% \end{figure*}

Both FRCNN models output bounding boxes around detected clump candidates and possible contaminating objects and apply a classification scheme with multiple classes that include object classes for clumps, foreground stars, fore-/background galaxies, bulges and the generic background. In addition to the classification, the models also assign a score that is called `objectness' to each bounding box. The objectness is a measure between 0.0 and 1.0 and indicates how certain the models are whether the bounding box contains an object or not.

% The detection completeness is well above 0.8 and close to 1.0 over most of the ranges of the physical properties (apparent magnitude, colours, contrast, stellar mass, distance from the galactic centre) for clumps that are brighter than the instrument-specific detection limit (Table \ref{tab:hsc_data_survey_overview}). The models also reliably detect the brighter clumps over the full redshift range of the sample of galaxies analysed in this study and the detection completeness appears to be independent of the stellar mass or sSFR of the host galaxy (see Paper I).

Purity and completeness of both models are well above 80 to 90\% based on measurements of a large sample of simulated clumps with fluxes above the average $5\sigma$ point-source depth limits of the filter bands (Table \ref{tab:hsc_data_survey_overview}). The detections of the 5-channel model are similar to the detections of the 6-channel model if applied to the same set of galaxies (Paper I). The models show similar clump detections with an overlap of $\gtrsim 80\%$, although the 5-channel model tends to detect fewer clumps due to the missing \textit{u}-band, specifically those with very recent star-formation. Furthermore, the majority of clump candidates that were detected either by only the 6-channel or only the 5-channel model have low objectness scores $\lesssim 0.2$, suggesting that those predictions unique to each model contain spurious detections (see Paper I). However, as we later select our sample of star-forming clumps from the full sample of detected clump candidates based on their inferred physical properties, we did not apply a cut on the objectness to keep the detections as complete as possible. 

We postprocessed the raw detection results as described in Paper I. Briefly, we first applied a non-maximum suppression (NMS) to all detections from each galaxy using a threshold for the Intersection over Union of $\mathrm{IoU} \geq 0.2$ and kept only those detections and the corresponding object class predictions that have the highest objectness from each subset of overlapping bounding boxes. Bounding boxes that do not overlap with any other bounding boxes in a galaxy image were left unchanged. We then removed all bounding boxes that were classified as potential clumps and that are larger than the 95th percentile threshold of the bounding box size distribution (corresponding to a maximum bounding box size of $\sim 7.30\,\mathrm{kpc}$). We also removed clump detections for which the centroid of the bounding box lies outside the target galaxy's segmentation mask and discarded those bounding boxes that are close to or coincide with the central bulge of a galaxy (i.e. if the distance of its midpoint to the flux-weighted centroid of the galaxy is less than 2\% of the Petrosian radius).

\begin{figure}[H]
    \centering
    \includegraphics[width=0.8\columnwidth]{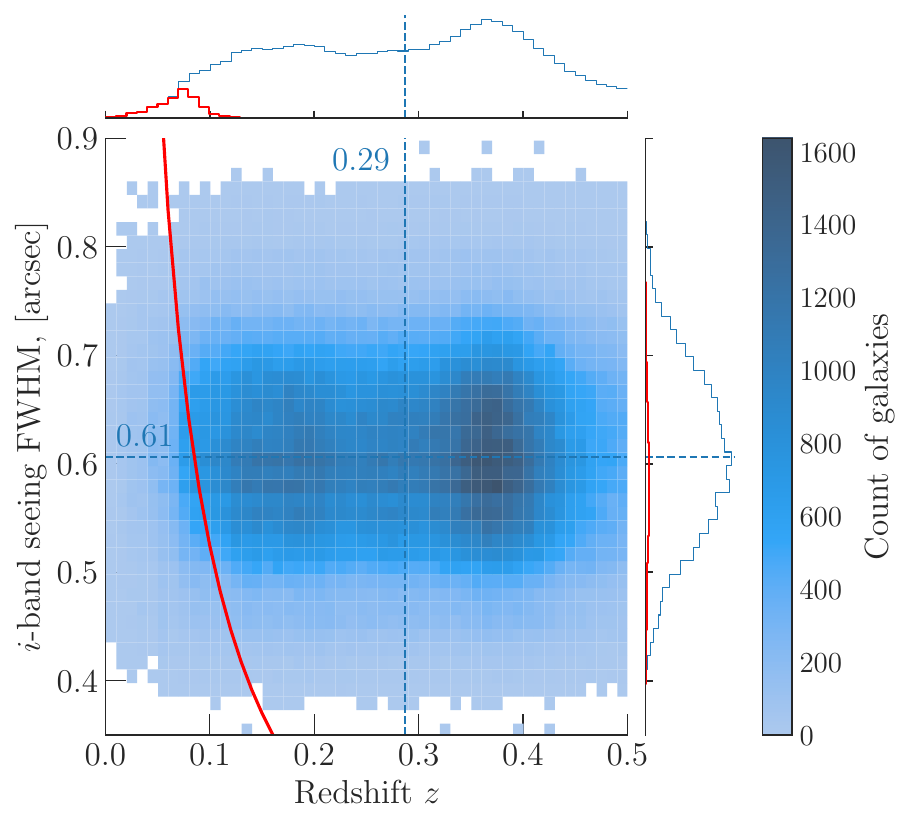}
    \caption[Redshift vs. \textit{i}-band seeing FWHM for HSC-SSP galaxies.]{Redshift vs. \textit{i}-band seeing FWHM for HSC-SSP galaxies and their image cutouts. Vertical and horizontal dashed lines mark the median redshift and \textit{i}-band seeing FWHM of the galaxy sample and are also shown as annotations in the plot. The red line plots the theoretical seeing FWHM that is required to resolve objects with 1.0 kpc in physical size. The marginal histograms on the top and right side of the two-dimensional histogram show the univariate distributions for all galaxies in blue and galaxies for which the \textit{i}-band seeing FWHM translates to a physical size of $\leq 1$ kpc in red.}
    \label{fig:seeing_redshift}
\end{figure}

With the given seeing, features with physical sizes $<1$ kpc can be theoretically resolved in only a small fraction of galaxies at $z\lesssim 0.1$. This is shown in Figure \ref{fig:seeing_redshift} where we plot the \textit{i}-band seeing full width at half maximum (FWHM) as a function of redshift for our galaxy sample. As clumps have reported physical sizes as low as $\sim$10-100 pc \citep[e.g.][]{Claeyssens2023,Claeyssens2025} and as the \textit{i}-band seeing is significantly better than the seeing of the other filter bands (Table \ref{tab:hsc_data_survey_overview}), the vast majority of the clumps are unresolved in our observations. Therefore, we extracted the local flux maxima or flux peaks within each potential clump bounding box and identified these peaks as marking the positions of our final sample of clump candidates which we treated as point-like objects that cannot be further resolved given the seeing limits. 

% However, ....mention clump complexes....
% Furthermore, we use the term `clump complex' to refer to a collection of clumps that were identified within a single bounding box of the FRCNN model detections. These clumps are likely to be blended into a single clump if observed from imaging data with lower spatial resolution. 
% The object detection model (Section \ref{sec:hsc_det_over_model_dev_postprocess}) resulted in bounding boxes around star-forming regions containing one or more clumps that form a clump complex (see also Section \ref{sec:hsc_det_over_peaks}). Individual clump candidates are detected from the peak flux locations within the bounding boxes (Section \ref{sec:hsc_det_over_peaks}).

% -----------------------------------------------
% == Clump photometry
\subsection{Clump photometry}\label{sec:clump_phot}
The details of our photometry measurements of the detected clump candidates are outlined in Paper II and we provide only a brief summary here. We used aperture photometry to recover the fluxes of our clump sample at their locations in the \textit{u}-band images, while holding the positions constant for the other \textit{g}-, \textit{r}-, \textit{i}-, \textit{z}- and \textit{y}-band images (forced photometry). We estimated the diffuse galaxy background light from an annulus around the aperture and subtracted the median value per pixel from the aperture pixel values. Furthermore, we masked the area outside the host galaxy extent and adjacent clump detections as their light would otherwise contaminate the background estimate if they are (partly) located within the annulus. The adjacent clump locations were masked out with a circular mask. 

We tested different sizes for the aperture radius, annulus radii and adjacent clump mask radius and determined those radii for which a high recovery accuracy was achieved based on simulated clumps and point-like catalogue objects with known fluxes (Paper II). The photometry of the detected clumps was finally measured using (in units of the filter band-specific seeing FWHM) an aperture with radius $r_{\mathrm{ap}}=0.25\, \mathrm{FWHM}_{\mathrm{seeing}}$, masking adjacent clump detections with a circular mask of radius $r_{\mathrm{mask}}=0.5\, \mathrm{FWHM}_{\mathrm{seeing}}$ and corrected for the extended shape of the PSF using existing and estimated PSF models. The underlying background light was estimated using an annulus of $r_{\mathrm{an}}=\{1.5, 2.0\}\, \mathrm{FWHM}_{\mathrm{seeing}}$ and subtracted from the flux in the aperture to minimise the contamination of the clump photometry measurements by adjacent objects and the underlying diffuse galaxy background. In relation to the total flux at each position of the clump candidates, the diffuse galaxy light that is measured in the annulus around each clump candidate and subtracted from the flux measured in the aperture varies between filter bands. The median fractions are $0.28$ in the \textit{u}-band, $0.40$ ($0.40$) in the \textit{g}-band, $0.40$ ($0.43$) in the \textit{r}-band, $0.52$ ($0.51$) in the \textit{i}-band, $0.46$ ($0.47$) in the \textit{z}-band and $0.51$ ($0.48$) in the \textit{y}-band for the clump candidates with six (five) filter band photometry. The fraction of the subtracted background light is lowest for the \textit{u}-band as the star-forming clumps are expected to be relatively bright in the \textit{u}-band compared to the galactic disk.

From the 30,636 (1,423,076) clump candidates in total with \textit{ugrizy} (\textit{grizy}) photometry, a valid flux value was measured for 28,814 (1,347,384) clump candidates but for 1,822 (75,692) clumps, or $5.95\%$ ($5.32\%$), the measurement returned either a negative flux value or the image contained missing per-pixel flux values at or around the pixel location of the measurement and were excluded from our sample. We also excluded all clump candidates with a measured magnitude that is fainter than the detection limit in at least one filter band (see Table \ref{tab:hsc_data_survey_overview}). As an additional postprocessing step, we applied corrections for the Galactic extinction to the measured magnitudes. The reddening $E(B-V)$ was calculated using the \citet{Schlegel1998} dust maps with an extinction to reddening ratio of $R_{V}=3.1$ for the Milky Way.

Table \ref{tab:hsc_phot_final_counts} lists all counts for each set with available photometry data. In total, combining the data with 6-channel and 5-channel photometry, we make use of 944,943 clump detections in 301,343 galaxies in this study.

% -----------------------------------------------
% == SED fitting
\subsection{SED fitting}\label{sec:clump_sed}
We derived the physical properties (stellar mass $M_\star$, stellar age $t_{\mathrm{age}}$, stellar metallicity $Z_\star$, dust attenuation $A_V$) by fitting stellar population synthesis (SPS) models to the observed broadband photometry of our detected clump candidates. The stellar masses inferred for the clump candidates are the surviving stellar mass since the initial formation. A SED was fit to each of the 944,943 clump candidates using \textsc{Prospector} \citep{Johnson2021}. \textsc{Prospector} is a Python-based SED fitting code to interpret photometric and/or spectroscopic data for wavelength ranges from the far-ultraviolet (FUV) to the far-infrared (FIR). It incorporates the libraries from the \textsc{FSPS}: Flexible Stellar Population Synthesis \citep{Conroy2009,Conroy2010} code for model definition and the inference process is based on Bayesian forward-modelling techniques that use nested sampling \citep{Skilling2004,Skilling2006} via \textsc{dynesty} \citep{Speagle2020} to explore the resulting parameter space.

The SED models were parametrised with a constant SFH model. Redshift was set fixed to the galaxy spectroscopic redshift where available and to the best photometric redshift, otherwise. Spectroscopic redshifts have been made available for the HSC-SSP objects from various public sources \citep{Aihara2022}, mainly SDSS \citep{Ahumada2020}. Photometric redshifts were provided by the HSC-SSP directly \citep{Tanaka2017}. For this study, we adopted the initial mass function (IMF) described by \citet{Chabrier2003} with an upper mass cutoff of $100\,M_\odot$ and the dust attenuation law from \citet{Calzetti2000}, for which we constrained $A_V$ between 0 and 4. The SED models were constructed using the stellar isochrone library \textsc{MIST} \citep{Paxton2010,Choi2016,Dotter2016} and stellar spectral library \textsc{MILES} \citep{SanchezBlazquez2006, Cenarro2007}. Nebular line and continuum emission was modelled using the \textsc{FSPS}-implementation of the \textsc{Cloudy} radiative transfer code \citep{Ferland1998,Ferland2013} and ionisation sources as described in \citet{Byler2017}. The ionisation parameter $U$ was set fixed to a value of $\log(U)=-2.0$ and a constant gas density of $n_{\mathrm{H}}=100\,\mathrm{cm}^{-3}$ was assumed.

We also included the \citet{Draine2007} dust emission model but set the three parameters necessary for modelling the dust emission fixed. The interstellar radiation field $J$ is set to a minimum value $J_{\mathrm{min}}=1.0$, expressed in units of the Milky Way radiation field \citep[$j_{\nu}^{\mathrm{MMP83}}$,][]{Mathis1983}. For the fraction of dust mass $1-\gamma$ that is exposed to the radiation intensity at $J_{\mathrm{min}}$, the value for $\gamma$ is set to $10^{-3}$ and the polycyclic aromatic hydrocarbon (PAH) fraction $q_{\mathrm{PAH}}$ to $4.0\%$.

The prior distributions of the main physical properties are motivated by reported observations of star-forming clumps from the literature. We allowed the stellar mass to vary between $10^4 - 10^{11}\,M_\odot$ but limited the upper bound for the stellar mass prior to not exceed $20\%$ of the host galaxy's stellar mass. The age prior was set to a uniform distribution between $0.001$ and $10$ Gyr to also cover the range needed for possible ex-situ clumps that are expected to be older than in-situ clumps which have reported ages between $<100\,\mathrm{Myr}$ for short-lived clumps and several hundreds Myr for long-lived clumps \citep[e.g][]{FoersterSchreiber2011,Guo2018,Zanella2019}. The prior distribution for metallicity was set to a uniform distribution of $\log(Z_\star/Z_\odot) = \mathcal{U}(-2.0, 0.19)$ and the gas metallicity $\log(Z_{\mathrm{gas}}/Z_\odot)$ was bound to the value of the stellar metallicity. All final prior distributions are listed in Table \ref{tab:hsc_sed_fitting_priors} and we describe the execution of our SED fitting process in Appendix \ref{sec:hsc_sed_fitting_exec}.

\begin{table}
	\centering
	\caption[Prior distributions of the model used for the SED fitting process.]{Prior distributions of the model used for the SED fitting process. The upper bound for the stellar mass prior is limited to not exceed $20\%$ of the host galaxy's stellar mass. The gas metallicity is bound to the prior values of the stellar metallicity and redshift is fixed to the redshift of the host galaxy. The interstellar radiation field $J$ is in units of the Milky Way radiation field \citep[$j_{\nu}^{\mathrm{MMP83}}$,][]{Mathis1983}.}
    \label{tab:hsc_sed_fitting_priors}
	\footnotesize
        \begin{tabular}{lllr}
		\hline
		Physical & Units & Distribution & Prior range \\
        parameter & & & \\
		\hline
        Redshift           & $z$                     & (fixed)     & $[0.005, 0.5]$ \\
        $M_\star$          & $M_\odot$               & log-uniform & $[10^4, 10^{11}]$ \\
        $t_{\mathrm{age}}$ & Gyr                     & uniform     & $[0.001, 10]$ \\
        $Z_\star$          & $\log(Z_\star/Z_\odot)$ & uniform     & $[-2.0, 0.19]$ \\
        $A_{V}$            & $m_{\mathrm{AB}}$       & uniform     & $[0.0, 4.0]$ \\
        $U$                & $\log(U)$               & (fixed)     & $-2.0$ \\
        $Z_{\mathrm{gas}}$ & $\log(Z_{\mathrm{gas}}/Z_\odot)$      & uniform & $[-2.0, 0.19]$ \\
        $J_{\mathrm{min}}$ & $j_{\nu}^{\mathrm{MMP83}}$ & (fixed)  & $1.0$ \\
        $q_{\mathrm{PAH}}$ & Percent                 & (fixed)     & $4.0$ \\
        $\gamma$           & Percent                 & (fixed)     & $10^{-3}$ \\
        \hline
	\end{tabular}
\end{table}

We fitted SPS models to the observed \textit{ugrizy} photometry of the detected clump candidates from the CLAUDS/HSC-SSP galaxy sample and to the \textit{grizy} photometry of the clump candidates from the HSC-SSP only sample. Based on the photometry measurements of the detected clump candidates (Section \ref{sec:clump_phot}), we included only those clump candidates that are brighter than the instrument-specific detection limit (Table \ref{tab:hsc_data_survey_overview}) in the SED fitting process. However, we also included the clump candidates that are fainter than the detection limit in the SED fitting process of the smaller 6-channel photometry clump sample to further assess the effect that the exclusion of the fainter clump candidates has on the total sample of clump detections.

Unless otherwise indicated, we use the maximum a posteriori (MAP) estimates to construct the best fit model spectra in this work. For reported values of individual physical properties, we use the MAP values together with 95\% credible intervals (CIs) constructed with the MAP as the central value. A catalogue containing the detected clump candidates and their measured and derived properties is published as \citet{Popp2026} and described in Paper II. The catalogue also contains the median and mode estimates of the marginalised posterior for each clump property in addition to the MAP estimates.

The sample of galaxies from HSC-SSP that have no crossmatch with the CLAUDS catalogue provides a sample of 925,179 clump candidates with measured magnitudes in the five \textit{grizy}-filter bands that are brighter than the instrument-specific detection limit (Table \ref{tab:hsc_phot_final_counts}). To explore whether the analysis of the physical properties of the clump candidates can be extended to this much larger sample with only five photometric data points, we also tested the SED models on the smaller \textit{ugrizy} dataset but excluded the \textit{u}-band photometry data point (Appendix \ref{sec:hsc_sed_eval_sed_GRIZY}). Using only the five \textit{grizy} instead of the six \textit{ugrizy}-filter bands to infer the physical properties of clumps does not introduce a strong bias. Specifically, inferring the stellar mass of a clump candidate from five photometry data points alone yields similar results to the estimates from all six photometry data points. However, age estimates need to be treated with caution. Even when the overlapping CIs confirm that the true age is very likely to be contained within both CIs, the point estimates (i.e. MAP) can vary considerably. Therefore, we present our results separately for both samples of clump detections in the following sections.

In addition to the model using the priors in Table \ref{tab:hsc_sed_fitting_priors}, we also constructed and fitted alternative SPS models with a delayed $\tau$ SFH model and a constant SFH model for which the metallicity was set fixed to solar metallicity $\log(Z_\star/Z_\odot)=0.0$. All other parameters remained unchanged. We use the results of these models in comparison with the main model to validate the robustness against different model assumptions. 

The clumps are expected to evolve differently to the inter-galactic environment they are embedded in. Therefore, changes in physical properties over time or galactocentric distance are expected to be different for the clump candidates themselves and their underlying stellar background. The background fluxes from all six \textit{ugrizy}-filter bands that were subtracted from the clump photometry (Section \ref{sec:clump_phot}) are an estimate of that underlying diffuse light. To infer the physical properties of the underlying stellar background of the clumps, we used the background measurements of all 19,764 clump candidates detected in the CLAUDS/HSC-SSP imaging data and fitted SED models with different priors to the photometric data points (Appendix \ref{sec:hsc_sed_eval_sed_background}).

% ================================================
% = Properties of star-forming regions
% ================================================
\section{Properties of star-forming regions}\label{sec:results}
In the following sections, we first describe the final selection of clumps from the set of postprocessed FRCNN model detections based on their measured and inferred properties. The term `clump' is used synonymously for a detected star-forming region that satisfies the empirically determined thresholds for its stellar mass and flux described in Section \ref{sec:hsc_phys_props_clump_sample}. If a detection of a possible clump does not meet these criteria we refer to it as a `clump candidate' instead to distinguish the FRCNN model detections from the final clump sample. Section \ref{sec:hsc_phys_props_clumps} provides a detailed analysis of the physical properties of the clumps, specifically their stellar mass distribution and the gradual changes of colour, stellar mass and age with distance from the galactic centre.

% -----------------------------------------------
% == Sample Selection of the Star-forming Clumps
\subsection{Sample selection of the star-forming clumps}\label{sec:hsc_phys_props_clump_sample}
\begin{figure}
    \centering
    \subfloat[\centering Distribution of clump-galaxy stellar mass ratio. \label{fig:hsc_phys_props_clump_sample1a}]{{\includegraphics[width=0.85\columnwidth]{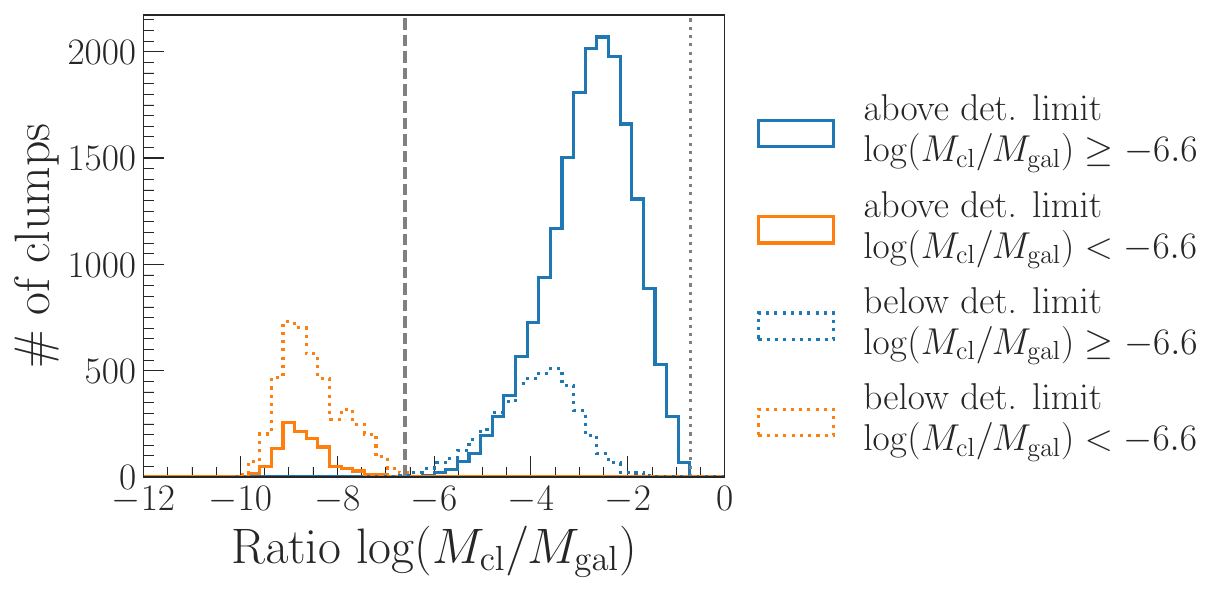} }}
    \\
    \subfloat[\centering Distribution of stellar mass. \label{fig:hsc_phys_props_clump_sample1b}]{{\includegraphics[width=0.5\columnwidth]{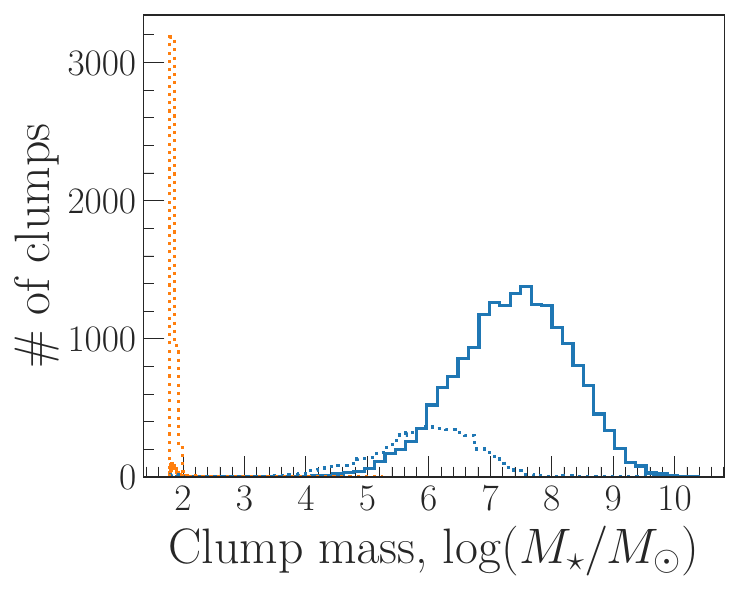} }}
    \subfloat[\centering Distribution of stellar age. \label{fig:hsc_phys_props_clump_sample1c}]{{\includegraphics[width=0.5\columnwidth]{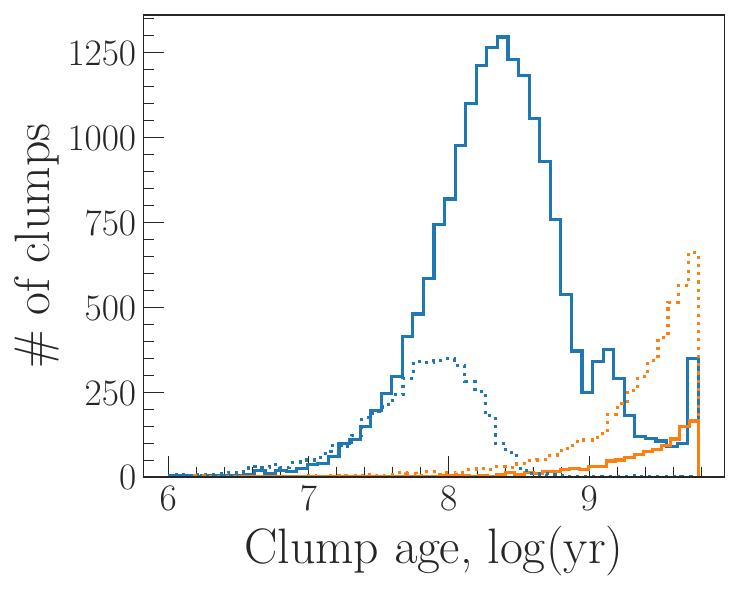} }}    
    \caption[Distribution of clump-galaxy stellar mass ratio, clump stellar mass and age of the clump candidates with 6-filter band photometry from the CLAUDS/HSC-SSP galaxy sample.]{Distribution of clump-galaxy stellar mass ratio, clump stellar mass and age of the clump candidates with 6-filter band photometry from the CLAUDS/HSC-SSP galaxy sample. The distributions are plotted for clumps with all six photometry measurements brighter (solid lines) and fainter than the instrument-specific detection limit (dotted lines). The dotted vertical line marks the prior boundary of $M_{\mathrm{cl}}/M_{\mathrm{gal}} \leq 0.2$ and the dashed vertical line the cutoff between non-clump and clump detections at $\log(M_{\mathrm{cl}}/M_{\mathrm{gal}})=-6.6$. The clump candidates with a clump-galaxy stellar mass ratio of $\log(M_{\mathrm{cl}}/M_{\mathrm{gal}}) < -6.6$ are drawn with orange lines and those greater than the cutoff are drawn with blue lines.}
    \label{fig:hsc_phys_props_clump_sample1}
\end{figure}

Figure \ref{fig:hsc_phys_props_clump_sample1a} shows the distribution of the clump candidates as a function of the ratio between the MAP estimate for the stellar mass of the clump candidate and the stellar mass of its host galaxy $\log$($M_{\mathrm{cl}}/M_{\mathrm{gal}}$). Two populations of clump candidates can be separated that have either a log mass ratio $< -6.6$ or $\geq -6.6$. Clump candidates above that threshold mainly have photometry measurements with fluxes larger than the detection limit whereas the majority of the clump candidates below the threshold have fluxes that are smaller than the detection limit. 

The corresponding stellar mass distribution is shown in Figure \ref{fig:hsc_phys_props_clump_sample1b} and the stellar age distribution of the clump candidates in Figure \ref{fig:hsc_phys_props_clump_sample1c}. The clump candidates with a mass ratio below the separation threshold have very low stellar masses of $M_\star \lesssim 10^{2.5}\,M_\odot$, which are too low to be considered for a star-forming clump. The estimated ages are also much higher for the clumps with a clump-galaxy stellar mass ratio of $\log$($M_{\mathrm{cl}}/M_{\mathrm{gal}}$) $< -6.6$, suggesting that these detections are non-clump detections and likely measurements of the galactic disk environment. 

Figure \ref{fig:hsc_phys_props_clump_sample1} also shows the distribution of clump candidates that are above the threshold of the clump-galaxy stellar mass ratio but below the detection limit in blue dotted lines. These clump candidates have similar stellar masses and ages to the sample above the detection limit but are excluded from the analysis due to the low SNR (Section \ref{sec:clump_phot}).

We excluded all clump candidates with an estimated clump-galaxy stellar mass ratio of $\log$($M_{\mathrm{cl}}/M_{\mathrm{gal}}$) $< -6.6$ as well as all clump candidates detected in host galaxies that are highly inclined or edge-on, i.e. elongation or ratio of semi-major to semi-minor axis greater than $3.0$ which we determined from our remeasurement of the morphological galaxy parameters as described in Paper II (Appendix A).

The detection of clumps becomes more difficult the closer clumps are to the galactic centre. The bright central regions of galaxies reduce the contrast differences required for the FRCNN model to detect clumps in the five or six different filter band images and the detection completeness starts to become much lower for the region that is within a radius of $0.3\,r_{\mathrm{eff}}$ from the central point of a galaxy (see Paper I and Paper II). Here, $r_{\mathrm{eff}}$ is the effective or half-light radius that we measured for each galaxy as described in Paper II (Appendix A). Hence, clump detections that are less than $0.3\,r_{\mathrm{eff}}$ from the galactic centre are also excluded. 

We also excluded clump detections with unrealistically high flux measurements in relation to the total galaxy flux, i.e. when measured clump fluxes are larger than half of the total flux of the host galaxy. In most cases these detections are artefacts or detections from neighbouring bright objects.

For this analysis, only clumps that are detected in the mass-complete sample of galaxies (Section \ref{sec:data}) are used. This further reduces the number of clumps by $\sim$15\% and the number of host galaxies by $\sim$22\%. The majority of the excluded clumps are hosted in galaxies with redshift between $0.32 < z \leq 0.5$. In this redshift range, the FRCNN model detections start to become less complete (see Paper I and Paper II). In effect, the limitation of the analysed clumps to only the mass-complete galaxy sample implies a soft redshift cut and ensures a high completeness of the detections throughout the whole sample.

Therefore, we define star-forming clumps based on the ratio of the clump's stellar mass $M_{\mathrm{cl}}$ to the host galaxy's stellar mass $M_{\mathrm{gal}}$ for this study. The stellar mass ratio of a clump candidate is required to be between $-6.6 \leq \log$($M_{\mathrm{cl}}/M_{\mathrm{gal}}$) $\leq -0.7$ (or $2.51 \times 10^{-7} \leq M_{\mathrm{cl}}/M_{\mathrm{gal}} \leq 0.2$) to be considered as a star-forming clump, without distinguishing between smaller clumps and GSFCs. This definition differs from other authors, who define GSFCs based on a clump-galaxy flux ratio in the restframe UV \citep{Guo2015} or SDSS u-band \citep{Adams2022}, while other studies base their definition on a stellar mass threshold \citep[e.g. $M_{\mathrm{cl}}>10^{7}\,M_\odot$,][]{HuertasCompany2020}. Instead of separating GSFCs from smaller regions of star-formation, our definition based on the clump-galaxy stellar mass ratio will allow for the analysis of star-forming regions that span the mass range between (large) local \textsc{HII} regions and GSFCs.

We list the full breakdown of the number of detected clump candidates after each exclusion step in Table \ref{tab:hsc_phot_final_counts}.

\begin{table}
	\centering
	\caption[Number of clumpy galaxies and clumps with available \textit{ugrizy} and \textit{grizy} photometry.]{Number of clumpy galaxies and clumps with available 6-filter band (\textit{ugrizy}, CLAUDS/HSC-SSP sample) and 5-filter band (\textit{grizy}, HSC-SSP only sample) photometry. The instrumental detection limits are shown in Table \ref{tab:hsc_data_survey_overview}. The mass-complete sample consists of host galaxies with $M_\star \geq 10^9 M_\odot$ and $z \leq 0.32$.}
    \label{tab:hsc_phot_final_counts}
	\footnotesize
        \begin{tabular}{lrrr}
		\hline
        Selection & \multicolumn{1}{l}{Galaxies} & \multicolumn{1}{l}{Clumps} & \multicolumn{1}{l}{Clumps} \\
                  &                              &                            & \multicolumn{1}{l}{per galaxy} \\
		\hline
        \multicolumn{4}{l}{\textbf{CLAUDS/HSC-SSP sample}} \\
        \hline
        Model detections                                   & 7,135 & 30,636 & 4.29 \\
        With valid flux                                    & 7,034 & 28,814 & 4.10 \\
        Above detection limit                              & 6,190 & 19,764 & 3.19 \\
        Galaxy elongation $<3.0$                           & 5,879 & 18,872 & 3.21 \\
        $\log(M_{\mathrm{cl}}/M_{\mathrm{gal}}) \geq -6.6$ & 5,688 & 17,759 & 3.12 \\
        Radial distance $\geq 0.3\,r_{\mathrm{eff}}$       & 5,599 & 17,341 & 3.10 \\
        Realistic \textit{ugrizy} flux                     & 5,515 & 17,122 & 3.10 \\
        Mass-complete sample                               & 4,162 & 14,358 & 3.45 \\
		\hline
        \multicolumn{4}{l}{\textbf{HSC-SSP only sample}} \\
        \hline
        Model detections                                   & 339,719 & 1,423,076 & 4.19 \\
        With valid flux                                    & 335,109 & 1,347,384 & 4.02 \\
        Above detection limit                              & 295,153 &   925,179 & 3.13 \\
        Galaxy elongation $<3.0$                           & 281,345 &   889,103 & 3.16 \\
        $\log(M_{\mathrm{cl}}/M_{\mathrm{gal}}) \geq -6.6$ & 273,568 &   843,101 & 3.08 \\
        Radial distance $\geq 0.3\,r_{\mathrm{eff}}$       & 268,387 &   821,056 & 3.06 \\
        Realistic \textit{grizy} flux                      & 267,561 &   818,635 & 3.05 \\
        Mass-complete sample                               & 207,085 &   696,949 & 3.37 \\
		\hline
        \multicolumn{4}{l}{\textbf{Total}} \\
        \hline
        Model detections                                   & 346,854 & 1,453,712 & 4.19 \\
        With valid flux                                    & 342,143 & 1,376,198 & 4.02 \\
        Above detection limit                              & 301,343 &   944,943 & 3.14 \\
        Galaxy elongation $<3.0$                           & 287,224 &   907,975 & 3.16 \\
        $\log(M_{\mathrm{cl}}/M_{\mathrm{gal}}) \geq -6.6$ & 279,256 &   860,860 & 3.08 \\
        Radial distance $\geq 0.3\,r_{\mathrm{eff}}$       & 273,986 &   838,397 & 3.06 \\
        Realistic (\textit{u)}\textit{grizy} flux          & 273,076 &   835,757 & 3.06 \\
        Mass-complete sample                               & 211,247 &   711,307 & 3.37 \\
        \hline
	\end{tabular}
\end{table}

% -----------------------------------------------
% == Clump Properties
\subsection{Clump properties}\label{sec:hsc_phys_props_clumps}
Investigating the physical properties of the clumps is one of the main goals of this study. In the following sections, we will generally compare the measured and inferred physical properties of the clumps detected in the CLAUDS/HSC-SSP galaxy sample with the clumps from the HSC-SSP only sample, the estimates for the diffuse underlying galaxy background and with our sample of simulated clumps from Paper I.

\begin{figure}
    \centering
    \subfloat[\centering Colour (\textit{u}-\textit{r}). \label{fig:hsc_phys_props_clumps_hist_UR}]{{\includegraphics[width=0.5\columnwidth]{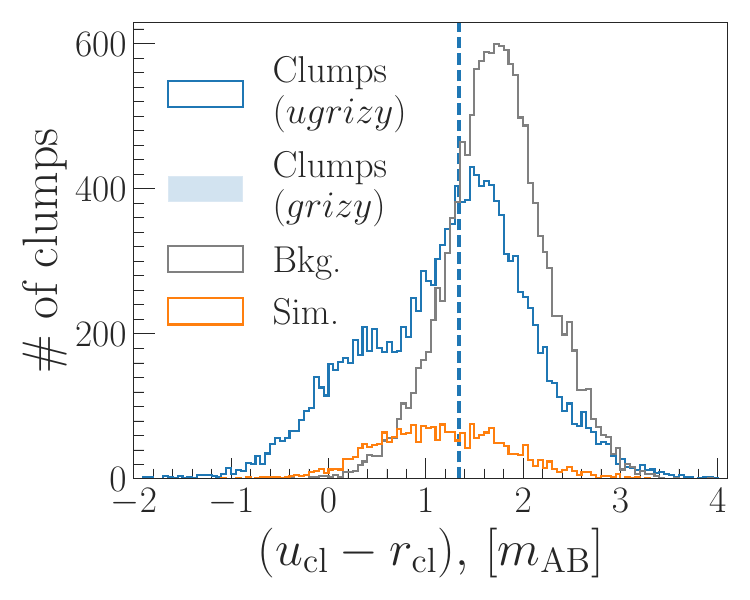} }}
    % \subfloat[\centering Colour (\textit{g}-\textit{r}). \label{fig:hsc_phys_props_clumps_hist_GR}]{{\includegraphics[width=0.5\columnwidth]{figures/sed_colour_GRIZY.pdf} }}
    \subfloat[\centering Stellar mass. \label{fig:hsc_phys_props_clumps_hist_mass}]{{\includegraphics[width=0.5\columnwidth]{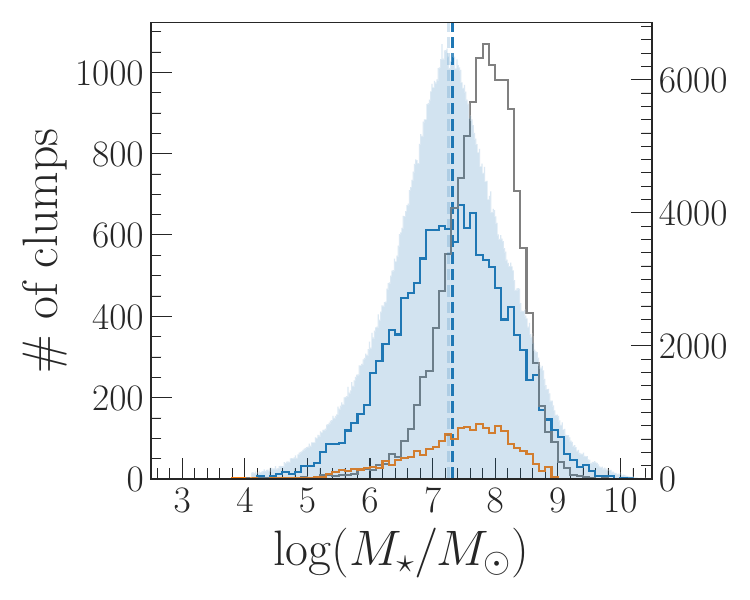} }}
    \\
    \subfloat[\centering Age. \label{fig:hsc_phys_props_clumps_hist_age}]{{\includegraphics[width=0.5\columnwidth]{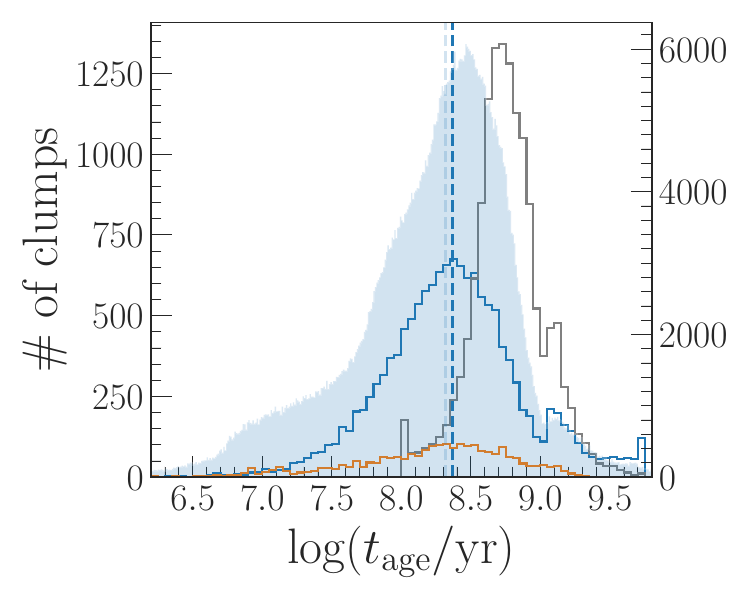} }}
    \subfloat[\centering SFR. \label{fig:hsc_phys_props_clumps_hist_SFR}]{{\includegraphics[width=0.5\columnwidth]{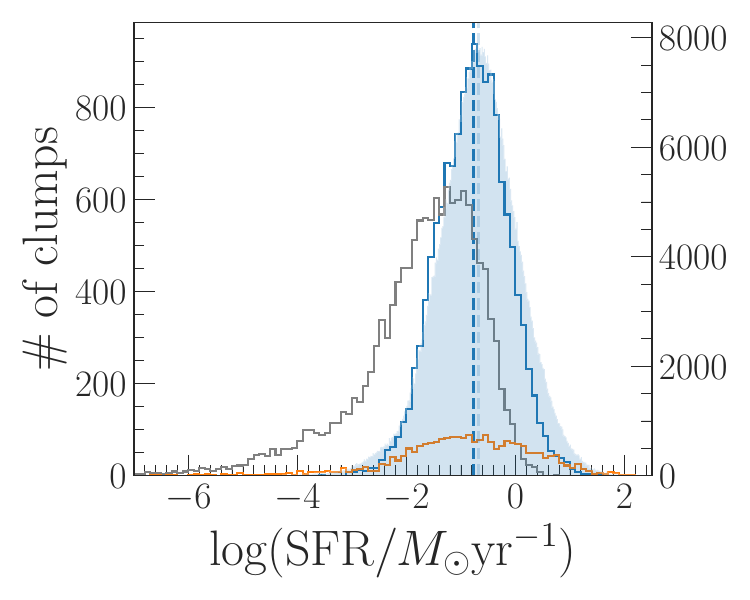} }}
    \\
    \subfloat[\centering sSFR. \label{fig:hsc_phys_props_clumps_hist_sSFR}]{{\includegraphics[width=0.5\columnwidth]{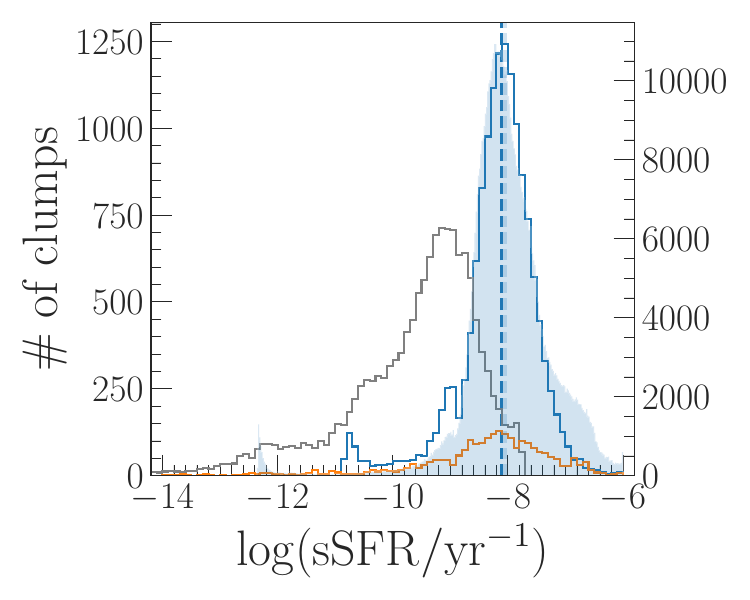} }}
    \subfloat[\centering Metallicity. \label{fig:hsc_phys_props_clumps_hist_metal}]{{\includegraphics[width=0.5\columnwidth]{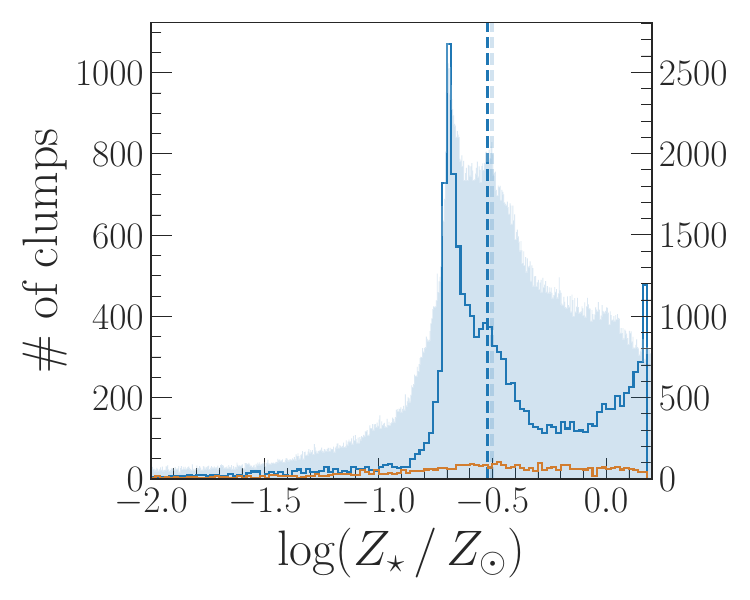} }}
    \\
    \subfloat[\centering Dust attenuation. \label{fig:hsc_phys_props_clumps_hist_dust}]{{\includegraphics[width=0.5\columnwidth]{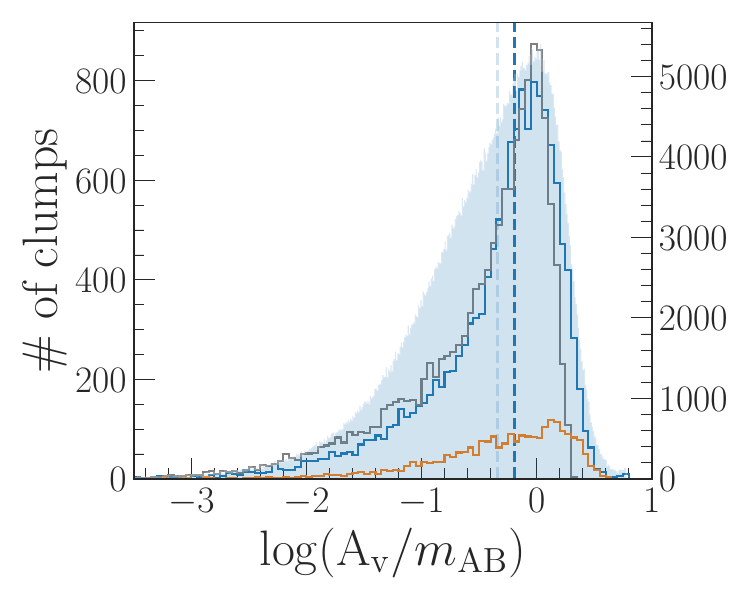} }}
    \caption[Histograms of the main physical properties of the detected clumps.]{Histograms of the main physical properties of the detected clumps: (a) colour (\textit{u}-\textit{r}), (b) stellar mass, (c) age, (d) SFR, (e) sSFR, (f) metallicity and (g) dust attenuation. The distribution of the physical properties of the clumps observed from the CLAUDS/HSC-SSP galaxy sample (\textit{ugrizy} clumps) are shown in blue and the properties of the clumps from the HSC-SSP only galaxy sample (\textit{grizy} clumps) as filled, light blue histograms on the second y-axis. The blue and light blue dashed lines are marking the corresponding medians for each distribution. For comparison, the distributions for the detected simulated clumps and the galaxy background of the clumps are also shown in orange and grey, respectively.}
    \label{fig:hsc_phys_props_clumps_hist}
\end{figure}

% -- Clump (U-R) colour ----------------------------
\subsubsection{Clump (\textit{u}-\textit{r}) colour}\label{sec:hsc_phys_props_colour}
The (\textit{u}-\textit{r}) colour distribution of the clumps is shown in Figure \ref{fig:hsc_phys_props_clumps_hist_UR}. The clumps from the combined CLAUDS/HSC-SSP galaxies are mostly bluer than the underlying galaxy background with a median for the (\textit{u}-\textit{r}) colour index of $1.34$ compared to $1.73$ for the galaxy background. Based on the (\textit{u}-\textit{r}) colour distribution it appears that there are two different clump populations, a bluer population with $(u-r)\lesssim 0.6$ and a redder population with $(u-r)\gtrsim 0.6$. 

\citet{Zanella2019} also found a bimodal colour distribution of their clumps and argue that the redder population consists of larger, more extended clumps, while the bluer population mainly consists of more compact and younger clumps. In our clump sample with \textit{u}-band fluxes, the median stellar mass and age for those clumps with $(u-r)\lesssim 0.6$ is $\log(M_\star/M_\odot) = 6.84$ and $\log(t_{\mathrm{age}}/\mathrm{yr})=8.06$, whereas the redder population ($(u-r)\gtrsim 0.6$) has a median stellar mass of $\log(M_\star/M_\odot) = 7.71$ and median age of $\log(t_{\mathrm{age}}/\mathrm{yr})=8.45$. %The difference in the median stellar masses indicates that the bluer population might be less extended than the redder population, but this needs to be confirmed with measurements of the physical sizes of the clumps. 

% The (\textit{g}-\textit{r}) colour index is closest to the (\textit{u}-\textit{r}) colour for the clumps with only \textit{grizy} photometry. The (\textit{g}-\textit{r}) colours for the \textit{ugrizy} clumps (median of $0.59$) differ from the \textit{grizy} clumps (median of $0.35$). This difference in the medians of the (\textit{g}-\textit{r}) colour distributions is statistically significant (Kolmogorov-Smirnov or KS-test, $p$-value of $1.43\times10^{-71}$) and is likely due to the missing \textit{u}-band data for the detection of the \textit{grizy} clumps. The photometry for each clump was measured at the position of the highest flux from the `bluest' filter band in the bounding box of the FRCNN model output (Section \ref{sec:clump_det}). Here, the `bluest' filter band is the \textit{u}-band for the \textit{ugrizy} clump detections and the \textit{g}-band for the \textit{grizy} clump detections. As the position of the highest flux in the \textit{u}-band does not necessarily coincide with the position of highest flux in the \textit{g}-band, the \textit{ugrizy} aperture may not be centred on the \textit{g}-band peak. The resulting flux loss means that the \textit{grizy} clumps tend to have higher \textit{g}-band fluxes resulting in smaller (\textit{g}-\textit{r}) colour indices.

% -- Clump stellar Mass and Age ----------------------------
\subsubsection{Clump stellar mass and age}\label{sec:hsc_phys_props_mass_age}
The stellar mass $M_\star$ and age $t_{\mathrm{age}}$ distributions of the clumps from both the combined CLAUDS/HSC-SSP and HSC-SSP only galaxy samples are similar but show slightly different medians in Figures \ref{fig:hsc_phys_props_clumps_hist_mass} and \ref{fig:hsc_phys_props_clumps_hist_age}, respectively. The median stellar mass for the \textit{ugrizy} (\textit{grizy}) clumps is $10^{7.32}M_\odot$ ($10^{7.22}M_\odot$) and the median age is $10^{8.37}\,\mathrm{yr}$ ($10^{8.34}\,\mathrm{yr}$). The clumps are significantly less massive and younger than the galaxy background with a median stellar mass of $10^{7.83}M_\odot$ and a median age of $10^{8.75}\,\mathrm{yr}$. The stellar mass and age distributions of the \textit{ugrizy} and \textit{grizy} clumps are similar but significantly different (KS-test, $p$-value of $6.76\times10^{-57}$). This is expected as the SED fitting process used different numbers of photometric data points for the two samples which results in similar but not identical fitting results (see also Appendix \ref{sec:hsc_sed_eval_sed_GRIZY}). %A small population of clumps with ages of $t_{\mathrm{age}} \gtrsim 10^9\,\mathrm{yr}$ is also visible for both samples in Figure \ref{fig:hsc_phys_props_clumps_hist_age}. These `old' clumps tend to be more massive than the younger clumps. Their distributions show median stellar masses of $10^{8.57}M_\odot$ and $10^{8.64}M_\odot$ for the combined CLAUDS/HSC-SSP and HSC-SSP galaxy samples, respectively. These older clumps might have a different origin, i.e. they may be ex situ formed clumps, or they might be more contaminated by older stars that formed outside the clump and have become part of the clump. %The flux measured for these clumps could also be more contaminated by stars of the host galaxy background due to a complex 3-dimensional geometrical setting that is not fully accounted for by the 2-dimensional observational identification method.

The inferred stellar mass distributions of the clumps overlap with the mass ranges of $10^7$-$10^{9.5}M_\odot$ that have been reported for GSFCs in the literature \citep[e.g][]{FoersterSchreiber2011a,DessaugesZavadsky2017,Guo2018,Zanella2019,HuertasCompany2020,Ambachew2022,Kalita2025} but also extend to lower mass ranges of $10^4$ to $10^7 M_\odot$. In the Milky Way and nearby galaxies, star-forming regions with stellar masses in this lower mass range exist mainly as globular star clusters (GCs) and super star clusters (SSCs). However, less massive clumps ($<10^7 M_\odot$) have also been observed in lensed galaxies \citep[e.g.][]{Vanzella2018,Mestric2022,Messa2022,Adamo2023,Claeyssens2023,Messa2025,Giunchi2025}. The findings from these authors and the stellar mass distribution of our clump sample suggest that these less massive clumps are probably GCs or progenitors of GCs and that there is a gradual transition from GSFCs via star-forming clumps and SSCs to proto-GCs.

The relatively low ages of the clumps compared to GCs and the low number of clumps with ages of $>10^9\,\mathrm{yr}$ found in this study indicates that the dissolution time scales for the clumps are less than $10^9\,\mathrm{yr}$. This is also in agreement with findings from clumps in lensed galaxies that span comparable stellar mass and age ranges of $10^5$-$10^9M_\odot$ and $10^6$-$10^9\,\mathrm{yr}$ \citep[][for example]{Claeyssens2023}.

% -- Clump SFR, sSFR, metallicity and dust attenuation ----------------------------
\subsubsection{Clump SFR, sSFR, metallicity and dust attenuation}\label{sec:hsc_phys_props_other}
The clumps from each of the galaxy samples are also very similar with respect to their SFR (Fig. \ref{fig:hsc_phys_props_clumps_hist_SFR}) and sSFR (Fig. \ref{fig:hsc_phys_props_clumps_hist_sSFR}). The median SFR (sSFR) is $10^{-0.76} M_\odot \mathrm{yr}^{-1}$ ($10^{-8.10}\,\mathrm{yr}^{-1}$) for the CLAUDS/HSC sample and $10^{-0.70} M_\odot \mathrm{yr}^{-1}$ ($10^{-8.07}\,\mathrm{yr}^{-1}$) for the HSC sample. This is higher than the median SFR inferred for the galaxy background (median SFR of $10^{-1.57} M_\odot \mathrm{yr}^{-1}$ and median sSFR of $10^{-9.37}\,\mathrm{yr}^{-1}$). However, the sSFR distribution of the clumps from the CLAUDS/HSC-SSP galaxy sample extends to much lower sSFRs than the clumps from the HSC-SSP sample with a small peak seen at $\log$($\mathrm{sSFR}/\mathrm{yr}^{-1}$) $\approx -10.7$.

\begin{figure*}
    \centering
    \includegraphics[width=0.9\textwidth]{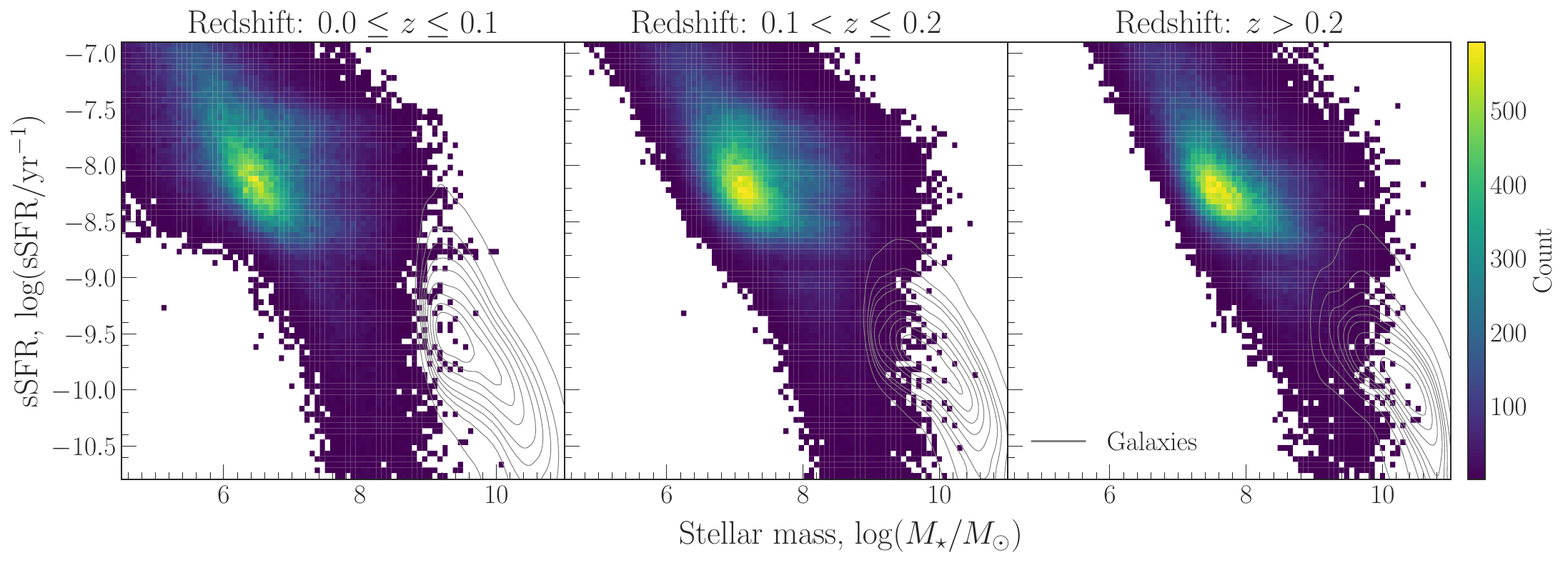}
    \caption[Clump and host galaxy sSFRs as a function of their stellar masses for clumps from the CLAUDS/HSC-SSP and HSC-SSP only galaxy sample and for different redshift bins.]{Clump and host galaxy sSFRs as a function of their stellar masses for clumps from the CLAUDS/HSC-SSP and HSC-SSP only mass-complete galaxy sample and for different redshift bins. The distributions of the clumps from this study are shown in coloured square bins and their host galaxies as grey contours.}
    \label{fig:hsc_phys_props_clump_sample_ssfr_mass}
\end{figure*}

The observed sSFRs of our clumps agree with values found for low-redshift clumps \citep[e.g.][]{Mehta2021} and high-redshift clumps \citep[e.g.][]{Guo2018,Zanella2019} and show that clumps are regions with elevated sSFRs within their host galaxies. Figure \ref{fig:hsc_phys_props_clump_sample_ssfr_mass} shows the 14,358 detected clumps with 6-filter band photometry from the CLAUDS/HSC-SSP galaxy sample in a sSFR/$M_\star$-diagram for redshift ranges $z \leq 0.1$, $0.1 < z \leq 0.2$ and $z > 0.2$. The clumps are star-forming and have elevated sSFRs compared to their host galaxies. Some massive clumps, however, have sSFRs that are similar to their host galaxies which is consistent with results from \citet{Wuyts2012,Guo2012,Guo2018}.

In contrast, the metallicity distributions (Fig. \ref{fig:hsc_phys_props_clumps_hist_metal}) are noticeably different for the clumps from each of the two galaxy samples. The median metallicities are similar (CLAUDS/HSC-SSP: median of $\log(Z_\star/Z_\odot)=-0.52$, HSC-SSP: median of $\log(Z_\star/Z_\odot)=-0.51$), but the shapes of the metallicity distributions are different. The number of clumps from the combined CLAUDS/HSC-SSP galaxy sample has peaks at $\log(Z_\star/Z_\odot)\approx -0.7$, $\log(Z_\star/Z_\odot)\approx -0.5$ and at the upper limit of the prior distribution ($\log(Z_\star/Z_\odot)\approx 0.19$). The metallicity distribution of the clumps from the HSC-SSP also peaks at $\log(Z_\star/Z_\odot)\approx -0.7$ and $\log(Z_\star/Z_\odot)\approx -0.5$ but declines toward the upper limit of the prior distribution, instead. However, metallicities are expected to be poorly constrained when using photometric data from only a few filter bands. Metallicities were not inferred for the galaxy background because a SED model with fixed solar metallicity was used for the background fits.

Figure \ref{fig:hsc_phys_props_clumps_hist_dust} shows the distribution of the clumps as a function of the dust attenuation. The histograms show similar peaks of the distributions at $\log(A_V/m_{\mathrm{AB}})\sim 0.0$ for the clumps from each of the two galaxy samples and also the galaxy background. The median dust attenuation for the CLAUDS/HSC-SSP clumps is $0.64\,m_{\mathrm{AB}}$ and for the HSC-SSP only clumps $0.48\,m_{\mathrm{AB}}$ and both distributions are significantly different (KS-test, $p$-value of $2.34\times10^{-144}$). The median dust attenuation of the galaxy background is $0.51\,m_{\mathrm{AB}}$.

% -----------------------------------------------
% == Clump Contribution to the Host Galaxy's Flux and stellar Mass ----------------------------
\subsection{Clump contribution to the host galaxy's flux and stellar mass}\label{sec:hsc_phys_props_contribution}
The flux from each of the available filter bands is a directly measured property of the detected clumps. The clumps contribute to varying degrees to the total galaxy flux and the ratio of the clump's flux to the galaxy's flux has been used by different authors to separate GSFCs from smaller areas of intense star-formation \citep[e.g.][]{Guo2015,Adams2022}. In a previous study, \citet{Guo2012} found that individual clumps contribute $1$-$10\%$ to the U- and V-band flux of the galaxy (median: $\sim 5\%$) and the combined contribution of all clumps peaks around $\sim 20\%$ in both filter bands. 

\begin{figure}
    \centering
    \subfloat[\centering CLAUDS/HSC-SSP galaxies. \label{fig:hsc_phys_props_contribution_flux_ratio_a}]{{\includegraphics[width=0.5\columnwidth]{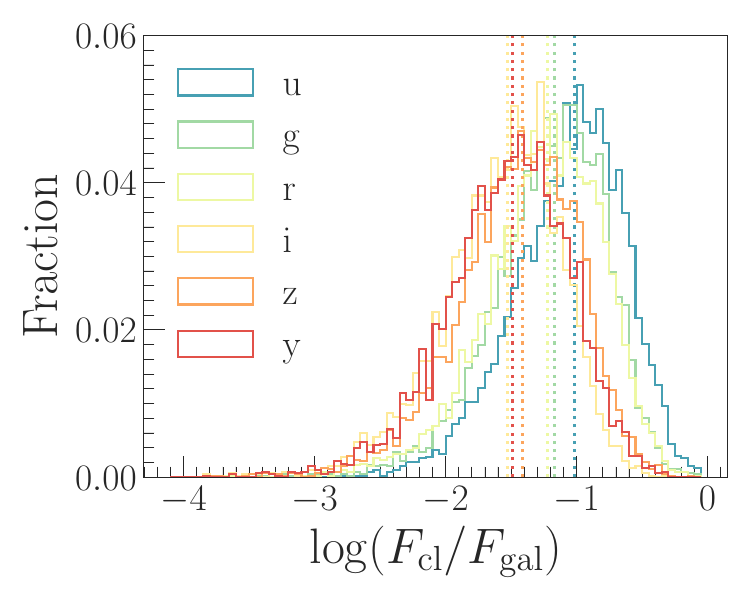} }}
    \subfloat[\centering HSC-SSP only galaxies. \label{fig:hsc_phys_props_contribution_flux_ratio_b}]{{\includegraphics[width=0.5\columnwidth]{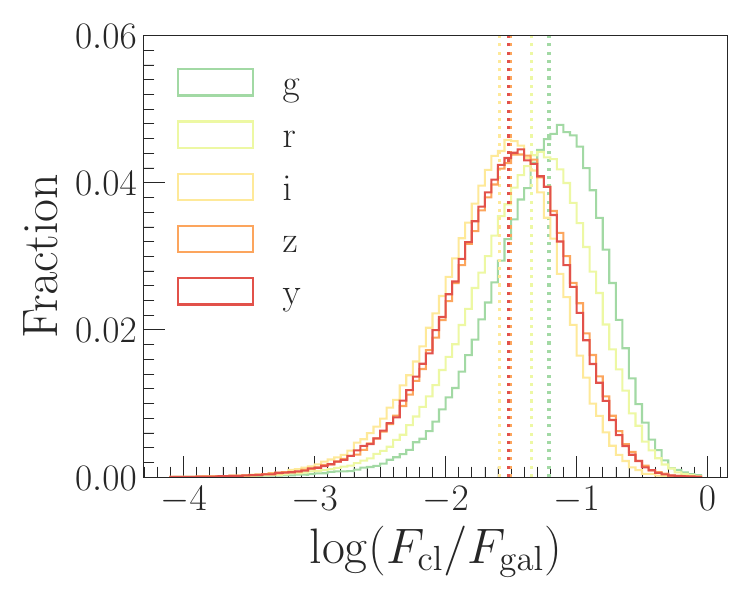} }}
    \caption[Fractional flux contributions of clumps to their host galaxies.]{Fractional flux contributions of clumps to their host galaxies. The histograms show the fraction of galaxies as a function of the ratio of the aggregated flux from all $n$ clumps within a galaxy $F_{\mathrm{cl}}=\sum_{i=1}^{n} F_{i,\mathrm{cl}}$ to the total flux of the galaxy $F_{\mathrm{gal}}$ in each filter band. Plot (a) shows the distributions of clumpy galaxies from the CLAUDS/HSC-SSP galaxy sample and plot (b) the distributions of clumpy galaxies from the sample of HSC-SSP only galaxies. Coloured vertical lines mark the median flux ratios for each filter band.}
    \label{fig:hsc_phys_props_contribution_flux_ratio}
\end{figure}

The distribution of the combined flux contribution for all detected clumps within each galaxy is shown in Figure \ref{fig:hsc_phys_props_contribution_flux_ratio_a} for the CLAUDS/HSC-SSP galaxies and in Figure \ref{fig:hsc_phys_props_contribution_flux_ratio_b} for the larger set of HSC-SSP only galaxies. The histograms show the fraction of galaxies from each sample as a function of the ratio of the aggregated flux from all $n$ clumps within a galaxy $F_{\mathrm{cl}}=\sum_{i=1}^{n} F_{i,\mathrm{cl}}$ to the total flux of the galaxy $F_{\mathrm{gal}}$ in each filter band. Individually, the clumps contribute only between $0.01\%$ and $10$-$20\%$ to the total galaxy flux. For the galaxies from the combined CLAUDS/HSC-SSP sample (HSC-SSP only sample), the combined flux contribution of the clumps to their host galaxies ranges between $0.02\%$ and $40$-$60\%$ with a median ratio of $9.87\%$ in the \textit{u}-band, $6.22\%$ ($6.33\%$) in the \textit{g}-band, $7.03\%$ ($4.65\%$) in the \textit{r}-band, $3.09\%$ ($2.66\%$) in the \textit{i}-band, $3.97\%$ ($3.18\%$) in the \textit{z}-band and $3.33\%$ ($3.12\%$) in the \textit{y}-band. The contribution of clumps to the \textit{u}-band flux of the host galaxy is higher than in the optical to near-infrared filter bands, which is expected from relatively young and star-forming regions within a galaxy.

The clumps from this study generally contribute less to the total galaxy fluxes when compared with the observed clumps from \citet{Guo2012} and peak far below a flux ratio of $\sim 20\%$. This difference is likely to be a selection effect related to different galaxy samples and the clump detection methods that were applied.

\begin{figure}
    \centering
    \subfloat[\centering CLAUDS/HSC-SSP galaxies. \label{fig:hsc_phys_props_contribution_ratio_z_a}]{{\includegraphics[width=0.5\columnwidth]{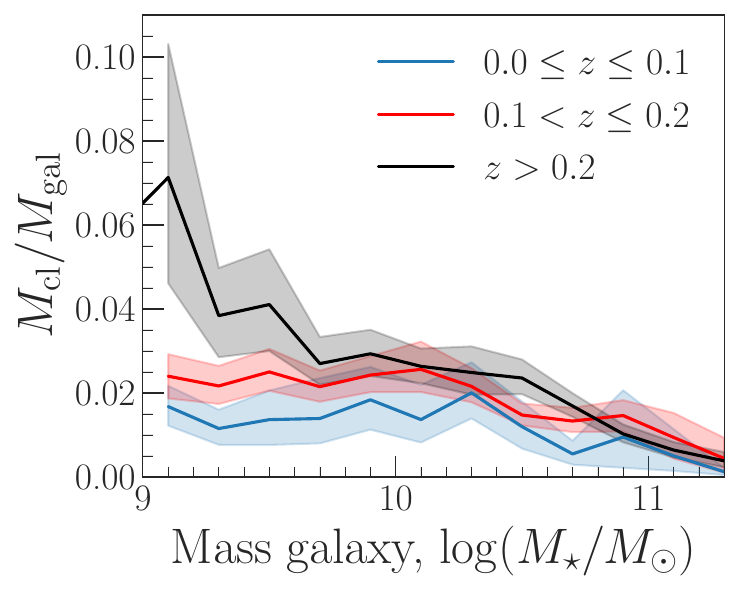} }}
    \subfloat[\centering HSC-SSP only galaxies. \label{fig:hsc_phys_props_contribution_ratio_z_b}]{{\includegraphics[width=0.5\columnwidth]{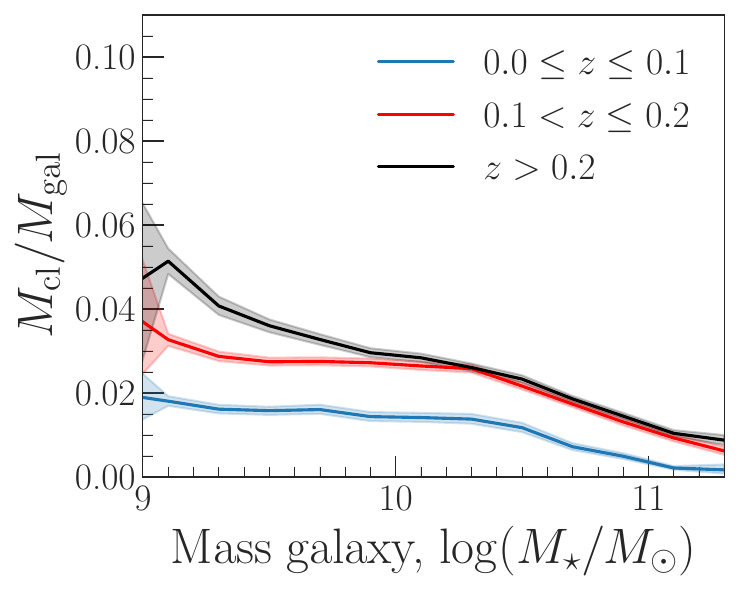} }}
    \caption[Clump-galaxy stellar mass ratio for clumps observed from the CLAUDS and HSC-SSP galaxies per redshift bin.]{Clump-galaxy stellar mass ratio for clumps observed from the CLAUDS and HSC-SSP mass-complete galaxy samples per redshift bin. Plot (a) shows the clumps from the CLAUDS/HSC-SSP galaxy sample and plot (b) the clumps from the sample of HSC-SSP only galaxies. Host galaxies at redshift $z \leq 0.1$ are shown in blue, at redshift $0.1 < z \leq 0.2$ in red and at redshift $z > 0.2$ in black. Shaded areas show the 95\% confidence interval.}
    \label{fig:hsc_phys_props_contribution_ratio_z}
\end{figure}

The contribution of clumps to the total stellar mass of the host galaxy decreases with increasing stellar mass of the galaxy. This has been suggested by previous observations \citep[e.g.][]{HuertasCompany2020} and we also find a correlation between the stellar mass confined to clumps and the total galaxy stellar mass. In Figure \ref{fig:hsc_phys_props_contribution_ratio_z}, we plot the ratio of the aggregated clump stellar mass to the galaxy stellar mass as a function of the total galaxy stellar mass for the combined CLAUDS/HSC-SSP (Fig. \ref{fig:hsc_phys_props_contribution_ratio_z_a}) and the HSC-SSP only galaxy samples (Fig. \ref{fig:hsc_phys_props_contribution_ratio_z_b}). The ratios are also plotted for different redshift bins, which shows that the stellar mass ratios are increased for galaxies at redshift $z > 0.1$.

The median fractional stellar mass confined to clumps declines from $\sim$2-6\% for galaxies with $M_\star = 10^9 M_\odot$ to $\sim$1.5\% for galaxies with $M_\star = 10^{11} M_\odot$. These values are in the range of findings from simulations of high-redshift galaxies \citep{Mandelker2016}. However, the stellar mass contribution by clumps is predicted by the simulations to increase with the stellar mass of the host galaxy, which is opposite to what we observe. The stellar mass ratios we observe are also lower than the average values of $10$-$15\%$ (galaxies with $M_\star \sim 10^9 M_\odot$) to $2$-$5\%$ (galaxies with $M_\star \gtrsim 10^{10} M_\odot$) reported from high-redshift galaxies from \citet{HuertasCompany2020}, even though those authors have also included non-clumpy SFGs in their sample to determine the fraction of stellar mass in clumps over all galaxies. Other authors have reported even higher mass fractions of $10$-$20\%$ \citep{FoersterSchreiber2011a,Zanella2019}, although only for samples of less than 50 galaxies.

\begin{figure}
    \centering
    \includegraphics[width=0.9\columnwidth]{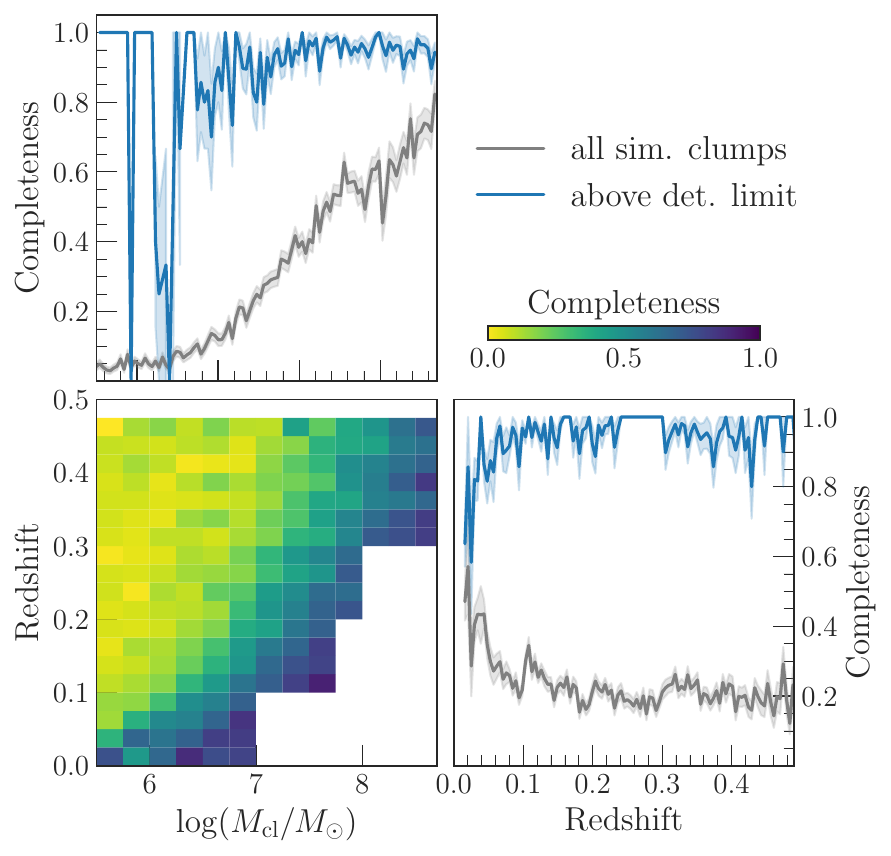}
    \caption[Detection completeness of the simulated clumps as a function of redshift and clump stellar mass.]{Detection completeness measured on the sample of simulated clumps as a function of redshift and clump stellar mass. The completeness over the range of the single variables are shown by the line plots in blue for detected simulated clumps with measured fluxes above the average 5$\sigma$ point-source depth limits of the filter bands and in grey without any flux cuts. The 1$\sigma$ errors are shown as shaded areas. The 2-dimensional histogram shows the completeness distribution for all detected simulated clumps without any flux cuts. The stellar masses of the simulated clumps were cut at an upper limit depending on the redshift of the host galaxy to specifically validate completeness limits extending into low stellar mass ranges.}
    \label{fig:sim_clumps_det_completeness_mass_z}
\end{figure}

However, even though the detection completeness for clumps that have measured fluxes above the average $5\sigma$ point-source depth limits of the filter bands (Table \ref{tab:hsc_data_survey_overview}) is high (see Paper I), the completeness of detected clumps that are below this magnitude cut decreases with decreasing stellar mass of the clumps (Fig. \ref{fig:sim_clumps_det_completeness_mass_z}). The completeness of our detections also decreases with increasing redshift of the host galaxies for those clumps. Therefore, the fractional stellar mass confined to clumps, which we measured from the magnitude-limited clump sample, represents only a lower limit of the true contribution of clumps to the total stellar mass of the host galaxy.

% -----------------------------------------------
% == Clump stellar Mass Function -------------------------------
\subsection{Clump stellar mass function}\label{sec:hsc_phys_props_cSMF}
The clump stellar mass function (cSMF) can help to understand the processes that led to the formation of clumps. For example, the shape and slope of the cSMF, if similar to the observed SMFs for stellar clusters in the Milky Way, can point towards turbulent and unstable gas dynamics within the disk of the host galaxy that lead to the formation of clumps \citep[e.g.][]{Elmegreen2006,DessaugesZavadsky2018,Elmegreen2018}.

\begin{figure}
    \centering
    \includegraphics[width=0.9\columnwidth]{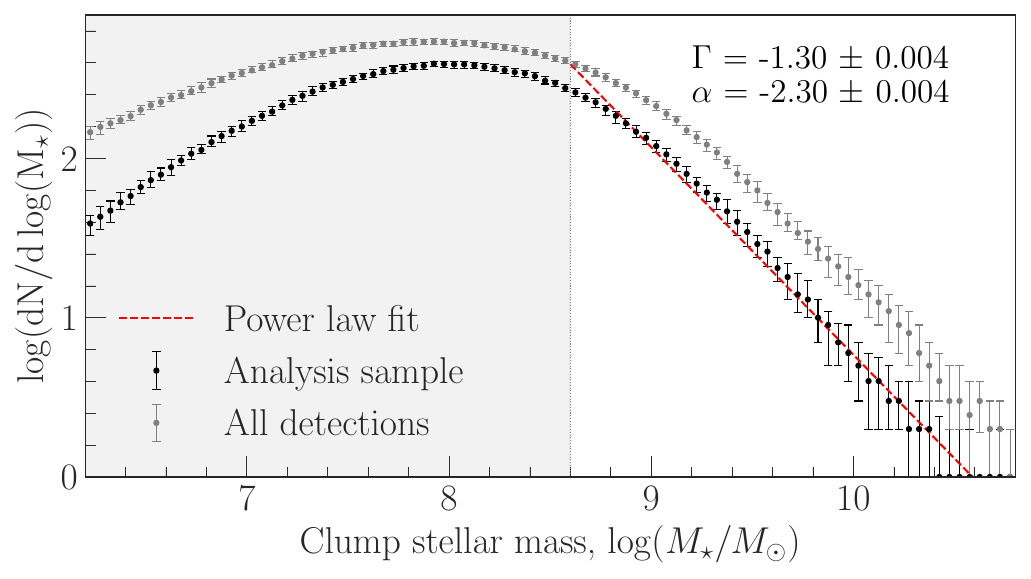}
    \caption[Clump stellar mass function.]{Clump stellar mass function (cSMF) for the clumps from the combined CLAUDS/HSC-SSP galaxy sample. The clump sample that was selected for the main analysis is shown in black and the clumps without any selection cuts are shown in grey. Error bars mark the 16 and 84\% credible region determined from 100 samples of the stellar mass posterior distribution of each clump. Also shown is a best power-law fit (red dashed line) of the form $\log(\mathrm{d}N/\mathrm{d}\log(M_\star)) = \Gamma \log(M_\star)+c$ for clumps with $\log(M_\star/M_\odot) \geq 8.6$. The fitted value for $\Gamma = \alpha + 1$ is shown as annotation in the plot. The grey shaded area for clump stellar masses lower than $\log(M_\star/M_\odot)<8.6$ separates the region for which the FRCNN model completeness drops below $0.8$ and the data was excluded from the power-law model fit.}
    \label{fig:hsc_phys_props_cSMF}
\end{figure}

As pointed out by \citet{DessaugesZavadsky2017,DessaugesZavadsky2018,HuertasCompany2020}, the completeness of the clump sample and the reliability of the derived stellar masses are crucial for the interpretation of the empirical cSMF. The sensitivity and spatial resolution of the imaging data affects the detection of clumps and the measurements of their physical properties. To ensure a reasonable mass-completeness when constructing the cSMF given the completeness of our clump detection model (Fig. \ref{fig:sim_clumps_det_completeness_mass_z}), we used only clump detections with $M_\star \geq 10^{8.6} M_\odot$ from the mass-complete CLAUDS/HSC-SSP galaxy sample (analysis clump sample) and show the cSMF and a power-law model fit in Figure \ref{fig:hsc_phys_props_cSMF}. However, we also extended the cSMF to lower stellar mass bins to illustrate the shape of the cSMF for the observed clump sample even though the detections are incomplete. Furthermore, we also plot a cSMF constructed from all detections without any clump or host galaxy selection cuts applied (all detections sample) for comparison. The 16-84\% credible intervals around the observed cSMFs were determined from 100 samples of the posterior distribution for the stellar mass of each clump.

Figure \ref{fig:hsc_phys_props_cSMF} shows that the selection cuts made for the analysis clump sample significantly reduce the relative number of low-mass clumps. The higher number of clumps in each mass bin from the unrestricted sample of all detections does not affect the high-mass tail of the slope of the cSMF. The cSMFs constructed from the analysis and unrestricted samples both peak at clump stellar masses of $\sim 10^8 M_\odot$ and decline for clumps with lower masses. A similar peak in the cSMF was also reported by \citet{HuertasCompany2020} based on observations of CANDELS galaxies \citep[The Cosmic Assembly Near-infrared Deep Extragalactic Legacy Survey,][]{Grogin2011}. However, those authors validated their clump detections against detections made on mock galaxy images generated using radiative transfer modelling for simulated galaxies \citep[CANDELS-like images from the same simulation analysed by][]{Mandelker2016,HuertasCompany2018} and also against the raw 3-dimensional simulation output. They show that the flattening of the cSMF observed for real and simulated clumps is due to the incompleteness of the detections. The cSMF that is derived directly from the 3-dimensional galaxy simulation output shows a power-law increase at lower masses without any flattening.

The incompleteness of our clump detections is very likely the cause for the flattening of the observed cSMF towards lower masses in Figure \ref{fig:hsc_phys_props_cSMF}. Therefore, we limit the clump sample to contain only clumps with $M_\star \geq 10^{8.6} M_\odot$ for the recovery of the intrinsic cSMF. We fitted a linear function (as stated in Equation \ref{eq:hsc_phys_props_cSMF_linear}, Appendix \ref{sec:cSMF_tables}) to the clumps with stellar masses between $[10^{8.6}, 10^{10.5}]\,M_\odot$ and estimated the power-law exponents $\alpha$ of the cSMF.

The best power-law model fit to the observed cSMF of our clump sample has $\alpha = -2.30 \pm 0.004$ (see also Table \ref{tab:hsc_phys_props_cSMF_clump_age}). This agrees to some degree with values of $\alpha \simeq -2.0$ reported by \citet{DessaugesZavadsky2018,Kalita2024} and is similar to the power-law slopes determined by \citet{Sok2026} from clumps in high-redshift galaxies at $0.5 < z < 5.0$ ($\alpha \sim -1.6$ to $-2.3$). However, values for $\alpha$ determined from simulations by \citet[][$\alpha = -1.35 \pm 0.15$]{HuertasCompany2020} and observations by \citet[][$\alpha = -1.40 \pm 0.07$]{Ambachew2022} differ significantly from our observations. Even larger values for $\alpha$ have been determined by \citet{HuertasCompany2020} from the observed CANDELS galaxies and by \citet{Kalita2025} from IR-clumps in high-redshift galaxies ($\alpha = -0.61 \pm 0.17$ and $\alpha=-0.45\pm 0.11$, respectively). The top-light cSMF observed for low-redshift clumps in this study includes fewer massive clumps like those that are preferentially found in studies of high-redshift galaxies. This is somewhat expected as star-formation is more intense in galaxies at higher redshift. 

\begin{figure}
    \centering
    \includegraphics[width=0.9\columnwidth]{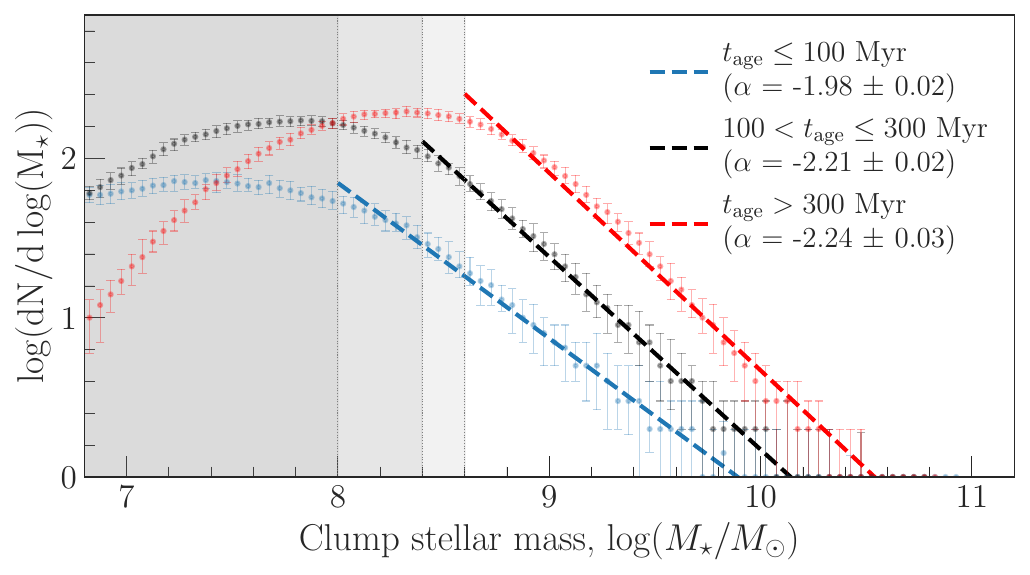}
    \caption[Clump stellar mass function for different clump stellar age bins.]{Similar to Fig. \ref{fig:hsc_phys_props_cSMF} but showing the cSMF for clumps with estimated ages of $t_{\mathrm{age}}\leq 100\,\mathrm{Myr}$, $100 < t_{\mathrm{age}}\leq 300\,\mathrm{Myr}$ and $t_{\mathrm{age}}>300\,\mathrm{Myr}$ from the mass-complete CLAUDS/HSC-SSP galaxy sample. The grey shaded areas for clump stellar masses lower than $\log(M_\star/M_\odot)< \{8.0, 8.4, 8.6\}$ separates the regions that were excluded from the power-law model fits.}
    \label{fig:hsc_phys_props_cSMF_clump_age}
\end{figure}

The observed cSMF includes clumps of different ages and is probably not representative of the initial cSMF as clumps may have been dissolved by stellar feedback or accumulated more stellar mass over time. For example, \citet{Fensch2021} showed in their galactic disk simulations that, depending on the galactic gas fraction, low-mass clumps are likely to be destroyed by stellar feedback within $<100\,\mathrm{Myr}$. To achieve a more realistic estimate for the initial cSMF, we separated the clump sample into three bins depending on the estimated stellar ages of the clumps: (1) clumps with estimated ages of $t_{\mathrm{age}}\leq 100\,\mathrm{Myr}$, (2) $100 < t_{\mathrm{age}}\leq 300\,\mathrm{Myr}$ and (3) $t_{\mathrm{age}}>300\,\mathrm{Myr}$. Figure \ref{fig:hsc_phys_props_cSMF_clump_age} shows the cSMF for each of the clump stellar age bins. While the cSMF slope of clumps that are younger than 100 Myr is close to the value of the SMF slope of stellar clusters, the older clump samples have steeper estimated slopes. The clump population appears to become enriched in medium- and high-mass clumps ($M_{\mathrm{cl}} > 10^8\,M_\odot$) and is shifted towards the more massive bins as the clumps age. A possible depletion in low-mass clumps is hinted at by the relative drop of the cSMF for stellar masses of $\lesssim 10^8\,M_\odot$, although we note that the clump detections are not complete for this stellar mass range so this may just be a selection effect.

We show a full list of all our estimated slopes $\alpha$ in Table \ref{tab:hsc_phys_props_cSMF_clump_age} in Appendix \ref{sec:cSMF_tables}, which also includes estimates of $\alpha$ for different splits of the host galaxy sample.

For example, the cSMFs for SFGs and non-SFGs differ substantially. Here, we separate SFGs from non-SFGs using a sSFR threshold such that galaxies with $\log$($\mathrm{sSFR/\mathrm{yr}^{-1}}$) $\geq-11.0$ are considered to be star-forming and are otherwise defined to be non star-forming \citep[e.g.][]{McGee2011,Wetzel2013}. The slope of the cSMF is less steep for clumps detected in non-SFGs and the value of $\alpha=-2.09\pm 0.01$ indicates that the clumps are on average more massive than the clumps detected in SFGs ($\alpha=-2.37\pm 0.01$). This also suggests that the clumps found in non-SFGs possibly include ex situ clumps that are expected to be more massive than in situ clumps \citep{Mandelker2014}.

If the host galaxies are split into different redshift and stellar mass bins, $\alpha$ increases with increasing galaxy stellar mass but decreases (in most cases) with increasing redshift (Table \ref{tab:hsc_phys_props_cSMF_clump_age}). Similar changes of the cSMF slopes that are dependent on the redshift and stellar mass of the host galaxy have been reported by \citet{Sok2026} from galaxies observed using JWST at redshifts $0.5 < z < 5.0$. 

The increase of $\alpha$ for more massive galaxies in our sample is likely linked to the clump-galaxy stellar mass ratio. Physically, because massive clumps contribute a significant fraction to the total stellar mass of the galaxy, very massive clumps might be limited to more massive galaxies where gas reservoirs are large enough to support their formation. %The steepening of the cSMF slope with redshift is a weaker effect than the flattening of the slope for more massive galaxies and could be related to the lower mass-completeness of the clump sample and the blending of otherwise separate clumps into larger clumps due to the lower spatial resolution at higher redshift. %The observed cSMF peaks at $\log(M_\star/M_\odot) \sim 7.2$ for clumps in galaxies at redshift $z \leq 0.1$, at $\log(M_\star/M_\odot) \sim 8.0$ for redshift $0.1 < z \leq 0.2$ and at $\log(M_\star/M_\odot) \sim 8.5$ for $z > 0.2$. This roughly traces the completeness limit of the cSMF and, while the upper limit is fixed for a given stellar mass of the host galaxy, leads to a steepening of the estimated cSMF slope.

\subsection{Radial gradients}\label{sec:clump_gradients}
In the following sections we describe the radial gradients that quantify the variations of the physical properties of the clumps as a function of the distance $\mathrm{d}_{\mathrm{cl}}$ from the galactic centre which is normalised by our measured effective radius $r_{\mathrm{eff}}$ of each galaxy as described in Paper II (Appendix A). The distribution of each property is plotted as a 2-dimensional histogram where darker colours indicate a higher number count. The median values of the clump properties in each radial distance bin are shown as a line with the standard error of the median indicated by shaded areas. The plots also show the median values of the diffuse galactic background estimated for each clump and the median values of the detected simulated clumps described in Paper I. These are used to compare the measurements of the real clumps to the randomly distributed set of simulated clumps and to the galactic environment the clumps are embedded in. Neither the simulated clumps, nor the galaxy background are expected to show strong radial gradients. 

To estimate the slope of the gradients, we fitted linear models to the distributions as a function of the normalised radial distance $\mathrm{d}_{\mathrm{cl}}/r_{\mathrm{eff}}$ for the clumps, the underlying galaxy background and the simulated clumps. We list the parameters of all linear models fitted to different splits of the host galaxy sample in Appendix \ref{sec:grad_tables}.

\subsubsection{Colour gradient}\label{sec:hsc_phys_props_gradient_colour}
If clumps that have been formed in the outer areas of the host galaxy migrate inwards and grow and evolve as they migrate, then the stellar populations within clumps should become older and more massive closer to the centre of their host galaxy. A directly measurable indicator of the clump evolution is the colour index. Older clumps are expected to appear redder than younger ones as the number of bright and short-lived blue stars decreases with increasing age of the stellar population.

\begin{figure}
    \centering
    \includegraphics[width=1.0\columnwidth]{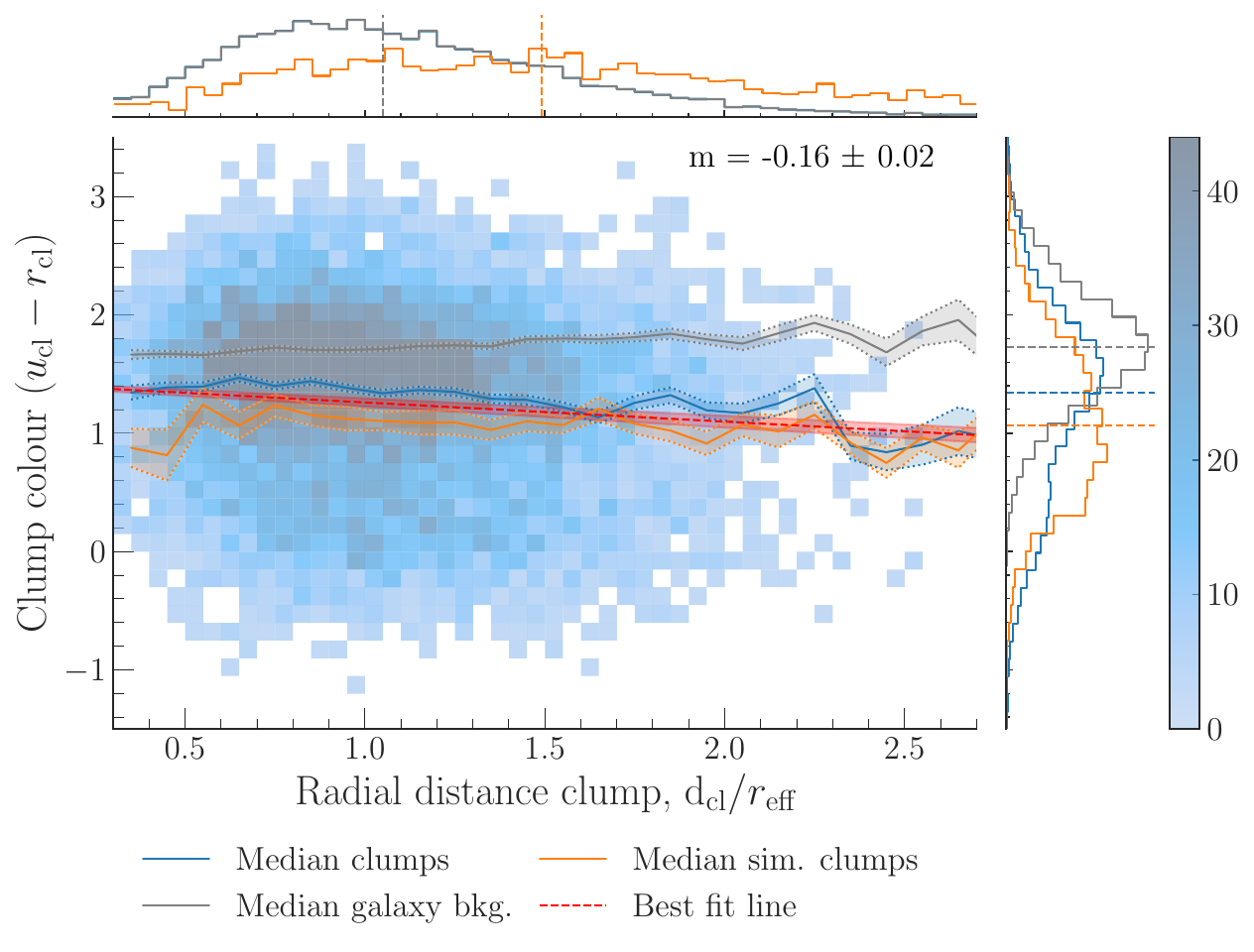}
    \caption[Clump (\textit{u}-\textit{r}) colour distribution as a function of the normalised galactocentric distance.]{Clump (\textit{u}-\textit{r}) colour distribution as a function of the normalised galactocentric distance for clumps observed from the CLAUDS/HSC-SSP galaxies (mass complete sample with $M_{\star,\mathrm{gal}}\geq 10^9 M_\odot$ and $z\leq0.32$). The radial distance $\mathrm{d}_{\mathrm{cl}}$ is normalised by the effective radius $r_{\mathrm{eff}}$ measured for each galaxy. The distribution and median of the observed clumps are shown in blue. The colour bar indicates the specific counts of clumps in each bin. For comparison, the median (\textit{u}-\textit{r}) colour of the underlying galaxy background (grey) and of the detected simulated clumps (orange) are also shown. The univariate distributions are shown separately on top of each axis, where vertical and horizontal lines mark the median for each sample. Shaded errors show the standard error of the medians. A linear model is fitted to the distribution of the observed clumps and shown as a red line with the standard error indicated by the red shaded area. The gradient $m$ of the linear model is shown as annotation in the plot.}
    \label{fig:hsc_phys_props_gradient_UR_reff}
\end{figure}

The median (\textit{u}-\textit{r}) colours of our sample decrease from $\sim 1.4\,m_{\mathrm{AB}}$ at a distance of $\mathrm{d}_{\mathrm{cl}} \approx 0.5\,r_{\mathrm{eff}}$ to $\sim 1.1\,m_{\mathrm{AB}}$ at $\mathrm{d}_{\mathrm{cl}} \approx 1.7\,r_{\mathrm{eff}}$ and then fluctuate around $\sim 1.0\,m_{\mathrm{AB}}$ at larger distances from the galactic centre (Fig. \ref{fig:hsc_phys_props_gradient_UR_reff}). A fitted linear model of the form $(u-r) = m (\mathrm{d}_{\mathrm{cl}}/r_{\mathrm{eff}})+c$ gives $m=-0.16 \pm 0.02$, indicating that clumps further away from the galactic centre are bluer than clumps closer to the galactic centre. The slope $m$ of the fitted model is significantly different from a model with a slope of $m=0.0$ as indicated by the $p$-value in Table \ref{tab:hsc_phys_props_gradient_UR}. In comparison, the galaxy background (\textit{u}-\textit{r}) colour index is almost flat and a fitted linear model only yields a small positive slope indicating that the intra-clump regions of the host galaxy are becoming slightly redder with radial distance, i.e. the opposite to the clumps. Similar colour gradients of the galaxy light have been observed from \citet{Liu2016,Tacchella2018,Wang2017} for SFGs at redshift $z \sim 0.5 -2.0$ as a result of increasing dust attenuation with increasing distance from the galactic centre. The (\textit{u}-\textit{r}) colour gradient for the simulated clump is slightly negative and also significantly different from a linear model with a slope of $m=0.0$, but the $p$-value is much higher than for the real clumps, which also have a steeper slope (Table \ref{tab:hsc_phys_props_gradient_UR}).

The colour gradient of the real clumps is even steeper in galaxies at redshift $z \leq 0.1$ ($m=-0.19$) but becomes flatter with increasing redshift ($m=-0.17$ and $m=-0.12$ for $0.1<z\leq0.2$ and $z>0.2$, respectively, Table \ref{tab:hsc_phys_props_gradient_UR}). As the spatial resolution is decreasing with increasing redshift, the clump flux measurements become more contaminated by the intra-clump galaxy light and overall the clumps are slightly shifted to redder colours. However, the gradient is still significant and the colour gradients are steeper than those that are estimated for the galaxy background light and the simulated clumps. The gradients of the simulated clumps are not significantly different from a flat gradient when observed in the different redshift bins.

The colour gradient is also steeper for clumps that are detected in more massive galaxies (Table \ref{tab:hsc_phys_props_gradient_UR}). For example, the gradient is $m=-0.18$ in low-mass galaxies ($9.0 \leq \log$($M_{\mathrm{gal}}/M_\odot$) $< 9.8$), which is half as steep as in high-mass galaxies ($\log$($M_{\mathrm{gal}}/M_\odot$) $\geq 10.6$) with an estimated slope of $m=-0.37$. A similar observation was made by \citet{Guo2018}, although the authors estimated less steep slopes for their sample of clumps from high-redshift galaxies. 

The galaxy background light and the simulated clumps show close-to-flat gradients in the two lower-mass galaxy bins. Only for the high-mass galaxies do the simulated clumps show a similar negative colour gradient to the real clumps. It is unclear why the colour gradient is significantly steeper for simulated clumps that are detected in high-mass galaxies, while the best fit models found in the other two mass bins have slopes that are not significantly different from 0.0.

The FRCNN models not only detected clumps in SFGs but also in non-SFGs. The clumps in the non-SFGs tend to be redder (median $(u-r)= 1.84\,m_{\mathrm{AB}}$) and also tend to be further from the galactic centre (median $\mathrm{d}_{\mathrm{cl}}/r_{\mathrm{eff}} = 1.27$) than the clumps from the SFGs (median $(u-r)= 1.30\,m_{\mathrm{AB}}$ and median $\mathrm{d}_{\mathrm{cl}}/r_{\mathrm{eff}} = 1.02$). The gradient of the (\textit{u}-\textit{r}) colour as a function of radial distance is also significantly steeper (Welch's T-test: $p\text{-value}= 0.006$) for the clumps in non-SFGs than for those in SFGs, even though the 95\% confidence intervals of the gradient slopes are close to each other (Table \ref{tab:hsc_phys_props_gradient_UR}). Furthermore, the clumps from non-SFGs are more comparable to the galactic background light in regards to (\textit{u}-\textit{r}) colour and clumps that are within $d_{\mathrm{cl}} \lesssim r_{\mathrm{eff}}$ of the host galaxy tend to have similar colours to their environment. This suggests that non-SFGs might contain a higher fraction of clumps that have formed and evolved differently (i.e. ex situ) to the clumps found in SFGs.

To compare the colour gradients of clumps from the CLAUDS/HSC-SSP galaxy sample and from the larger HSC-SSP only sample, we used the (\textit{g}-\textit{r}) colour as it is closest to the (\textit{u}-\textit{r}) colour. The estimated gradients of the (\textit{g}-\textit{r}) colours are generally shallower than the (\textit{u}-\textit{r}) gradients for the real clump samples as well as for the splits by redshift and galaxy stellar mass bins (Table \ref{tab:hsc_phys_props_gradient_GR}). For all real clumps, the (\textit{g}-\textit{r}) colour decreases slightly with distance from the galactic centre with `bluer' clumps further away from the centre and becomes steeper with redshift and/or stellar mass of the host galaxy. The gradients are significantly steeper than the gradients estimated for the galaxy background and the simulated clumps, both of which show a colour distribution that is unrelated to the radial distance of the simulated clump or the position for which the galaxy background was measured.

% -- Stellar Mass Gradient -------------------------------
\subsubsection{Stellar mass gradient}\label{sec:hsc_phys_props_gradient_mass}
If star-forming clumps live long enough to spiral in towards the galactic centre and they continue to accrete matter from the surrounding disk, they should grow in stellar mass. Under this assumption, the stellar mass distribution of clumps should follow a negative gradient as a function of distance from the host galaxy centre such that clumps further from the galactic centre have lower stellar masses than clumps closer to the centre. This has been predicted from analytical galaxy models \citep{Dekel2022}, cosmological simulations \citep{Mandelker2014,Mandelker2016} and supported by observations at high redshift \citep[e.g.][]{Guo2018}.

\begin{figure}
    \centering
    \subfloat[\centering CLAUDS/HSC-SSP galaxies. \label{fig:hsc_phys_props_gradient_mass_reff_a}]{{\includegraphics[width=1\columnwidth]{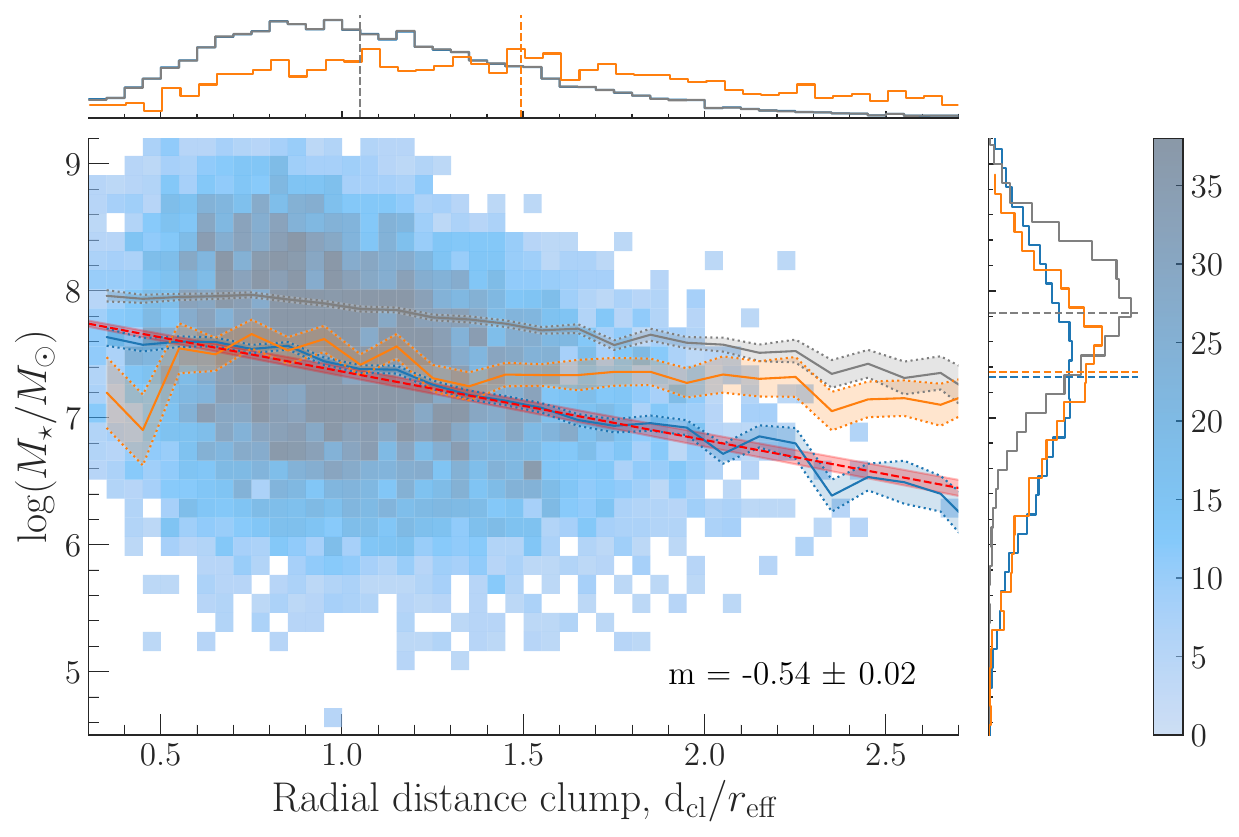} }}
    \\
    \subfloat[\centering HSC-SSP only galaxies. \label{fig:hsc_phys_props_gradient_mass_reff_b}]{{\includegraphics[width=1\columnwidth]{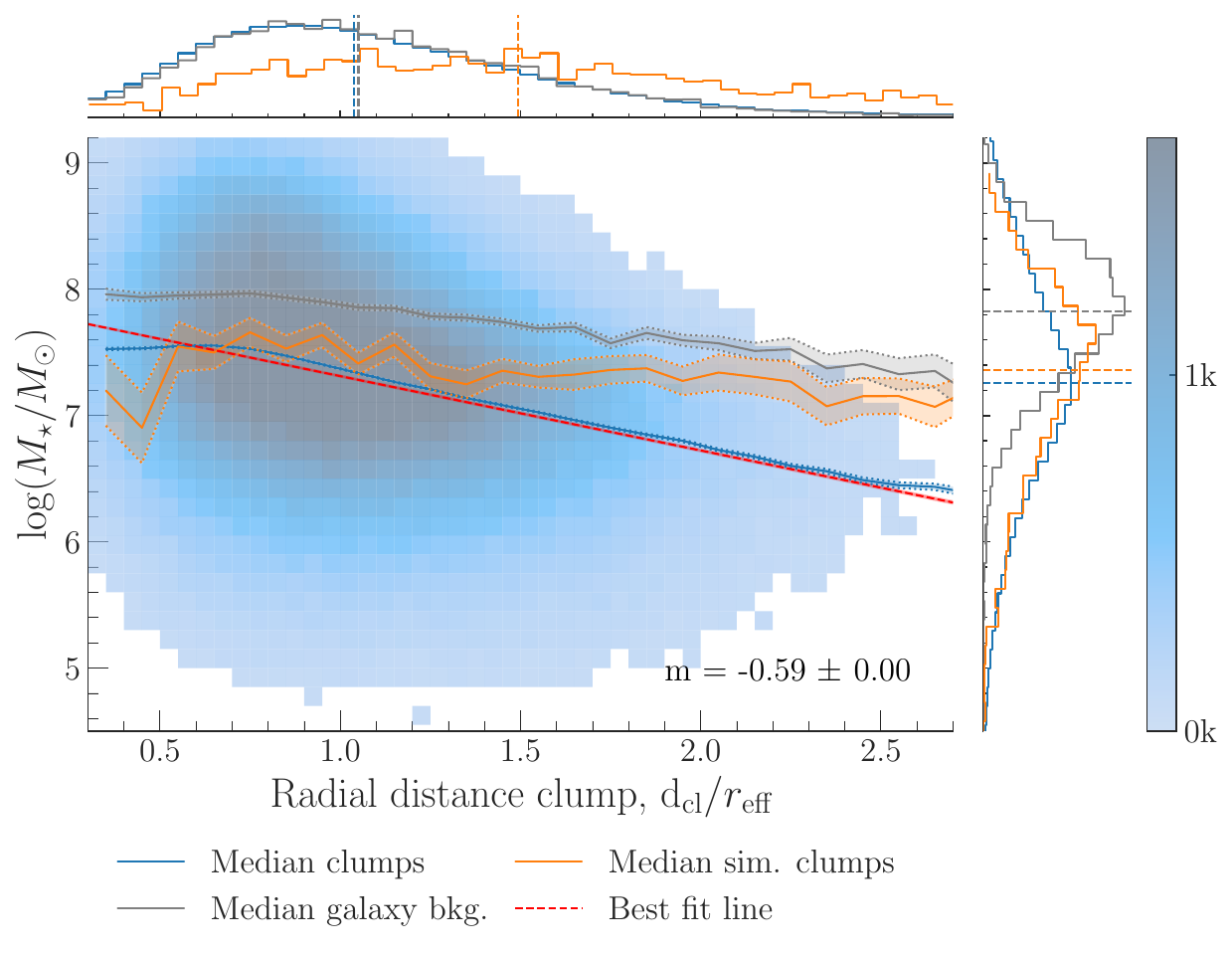} }}
    \caption[Clump stellar mass distribution as a function of the normalised galactocentric distance.]{Similar to Fig. \ref{fig:hsc_phys_props_gradient_UR_reff} but showing the clump stellar mass distribution as a function of the normalised galactocentric distance for the combined CLAUDS/HSC-SSP galaxy sample (a) and the HSC-SSP only galaxy sample (b).}
    \label{fig:hsc_phys_props_gradient_mass_reff}
\end{figure}

In Figure \ref{fig:hsc_phys_props_gradient_mass_reff}, we show the observed stellar mass distribution as a function of the galactocentric distance for the clumps from the CLAUDS/HSC-SSP and the larger HSC-SSP only galaxy sample. We also fitted a linear model of the form $\log(M_{\mathrm{cl}}/M_\odot) = m (\mathrm{d}_{\mathrm{cl}}/r_{\mathrm{eff}})+c$ to the real clump distribution. The stellar masses of the observed clumps are generally lower further away from the galactic centre. The medians of the distributions and the slopes of the fitted linear models are clearly decreasing with increasing radial distance for clumps from both of the galaxy samples. The stellar masses of both samples are similarly distributed as a function of the radial distance and the gradients have similar values of $m=-0.54$ and $m=-0.59$ for the clumps from the CLAUDS/HSC-SSP and the larger HSC-SSP galaxy sample, respectively. The model fits are also significantly better than a model with no gradient or a slope of $m=0.0$ (Table \ref{tab:hsc_phys_props_gradient_mass}). The inferred stellar masses of the galaxy background and the detected simulated clumps also decrease with the distance from the galactic centre but the gradients are less steep compared to the real clumps. A decreasing stellar mass gradient is expected for the galaxy background as the surface mass density of the host galaxy also generally decreases with distance from the centre. 

However, the stellar mass gradient of the simulated clumps is not expected to be dependent on the radial distance as their positions and physical properties have been randomly assigned. The significant negative gradient of the simulated clumps (Table \ref{tab:hsc_phys_props_gradient_mass}) appears to be caused by a decreasing detection completeness of clumps with lower masses closer to the galactic centre. As the overall brightness of the galaxy increases towards the centre, lower-mass clumps are likely to blend with the surroundings they are embedded in and are more difficult to detect using the FRCNN models (see Paper I). Hence, the estimated stellar mass gradients of the real clumps are considered to be a lower limit for the true gradients, i.e. the true gradient could be shallower (less negative).

\begin{figure*}
    \centering
    \includegraphics[width=0.8\textwidth]{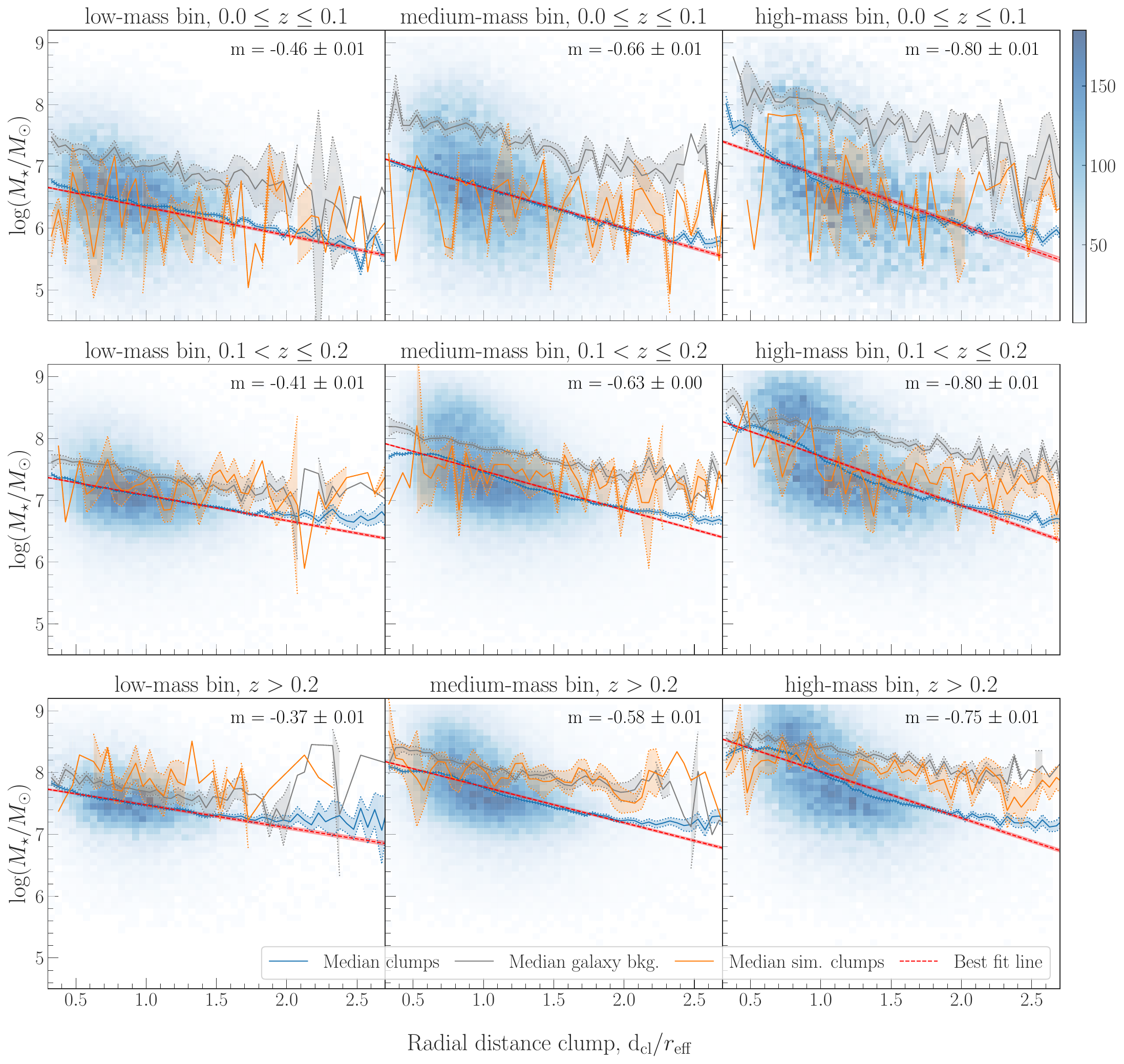}
    \caption[Clump stellar mass distribution as a function of the normalised galactocentric distance per redshift and galaxy stellar mass bins.]{Similar to Fig. \ref{fig:hsc_phys_props_gradient_UR_reff} but showing the clump stellar mass distribution as a function of the normalised galactocentric distance per redshift (rows) and host galaxy stellar mass bins (columns) for the mass-complete sample of CLAUDS and HSC-SSP galaxies. The stellar mass bins for the galaxies with redshift $z\leq 0.32$ are: $9.0 \leq \log(M_{\mathrm{gal}}/M_\odot) < 9.8$ (low), $9.8 \leq \log(M_{\mathrm{gal}}/M_\odot) < 10.6$ (medium) and $10.6 \leq \log(M_\star/M_\odot) < 11.4$ (high).}
    \label{fig:hsc_phys_props_gradient_mass_reff_z_massBin}
\end{figure*}

The stellar mass gradient of the clumps varies with redshift and stellar mass of the host galaxy (Fig. \ref{fig:hsc_phys_props_gradient_mass_reff_z_massBin} and Table \ref{tab:hsc_phys_props_gradient_mass}). The gradient tends to become flatter with increasing redshift but steeper with increasing mass of the host galaxy. Also, the stellar mass dependence on the radial distance does not appear to be purely linear as the fitted model $\log(M_{\mathrm{cl}}/M_\odot) = m (\mathrm{d}_{\mathrm{cl}}/r_{\mathrm{eff}})+c$ deviates noticeably from the median curve, especially for galaxies more massive than $\log$($M_{\mathrm{gal}}/M_\odot$) $> 9.8$. Closer to the centre of the galaxy ($\mathrm{d}_{\mathrm{cl}}/r_{\mathrm{eff}} \lesssim 0.6$), the clump stellar masses show almost no dependence on the distance and asymptotically approach a median stellar mass limit of $\sim 10^6$ to $\sim10^7M_\odot$. This limit is likely related to the detection limits of the FRCNN model as low-mass clumps might remain undetected if their fluxes are below the limiting magnitude of the CLAUDS and HSC-SSP surveys.

While the combined stellar mass contribution of all clumps to a host galaxy's mass is highest for low-mass galaxies (Fig. \ref{fig:hsc_phys_props_contribution_ratio_z}), the stellar masses of individual clumps tend to be higher in more massive galaxies, especially if the clumps are located closer to the centre of the galaxy. The median stellar mass of the clumps is larger for high-mass galaxies in the radial distance bins $\lesssim 0.8\,\mathrm{d}_{\mathrm{cl}}/r_{\mathrm{eff}}$ than for medium- and low-mass galaxies (Fig. \ref{fig:hsc_phys_props_gradient_mass_reff_z_massBin}). Physically, the stellar mass of a clump is limited by the host galaxy stellar mass as massive clumps account for a considerable fraction of the galaxy's total mass. The absence of massive clumps in galaxies from the low-mass bin ($9.0 \leq \log$($M_{\mathrm{gal}}/M_\odot$) $< 9.8$) and very massive clumps in galaxies from the medium-mass bin ($9.8 \leq \log$($M_{\mathrm{gal}}/M_\odot$) $< 10.6$) results in flatter gradients compared to the stellar mass gradient of the clumps detected in high-mass galaxies ($\log$($M_{\mathrm{gal}}/M_\odot$) $\geq 10.6$).

\begin{figure}
    \centering
    \subfloat[\centering Star-forming galaxies. \label{fig:hsc_phys_props_gradient_mass_reff_is_sf}]{{\includegraphics[width=1\columnwidth]{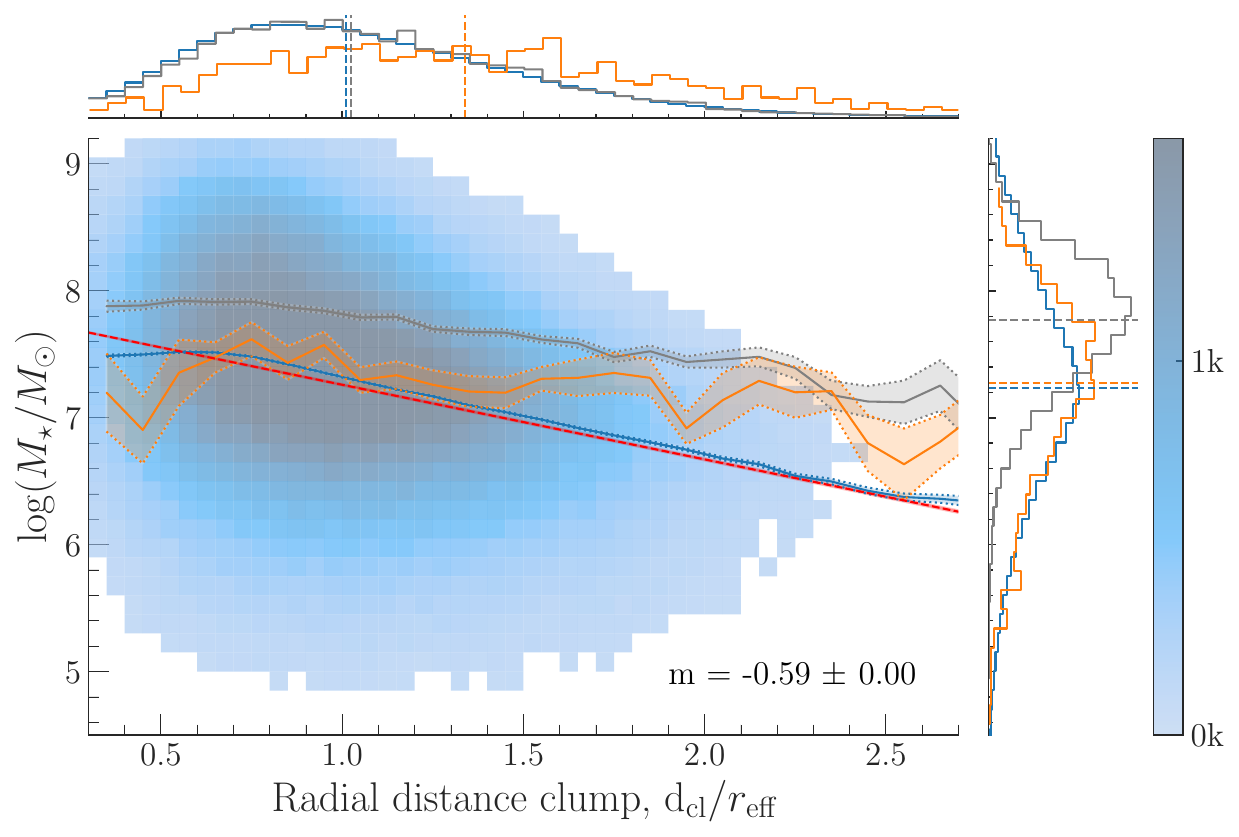} }}
    \\
    \subfloat[\centering Non star-forming galaxies. \label{fig:hsc_phys_props_gradient_mass_reff_not_sf}]{{\includegraphics[width=1\columnwidth]{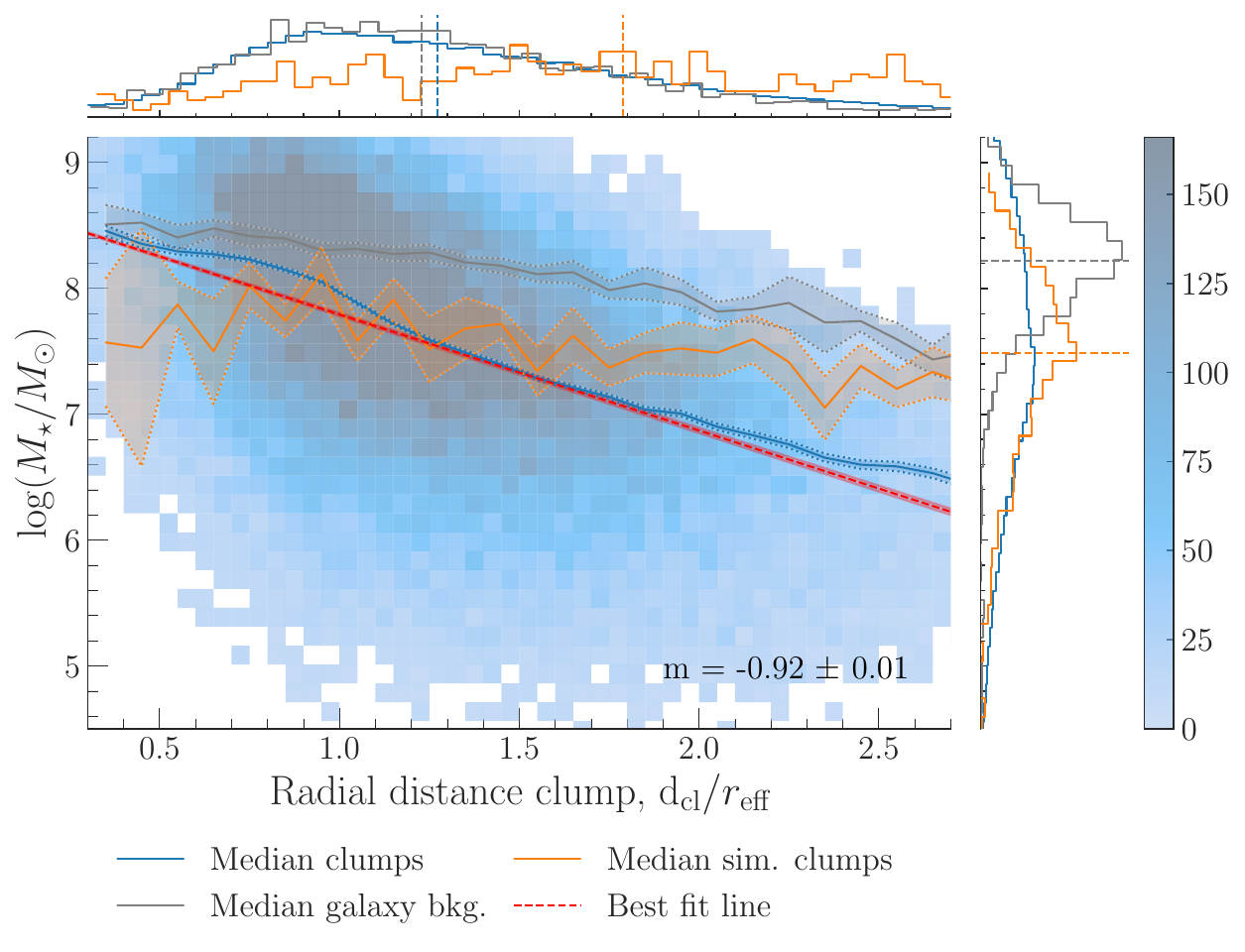} }}
    \caption[Clump stellar mass distribution as a function of the normalised galactocentric distance for SFGs and non-SFGs.]{Similar to Fig. \ref{fig:hsc_phys_props_gradient_UR_reff} but showing the clump stellar mass distribution as a function of the normalised galactocentric distance for SFGs (a) and non-SFGs (b) from the mass-complete sample of CLAUDS and HSC-SSP galaxies. A SFG is defined as having a sSFR of $\log$(${\mathrm{sSFR}}/\mathrm{yr}^{-1}$) $\geq -11.0$.}
    \label{fig:hsc_phys_props_gradient_mass_reff_sf}
\end{figure}

Clumps detected in galaxies that are not star-forming (galaxy sSFR is below $\log$($\mathrm{sSFR}/\mathrm{yr}^{-1}$) $<-11.0$) tend to be more massive (median stellar mass of $\log(M_{\mathrm{cl}}/M_\odot)=7.61$) and their locations tend to extend to areas further away from the galactic centre (median normalised distance of $\mathrm{d}_{\mathrm{cl}}/r_{\mathrm{eff}}=1.27$) than clumps in SFGs (median stellar mass of $\log(M_{\mathrm{cl}}/M_\odot)=7.34$ and median normalised distance of $\mathrm{d}_{\mathrm{cl}}/r_{\mathrm{eff}}=1.01$). The median stellar mass as a function of the radial distance, shown for the clumps in SFGs in Figure \ref{fig:hsc_phys_props_gradient_mass_reff_is_sf} and the clumps in non-SFGs in Figure \ref{fig:hsc_phys_props_gradient_mass_reff_not_sf}, also declines more steeply for the clumps detected in non-SFGs than for clumps in SFGs. The slopes of the fitted linear models to the stellar mass/radial distance distributions (Table \ref{tab:hsc_phys_props_gradient_mass}) are also significantly different (Welch's T-test: $p\text{-value}= 6.67 \times 10^{-6}$).

The gradient for the galaxy background is also slightly steeper in non-SFGs ($m = -0.43 \pm 0.01$) than in SFGs ($m = -0.36 \pm 0.01$) and a similar decrease in the slope value is observed from the detected simulated clumps (non-SFGs: $m = -0.29 \pm 0.05$, SFGs: $m = -0.23 \pm 0.05$). However, the difference between the slopes for the real clumps in SFGs and non-SFGs is larger than the slope difference between the real clumps and the simulated clumps or galaxy background in either the SFGs or non-SFGs in isolation. This seems to corroborate evidence from the (\textit{u}-\textit{r}) colour gradient differences that SFGs and non-SFGs may host two populations of physically different clumps.

We also tested our results against different choices of the radial distance measure, i.e. the different half-light radii, selecting galaxies with spec-z/photo-z measurements or different assumptions of the SFH (Appendix \ref{sec:hsc_phys_props_gradient_robust}). The observed negative stellar mass gradients vary with different choices but the existence of a negative stellar mass gradient is not changed.

% -- Radial Age Gradient -------------------------------
\subsubsection{Radial age gradient}\label{sec:hsc_phys_props_gradient_age}
The evolution of clumps and their longevity is strongly debated in the literature. Most observational searches for a possible clump migration towards the gravitational centre of the host galaxies are from high-redshift galaxies \citep[$1 < z < 3$, e.g.][]{Bournaud2007,Elmegreen2008,Ceverino2010,Bournaud2013,Mandelker2014} and only a few have observed clumps in low-redshift or nearby galaxies with a focus on clump migration \citep[e.g.][]{Mehta2021}. A negative stellar age gradient as a function of radial distance from the centre of the host galaxy is predicted from simulations for clumps that live long enough to migrate towards the centre of the galaxy \citep{Ceverino2010,Mandelker2014,Mandelker2016,Dekel2022}. Younger clumps should be preferentially located in the outer regions of a galaxy from where they spiral into the galactic centre, whereas older clumps should be found closer to the galactic centre after their inward migration over a few orbital timescales. 

\begin{figure}
    \centering
    \subfloat[\centering CLAUDS/HSC-SSP galaxies. \label{fig:hsc_phys_props_gradient_age_reff_a}]{{\includegraphics[width=1\columnwidth]{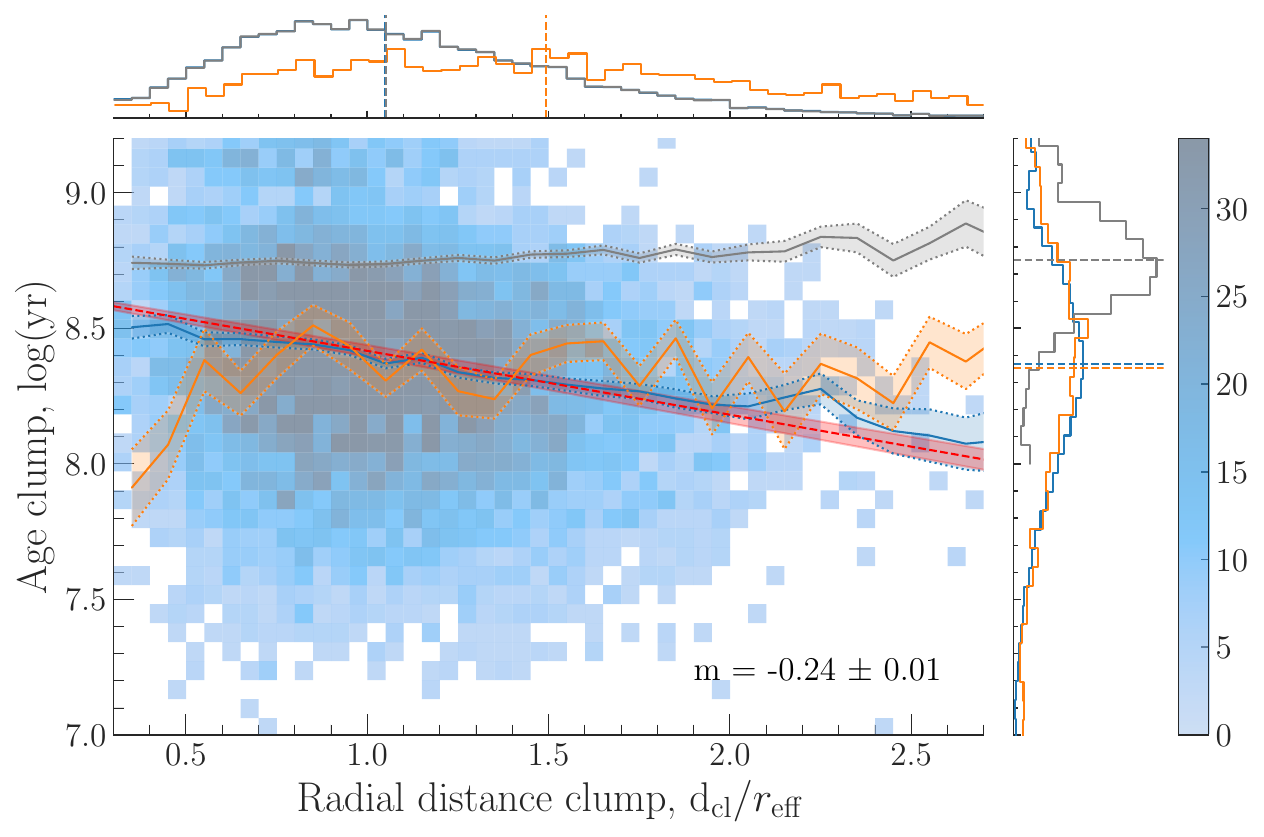} }}
    \\
    \subfloat[\centering HSC-SSP only galaxies. \label{fig:hsc_phys_props_gradient_age_reff_b}]{{\includegraphics[width=1\columnwidth]{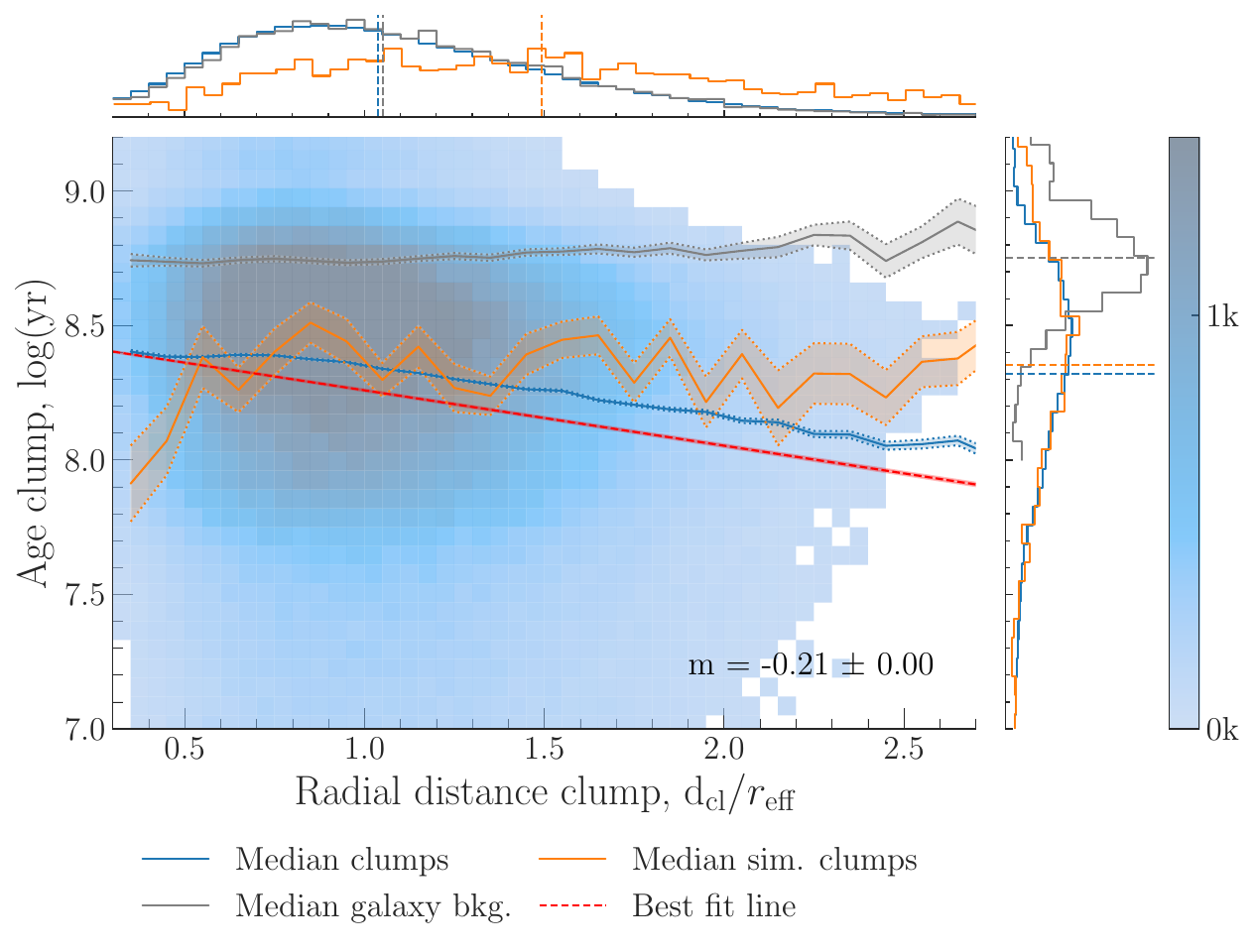} }}
    \caption[Clump stellar age distribution as a function of the normalised galactocentric distance.]{Similar to Fig. \ref{fig:hsc_phys_props_gradient_UR_reff} but showing the clump stellar age distribution as a function of the normalised galactocentric distance for the CLAUDS/HSC-SSP galaxy sample (a) and the HSC-SSP only galaxy sample (b).}
    \label{fig:hsc_phys_props_gradient_age_reff}
\end{figure}

In Figure \ref{fig:hsc_phys_props_gradient_age_reff}, we plot the stellar age distribution as a function of radial distance. The two panels of the figure show the age distribution for the \textit{ugrizy} clumps from the CLAUDS/HSC-SSP galaxy sample (Figure \ref{fig:hsc_phys_props_gradient_age_reff_a}) and for the \textit{grizy} clumps from the HSC-SSP only galaxy sample (Figure \ref{fig:hsc_phys_props_gradient_age_reff_b}). The median age of the real clumps decreases with increasing distance from the centre of the galaxy and the gradient $m$, estimated from a fitted linear model $\log(t_{\mathrm{age}}/\mathrm{yr}) = m (\mathrm{d}_{\mathrm{cl}}/r_{\mathrm{eff}})+c$, is slightly steeper for the \textit{ugrizy} clumps ($m=-0.24 \pm 0.01$) than for the \textit{grizy} clumps ($m=-0.21 \pm 0.00$). Both estimated slopes are significantly different from zero. In contrast, the age gradient of the simulated clumps is not significantly different from zero (Table \ref{tab:hsc_phys_props_gradient_age}). The median age of the galaxy background is significantly larger than the clumps (see also Fig. \ref{fig:hsc_phys_props_clumps_hist_age}) and shows only a weak dependence on the radial distance from the galactic centre.

\begin{figure*}
    \centering
    \includegraphics[width=0.8\textwidth]{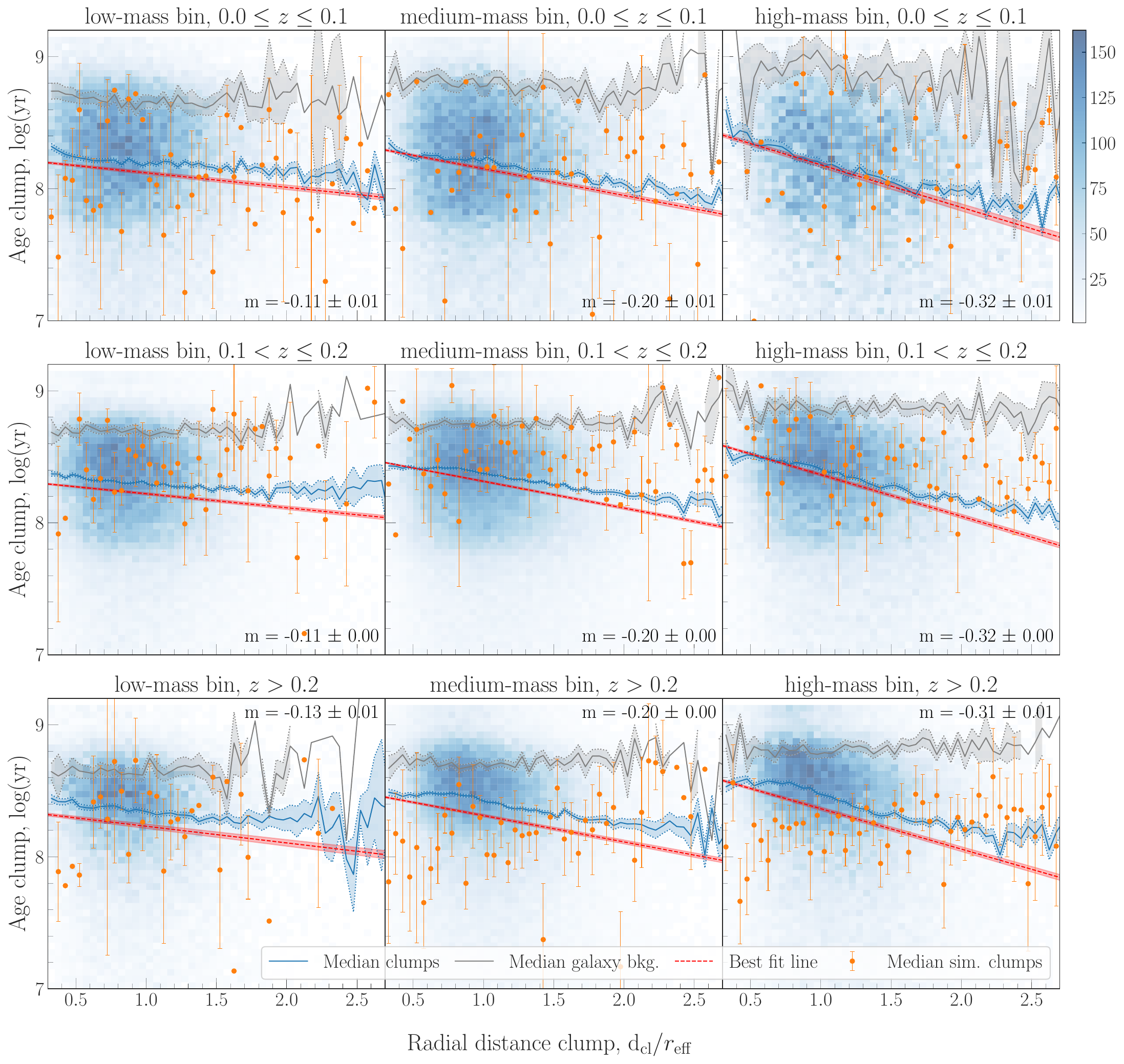}
    \caption[Clump stellar age distribution as a function of the normalised galactocentric distance per redshift and galaxy stellar mass bins.]{Similar to Fig. \ref{fig:hsc_phys_props_gradient_UR_reff} but showing the clump stellar age distribution as a function of the normalised galactocentric distance per redshift (rows) and host galaxy stellar mass bins (columns) for the mass-complete sample of CLAUDS and HSC-SSP galaxies. The median and standard error of the median for the simulated clumps are shown as orange points with error bars for clarity.}
    \label{fig:hsc_phys_props_gradient_age_reff_z_massBin}
\end{figure*}

If the host galaxies are split by redshift and stellar mass bins, the dependence of the estimated stellar ages of the galaxy background varies very weakly around $m\simeq 0.0$. The estimated gradients of the simulated clumps are not significantly different from zero for each redshift or stellar mass bin (Table \ref{tab:hsc_phys_props_gradient_age}). We note that the number of simulated clumps in each bin is relatively small, so the variance of the ages is large. In contrast, the stellar age gradients of the real clumps are all negative and significantly non-zero. The gradients do not depend on redshift but tend to steepen with increasing stellar mass of the host galaxy (Figure \ref{fig:hsc_phys_props_gradient_age_reff_z_massBin} and Table \ref{tab:hsc_phys_props_gradient_age}). Older clumps tend to be less frequent in less massive galaxies and the oldest clumps might only be present in the high-mass galaxies.

\begin{figure}
    \centering
    \subfloat[\centering Star-forming galaxies. \label{fig:hsc_phys_props_gradient_age_reff_is_sf}]{{\includegraphics[width=1\columnwidth]{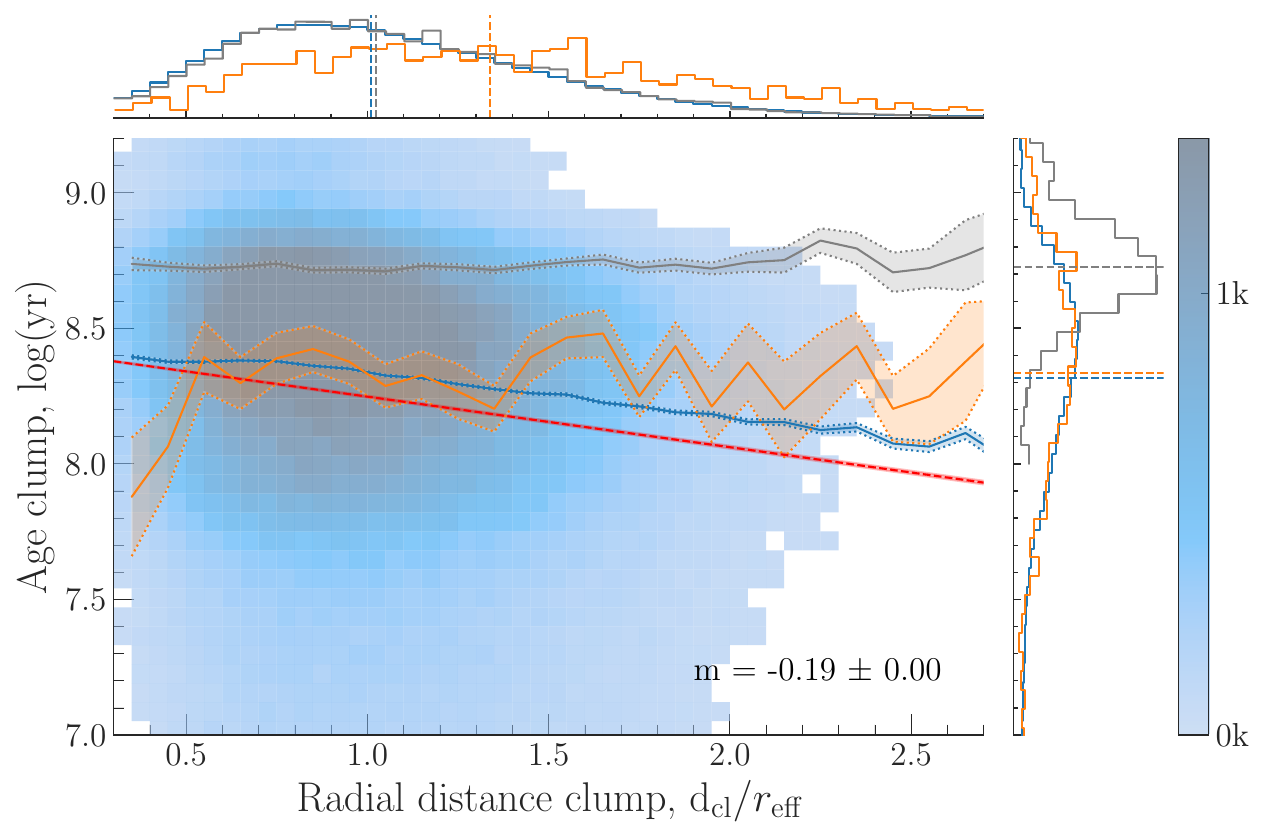} }}
    \\
    \subfloat[\centering Non star-forming galaxies. \label{fig:hsc_phys_props_gradient_age_reff_not_sf}]{{\includegraphics[width=1\columnwidth]{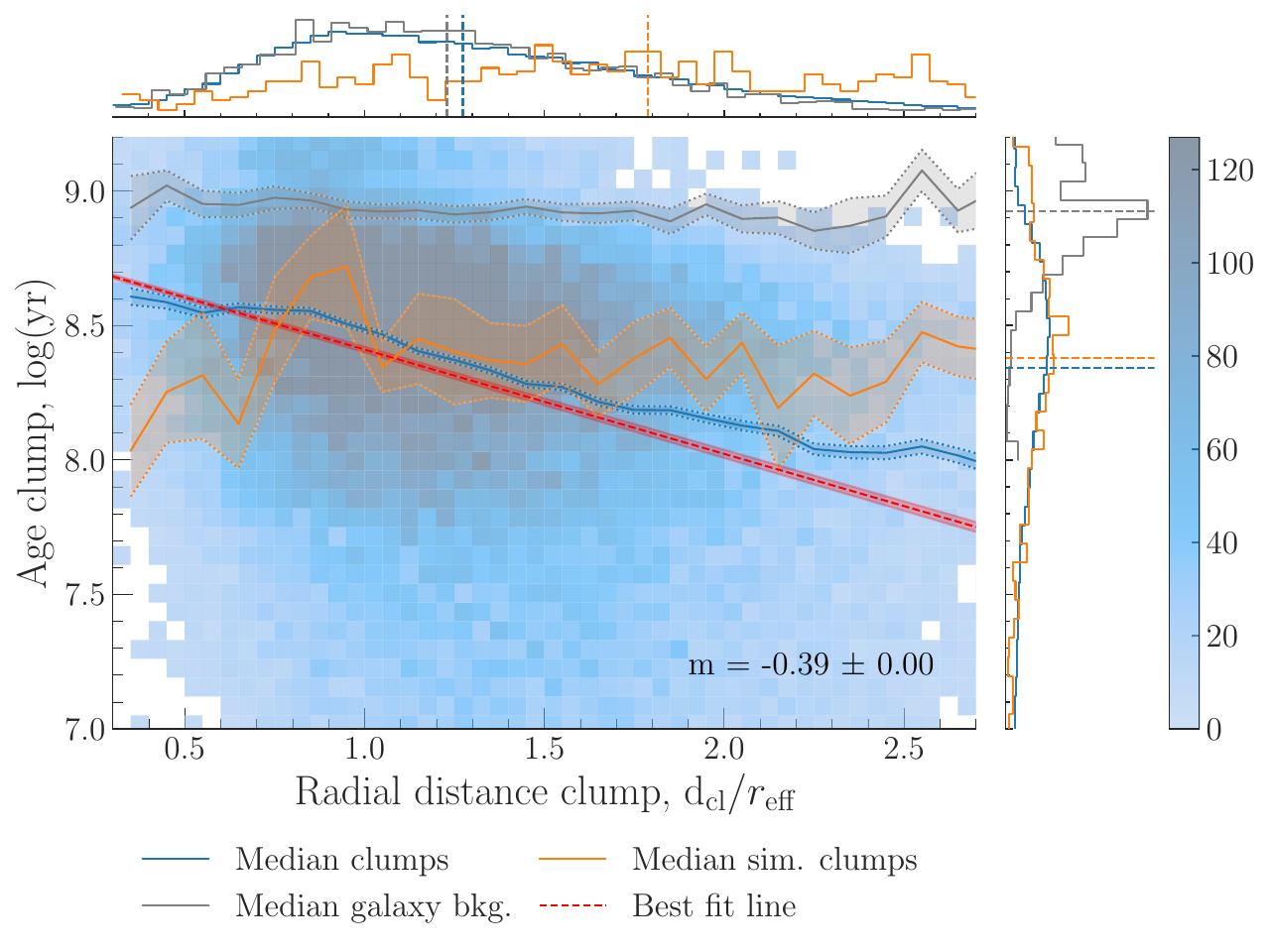} }}
    \caption[Clump stellar age distribution as a function of the normalised galactocentric distance for SFGs and non-SFGs.]{Similar to Fig. \ref{fig:hsc_phys_props_gradient_UR_reff} but showing the clump stellar age distribution as a function of the normalised galactocentric distance for SFGs (a) and non-SFGs (b) from the mass-complete sample of CLAUDS and HSC-SSP galaxies. A SFG is defined as having a sSFR of $\log$(${\mathrm{sSFR}}/\mathrm{yr}^{-1}$) $\geq -11.0$.}
    \label{fig:hsc_phys_props_gradient_age_reff_sf}
\end{figure}

Figure \ref{fig:hsc_phys_props_gradient_age_reff_sf} reveals evident differences in stellar age gradients between SFGs and non-SFGs. While the median age $t_{\mathrm{age}} = 10^{8.32}\,\mathrm{yr}$ of the clumps in SFGs is nearly identical to the median age $t_{\mathrm{age}} = 10^{8.36}\,\mathrm{yr}$ of the clumps in non-SFGs, the age gradients as a function of the radial distance differ. The gradient for the clumps in SFGs is flatter ($m=-0.19 \pm 0.00$) than the gradient for the non-SFGs ($m=-0.39 \pm 0.01$, Table \ref{tab:hsc_phys_props_gradient_age}). The non-SFGs might contain a larger fraction of ex situ clumps that have not been formed inside the host galaxy but instead acquired via minor mergers. Such ex situ clumps tend to be older, more massive and appear concentrated near the outer disc \citep{Mandelker2016}. The clumps detected in non-SFGs are, on average, further away from the galactic centre (median normalised distance non-SFGs: $\mathrm{d}_{\mathrm{cl}}/r_{\mathrm{eff}}=1.27$, SFGs: $\mathrm{d}_{\mathrm{cl}}/r_{\mathrm{eff}}=1.01$) and we showed in Section \ref{sec:hsc_phys_props_gradient_mass} that they also tend to be more massive. While the overall median age is comparable to the clumps in SFGs, the clumps in non-SFGs that are within a distance of $\mathrm{d}_{\mathrm{cl}} \lesssim 1.0\,r_{\mathrm{eff}}$ from the galactic centre are older (the median age is $\sim 0.2\,\mathrm{dex}$ higher), which causes the age gradient to become steeper.

The median age of the galaxy background in non-SFGs is $\sim 0.2\,\mathrm{dex}$ higher than the background in SFGs over the entire radial distance range. The gradient of the galaxy background is also slightly steeper in non-SFGs ($-0.06 \pm 0.01$) than in SFGs for which the slope is consistent with zero. The slope values that are observed from the detected simulated clumps are almost identical and not significantly different from zero (non-SFGs: $-0.02 \pm 0.04$, SFGs: $-0.03 \pm 0.03$).

% ================================================
% = Discussion
% ================================================
\section{Discussion}\label{sec:discussion}
% -----------------------------------------------
% == Diffuse background subtraction
\subsection{Diffuse background subtraction}\label{sec:hsc_discussion_photometry}
The subtraction of the diffuse galaxy background has the biggest impact on the photometry measurements and the resulting estimates of the physical properties of the clumps. The applied aperture photometry method is capable of accurately recovering the fluxes in all six \textit{ugrizy} filter bands from objects that are not embedded in a complex background setting (see Paper II). However, in a setting with a complex background measured photometry values can differ significantly from the true values, especially for objects with flux values that are close to the detection limits of the instruments. The value of the background subtraction varies with the size of the annuli chosen to determine the median per-pixel flux coming from the background (Section \ref{sec:clump_phot}) and ranges between $\sim$28\% and $\sim$51\% (depending on the filter band) of the measured flux before subtraction. This can affect the gradients of the colour, stellar mass and age of the clumps. To test the robustness of the background subtraction, we repeated the estimation of the physical properties for a sample of \textit{ugrizy} clumps using different annuli for the background estimation and also included the case where the diffuse galaxy background was not subtracted from the measurements. Here, we randomly selected $\sim 10\%$ of the detected clump candidates from the sample of CLAUDS/HSC-SSP galaxies (708 galaxies with 3,066 clump candidates) and ran the SED fitting process as described in Section \ref{sec:clump_sed}.

\begin{figure}
    \centering
    \subfloat[\centering Clump stellar age. \label{fig:hsc_discussion_photometry_b}]{{\includegraphics[width=0.7\columnwidth]{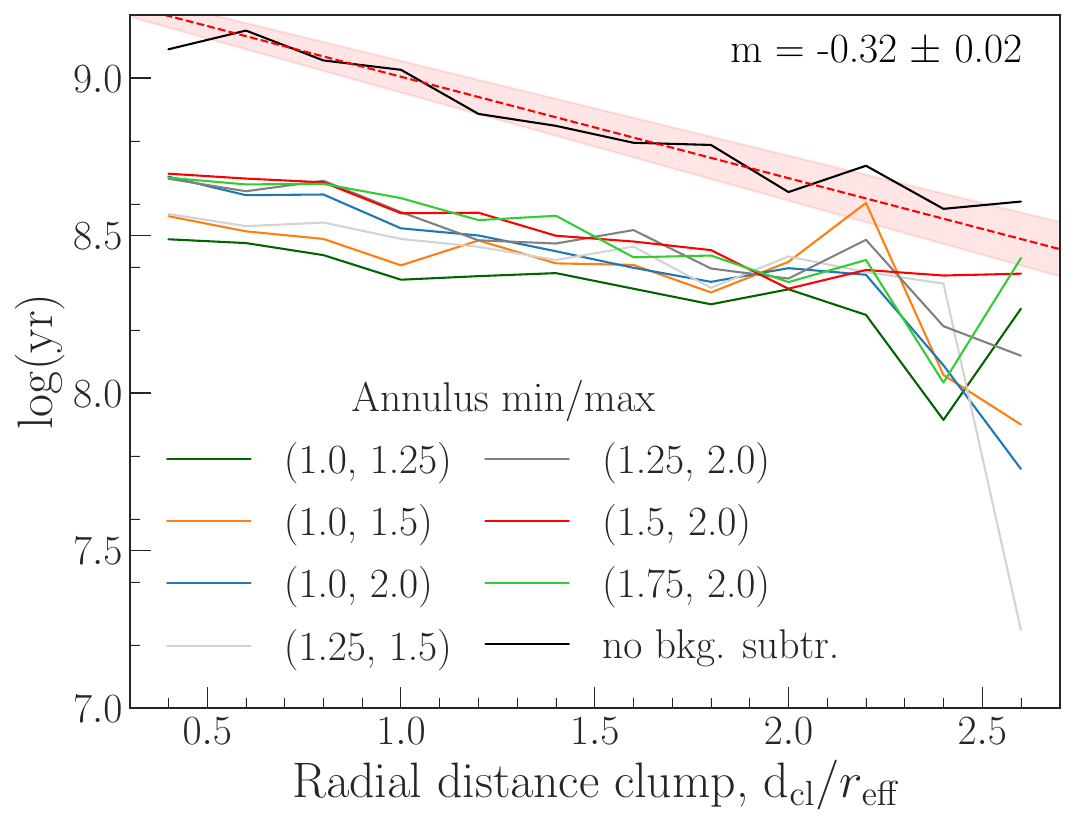} }}
    \\
    \subfloat[\centering Clump stellar mass. \label{fig:hsc_discussion_photometry_c}]{{\includegraphics[width=0.7\columnwidth]{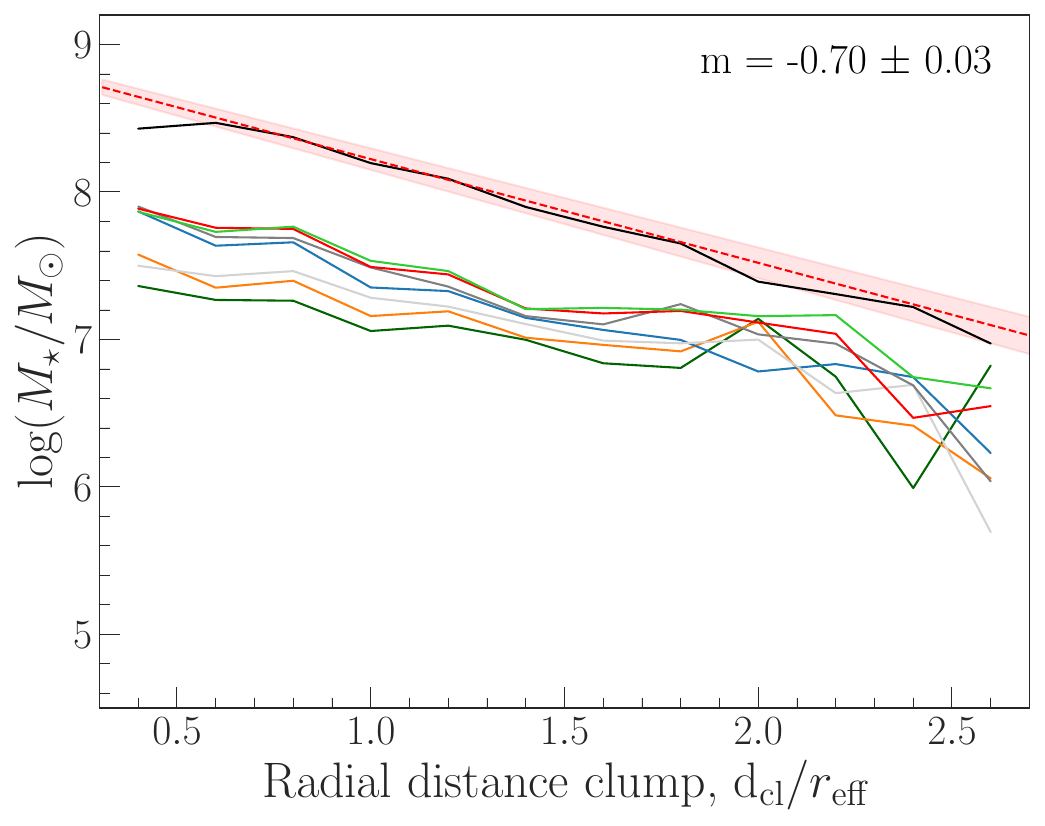} }}    
    \caption[Median of the stellar mass and age distribution as a function of the normalised galactocentric distance for different background subtraction methods.]{Median of the stellar age (a) and mass distribution (b) as a function of the normalised galactocentric distance for different background subtraction methods. The medians were determined for each radial distance bin from a sample of $\sim 10\%$ of the detected \textit{ugrizy} clump candidates. For clarity, only the medians without standard errors are plotted for the background subtraction methods with different annuli sizes (in units of the seeing FWHM, see also Section \ref{sec:clump_phot}). The medians for values that are estimated from the observed photometry with no subtraction of the diffuse galaxy background are shown as black lines. Linear models are fitted to the distributions of the observed values with no background subtraction and are shown as dashed red lines with the standard errors indicated by the red shaded areas. The gradients $m$ of the linear models are shown as annotation in the plot.}
    \label{fig:hsc_discussion_photometry}
\end{figure}

Figure \ref{fig:hsc_discussion_photometry} shows the estimated median stellar age (Fig. \ref{fig:hsc_discussion_photometry_b}) and median stellar mass (Fig. \ref{fig:hsc_discussion_photometry_c}) of the clump sample as a function of the galactocentric distance. We also show a linear model fitted to the values that are estimated using the observed photometry with no background subtraction in each plot of Figure \ref{fig:hsc_discussion_photometry}.

Irrespective of the background subtraction method, the gradients of the clump properties with distance from the galactic centre are preserved. A more aggressive background subtraction with a small annulus, e.g. from 1.0 to 1.25 times the seeing FWHM, leads to generally lower stellar masses and ages of the clumps in comparison with a more conservative subtraction with a wider annulus but does not change the general negative trend with increasing distance from the galactic centre. The estimated slope of the stellar mass (age) gradient with no background subtraction is $-0.70$ ($-0.32$) whereas the same gradient estimated from the background subtracted fluxes of the sample of \textit{ugrizy} clumps is $-0.54$ ($-0.24$, Tables \ref{tab:hsc_phys_props_gradient_mass} and \ref{tab:hsc_phys_props_gradient_age}).

Two observations can be made from this robustness test. First, the existence of the negative gradients, which we observe for our sample of clumps with galactocentric distance, is not changed by the background subtraction methods, although the choice of method may alter the values of the observed colours and estimates for stellar mass and age of the clumps. Second, if the diffuse galaxy background is not subtracted from the detected clumps, then the clump fluxes seem to contain some contamination from galactic disk stars. Those background stars are older than the stars formed in the clumps and adding the flux of those stars makes the clumps appear redder and older (Fig. \ref{fig:hsc_discussion_photometry_b}).

Hence, we argue that it is essential to subtract the diffuse galaxy background light from the clump fluxes. By using an aperture with radius $r_{\mathrm{ap}}=0.25\, \mathrm{FWHM}_{\mathrm{seeing}}$, masking adjacent clump detections with a circular mask of radius $r_{\mathrm{mask}}=0.5\, \mathrm{FWHM}_{\mathrm{seeing}}$ and estimating the background light using an annulus of $r_{\mathrm{an}}=\{1.5, 2.0\}\, \mathrm{FWHM}_{\mathrm{seeing}}$ (Section \ref{sec:clump_phot}), contamination of the clump photometry measurements by adjacent objects and the underlying diffuse galaxy background was minimised (Paper II).

% % -----------------------------------------------
% % == Completeness of the Clump Detections
\subsection{Completeness of the clump detections}\label{sec:hsc_discussion_complete}
The detection completeness of our clump sample, which is based on a definition of star-forming clumps that have a clump-galaxy mass ratio of $\log$($M_{\mathrm{cl}}/M_{\mathrm{gal}}$) $\geq -6.6$, is high for all clumps above the $5\sigma$ point-source depth limits of the instruments (Paper I). However, there is a fraction of clumps that fall below the detection limit but are above the clump-galaxy mass ratio threshold and are possibly genuine star-forming clumps (Fig. \ref{fig:hsc_phys_props_clump_sample1b} and \ref{fig:hsc_phys_props_clump_sample1c}). This indicates that our clump sample is not fully complete for clumps with stellar masses $\lesssim 10^{7.6}\,M_\odot$, which is approximately the upper limit of the distribution of clumps that pass the clump-galaxy mass ratio threshold but are below the detection limit. Furthermore, Figure \ref{fig:sim_clumps_det_completeness_mass_z} shows that the completeness, which is measured on the sample of simulated clumps, drops below $0.8$ for a clump stellar mass of $\lesssim 10^{8.6}\,M_\odot$. 

The vast majority ($99.9\%$) of the detected clumps, which pass the clump-galaxy mass ratio threshold but are below the detection limit, are below the $5\sigma$ point-source depth limits of the \textit{grizy}-filter bands from HSC-SSP, specifically below the limits of the \textit{y}-filter band. We could have relaxed the restriction from the \textit{y}-filter band to include those clumps in the final sample but doing so would have meant either leaving out one photometric data point or including \textit{y}-band data that could be less accurate for some of the galaxy sample when performing the SED fitting. Neither option is robust against systematic effects and we kept the restriction on all filter bands for consistency while we were extending the analysis from the sample of \textit{ugrizy} clumps to the larger sample of \textit{grizy} clumps for which only HSC-SSP data is available.

The power-law slopes of the cSMFs were only estimated in a stellar mass range for which the clump sample is $>80\%$ complete (see also Section \ref{sec:hsc_discussion_form_evol_csmf}). However, the estimated slopes of the radial gradients (see also Section \ref{sec:hsc_discussion_form_evol_gradients}) are affected by the mass-incompleteness. Based on the properties of the detected clumps that pass the clump-galaxy mass ratio threshold but are below the detection limit, the subset that is probably missing from a complete clump sample is on average bluer (median (\textit{u}-\textit{r}) colour of $0.93$), less massive (median stellar mass of $10^{6.07}\,M_\odot$), younger (median age of $10^{7.84}\,\mathrm{yr}$) and located further away from the galactic centre (median distance of $1.45\,r_{\mathrm{eff}}$). A mass-complete clump sample would therefore most likely result in steeper (\textit{u}-\textit{r}) colour, stellar mass and age gradients than those estimated in Section \ref{sec:hsc_phys_props_gradient_colour}, \ref{sec:hsc_phys_props_gradient_mass} and \ref{sec:hsc_phys_props_gradient_age}, respectively. Therefore, the slopes of the radial gradients presented in this work are a conservative estimate as the true gradients are likely to be steeper.

% -----------------------------------------------
% == Clump Stellar Mass Function
\subsection{Clump stellar mass function}\label{sec:hsc_discussion_form_evol_csmf}
The SMFs of stellar clusters are found to have power-law slopes $\alpha = -2.0 \pm 0.3$ at their massive ends \citep[e.g.][]{Gieles2006,Adamo2013,Chandar2014,Mandelker2016}. If clumps were formed in situ through gravitational instabilities or VDI, the cSMF is expected to have similar $\alpha$ values, as discussed by \citet{Elmegreen1997,DessaugesZavadsky2018}. 

The power-law slopes of the cSMF we estimated from our sample of clumps are steeper. Depending on redshift and the stellar mass of the host galaxy, in most cases $\alpha \simeq [-2.8, -2.2]$ (Table \ref{tab:hsc_phys_props_cSMF_clump_age}). Compared to a typical SMF of stellar clusters, the observed top-light cSMFs reflect populations that contain either relatively few high-mass objects or relatively more low- and intermediate-mass objects. A steeper slope of the cSMF could reflect either an intrinsically top-light distribution during the formation of the clumps or that the cSMF has evolved from an initially different distribution into a top-light distribution.

Power-law slopes of the cSMFs estimated for clumps in high-redshift galaxies mostly agree with the VDI formation mode of the clumps that are reflected by estimated slopes of $\alpha\lesssim -2.0$ \citep[e.g.][]{DessaugesZavadsky2018,HuertasCompany2020,Kalita2024}. The higher gas fraction and more intense star-forming activity in the high-redshift galaxies compared to low-redshift galaxies could have facilitated the formation of more high-mass clumps. From their simulation study, \citet{Renaud2024a} predicted that higher gas fractions lead to less steep and top-heavy cSMFs with $\alpha$ values of -2.0, -1.5 and -1.6 for gas fractions of 10\%, 25\% and 40\%, respectively. The much lower gas fractions in low-redshift galaxies \citep[e.g.][]{Wang2022} could explain steeper slopes of the cSMF during clump formation that are observed from our sample of clumps. Also, the flattening of the cSMF that we measured between low-mass and high-mass galaxies (Table \ref{tab:hsc_phys_props_cSMF_clump_age}) could be explained by an increasing gas fraction as more massive galaxies have, at least locally, larger total gas reservoirs.

When split into three different stellar age bins ($t_{\mathrm{age}}\leq 100\,\mathrm{Myr}$, $100 < t_{\mathrm{age}}\leq 300\,\mathrm{Myr}$ and $t_{\mathrm{age}}>300\,\mathrm{Myr}$), the cSMF for clumps younger than 100 Myr has a power-law slope of $\alpha=-1.98$ (Fig. \ref{fig:hsc_phys_props_cSMF_clump_age}), which, if representative for an initial cSMF of the clumps, suggest in situ clump formation by VDI processes. As the clumps age, the slopes become steeper, indicating that clumps become more massive either by star-formation from a constant supply of gas or by merging with other, smaller clumps. This growth might not affect all clumps equally as the stellar mass of a clump might be limited if an equilibrium between gas accretion and gas loss due to radiative stellar feedback is reached, the instability criteria are no longer met, or the clump has fully migrated inwards \citep[e.g.][]{Dekel2009}. Massive clumps might also fragment into smaller clumps or lose stellar mass into the galactic disk over time.

Our observed cSMF still remains unconstrained for low-mass clumps. For stellar masses below the completeness limit ($\sim 10^8\,M_\odot$), the fate of these clumps is unclear. Low-mass clumps ($\lesssim 10^7\,M_\odot$) could either grow in mass similarly to the intermediate- ($\sim 10^7$ to $10^9\,M_\odot$) and high-mass clumps ($\gtrsim 10^9\,M_\odot$) or they might be destroyed soon after formation by internal stellar feedback and tidal forces in the host galaxy. \citet{Fensch2021} argued that low-mass clumps are likely to be quickly destroyed even by weak stellar feedback as they are mostly gravitationally unbound due to their low stellar mass. Figure \ref{fig:hsc_phys_props_cSMF_clump_age} may provide evidence for such a scenario. The relative number of clumps with stellar masses $<10^8\,M_\odot$ is much larger for the sample of clumps that are younger than 100 Myr than for the sample that contains clumps with ages of $>300\,\mathrm{Myr}$. We note, however, that the clump samples are not mass-complete for these stellar mass ranges and without more sensitive observations, the cSMF cannot be reliably estimated.

\begin{figure}
    \centering
    \includegraphics[width=0.9\columnwidth]{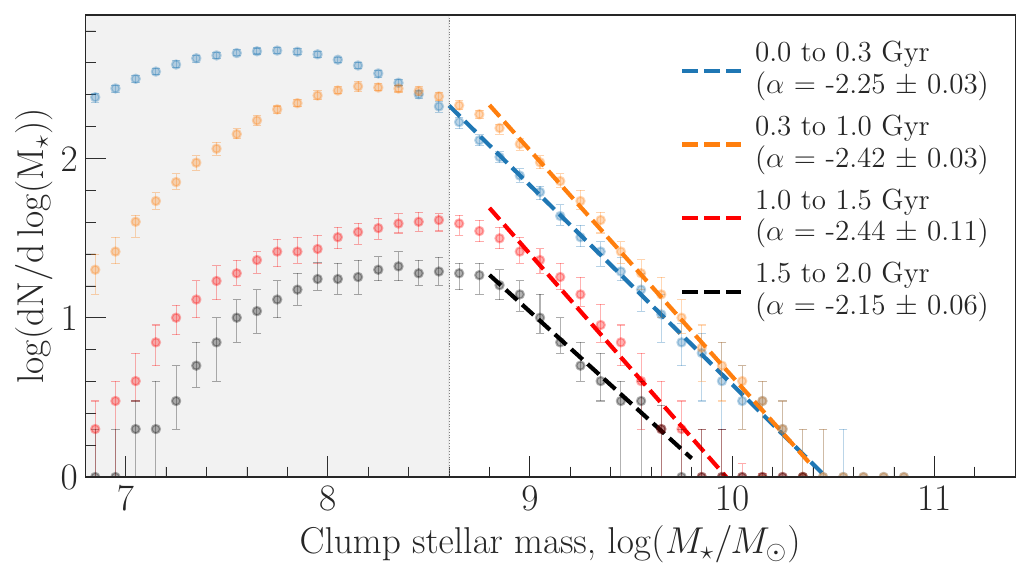}
    \caption[Evolution of the clump stellar mass function.]{Evolution of the clump stellar mass function for the clumps from the combined CLAUDS/HSC-SSP galaxy sample. The cSMFs are shown for different age bins of the clumps: $<0.3\,\mathrm{Gyr}$ (blue), 0.3-1.0 Gyr (orange), 1.0-1.5 Gyr (red) and 1.5-2.0 Gyr (black). The error bars show the 16 and 84\% credible interval determined from 100 samples of the stellar mass posterior distribution of each clump. Also shown is a best power-law fit for each subset in the same colours. The fitted values for $\alpha$ are shown as annotations. The grey shaded area for clump stellar masses lower than $\log(M_\star/M_\odot)<8.6$ separates the region for which the FRCNN model completeness drops below $0.8$.}
    \label{fig:hsc_discussion_form_evol_csmf}
\end{figure}

Further indications that the cSMF evolves with increasing clump age is shown in Figure \ref{fig:hsc_discussion_form_evol_csmf}. Here, four separate cSMFs have been constructed for clumps that are $<0.3\,\mathrm{Gyr}$, 0.3-1.0 Gyr, 1.0-1.5 Gyr and 1.5-2.0 Gyr old. The shapes of the cSMFs at four different clump ages could indicate a possible evolution of the clumps that starts with clump formation through VDI processes and is followed by continuing growth in stellar mass for the more massive clumps while low-mass clumps are disrupted and dissolved into the galactic disk early in their lifetime. The age range between $\sim$0.0 to 0.3 Gyr, when clumps are vulnerable to disruption, is represented by the blue cSMF in Figure \ref{fig:hsc_discussion_form_evol_csmf}. In theoretical models, the growth of intermediate- and high-mass clumps continues until the clumps have either migrated inwards to contribute to the build-up of the galactic bulge or the local environmental conditions in the disk have changed so that star-formation or other mass accumulation has ceased (orange cSMF for clump ages between 0.3-1.0 Gyr in Fig. \ref{fig:hsc_discussion_form_evol_csmf}). Older clumps that are still located in the disk continue to lose mass to the surrounding disk or are dissolved, which could explain the shape of the red and black cSMF for clump ages of 1.0-1.5 Gyr and 1.5-2.0 Gyr, respectively. Both cSMFs are depleted of intermediate- and high-mass clumps in comparison with the cSMFs of the younger clumps from the same sample of galaxies and the depletion becomes more prominent the older the clumps are (Fig. \ref{fig:hsc_discussion_form_evol_csmf}).

A study of high redshift clumps ($0.5 < z < 5.0$) by \citet{Sok2026}, that analysed the stellar mass distribution of clumps at different clump ages, also provides observational support for a scenario in which the cSMF evolves as the clumps age. In contrast with our findings, the authors found that the cSMF for 1.0-1.5 Gyr old clumps is flatter than the cSMF for younger, $<300\,\mathrm{Myr}$ old, clumps. They argue that this is caused by the more frequent disruption of less massive clumps while the more massive clumps are likely to survive. This would indicate that the clumps at the very massive end of the cSMF would simply continue growing in mass and would not get disrupted or lose stellar mass even after 1.0-1.5 Gyr.

Another possible explanation for an older clump population having a flatter cSMF is the presence of ex situ clumps. A top-light cSMF could appear `normal' or even top-heavy, if contamination by ex situ clumps is high. For example, the clumps in non-SFGs show a shallower cSMF compared with clumps detected in SFGs (Table \ref{tab:hsc_phys_props_cSMF_clump_age}). Non-SFGs are not expected to host massive star-forming clumps and it is likely that many of the clumps that contribute to the very massive end of the cSMF, leading to a shallower slope, are ex situ clumps.

Ex situ clumps are still present in our detected clump sample and they need to be properly separated from the population of in situ clumps before testing any scenario of a possible evolutionary trend of the cSMF. Some properties of ex situ clumps are predicted by \citet{Mandelker2014}, but observational studies that compare both populations in detail are still lacking. Further analysis through high-resolution simulations and observations is required to lower the mass-completeness limit of the observed clumps and to better constrain the cSMF for low-mass clumps. Our observations suggest that the cSMF is not only determined by the mechanisms that operate during the formation of the clumps but also by the growth in stellar mass of the clumps and disruption processes that affect low-mass clumps in early and more massive clumps in later phases of their evolution. This provides corroboration for many recent theoretical models of clump formation and evolution.

% -----------------------------------------------
% == Radial Gradients
\subsection{Radial gradients}\label{sec:hsc_discussion_form_evol_gradients}
Measuring the distribution of the physical clump properties as a function of their distance from the galactic centre is crucial to test different theoretical models for clump evolution. Theoretical models and simulations that are based on an inward migration scenario of the clumps \citep[e.g.][]{Ceverino2010,Mandelker2014,Mandelker2016,Dekel2022} predict negative gradients for the clump colours, stellar masses and ages. In such a scenario, clumps form preferentially in the outer regions of the galactic disk where the young clumps appear bluer than the galactic background and then migrate inwards while they grow in stellar mass and become progressively redder in colour as they age. 

The decreasing gradients of the clump stellar mass with increasing distance from the galactic centre that we measure confirm results from previous studies \citep[e.g.][]{Guo2018,Zanella2019,Ambachew2022,Dekel2022,Kalita2025} as well as predictions from simulations \citep[e.g.][]{Mandelker2014,Mandelker2016,Dekel2022}. The negative gradients are robust against systematic uncertainties introduced by the two different redshift measurement methods, the determination of the effective radius used for normalising the radial distance and changes to the assumed SFH of the clumps (Appendix \ref{sec:hsc_phys_props_gradient_robust}). %However, the decreasing spatial resolution with increasing redshift could introduce an additional bias to the observed gradients due to blending of unresolved clumps and/or contamination by stars from the intra-clump regions of the galactic disk. Figure \ref{fig:hsc_discussion_form_evol_gradients_mass_red} shows the stellar mass distribution of the observed clumps from the mass-complete sample of CLAUDS and HSC-SSP galaxies in comparison to the distribution of the sample of simulated clumps. The inferred stellar masses of the real clumps tend to increase with increasing redshift. However, a similar trend can also be observed from the simulated clumps, despite the fact that their stellar mass gradients are flatter, and we argue that the observed negative stellar mass gradients of the real clumps are only partially affected by the decreasing spatial resolution.

% \begin{figure}
%     \centering
%     \subfloat[\centering Clump stellar mass as a function of redshift. \label{fig:hsc_discussion_form_evol_gradients_mass_red}]{{\includegraphics[width=0.8\columnwidth]{figures/clump_6_redshift_mass.pdf} }}
%     \\
%     \subfloat[\centering Clump stellar age as a function of redshift. \label{fig:hsc_discussion_form_evol_gradients_age_red}]{{\includegraphics[width=0.8\columnwidth]{figures/clump_6_redshift_age.pdf} }}
%     \caption[Clump stellar mass and age distributions as a function of redshift.]{Clump stellar mass (a) and age (b) distributions as a function of redshift. Also shown is the distribution of the simulated clumps as black contours and linear models that are fitted to the distribution of the observed (red line) and simulated clumps (orange line). The gradients $m$ of the linear models are shown as annotation in the plot. Vertical and horizontal dashed lines mark the median stellar mass/age and redshift for the clumps from each sample in the same colours. Only clumps observed from the CLAUDS/HSC-SSP galaxy sample (mass complete sample with $M_{\star,\mathrm{gal}}\geq 10^9 M_\odot$ and $z\leq0.32$) are shown.}
%     \label{fig:hsc_discussion_form_evol_gradients_mass_age_red}
% \end{figure}

The abundance of clumps with measured ages $>10^8\,\mathrm{yr}$ suggests that clumps are not always the short-lived structures proposed by \citet{Buck2017} and \citet{Oklopcic2016} using simulations. Rather, some clumps appear to have survived over a few orbital timescales \citep[$\sim 10^8\,\mathrm{to}\,10^9\,\mathrm{yr}$,][]{Dekel2009,Ceverino2010,Dekel2022} making it possible for them to migrate towards the galactic centre. An important indicator for the inward migration of clumps, which is predicted from clump evolution models and simulations \citep{Bournaud2007,Elmegreen2008,Ceverino2010,Bournaud2013,Mandelker2014}, is an observed negative age gradient as a function of the distance from the galactic centre. The clump age is expected to decrease with increasing galactocentric distance and the age difference between the outer and inner clumps should reflect the inward migration time of a few hundred Myr. For our sample, the median age for inner clumps is $t_{\mathrm{age}} \approx 10^{8.4}$ at a galactocentric distance of $\mathrm{d}_{\mathrm{cl}}/r_{\mathrm{eff}} \approx 0.3$ and drops to $t_{\mathrm{age}} \approx 10^{8.2}$ for outer clumps at $\mathrm{d}_{\mathrm{cl}}/r_{\mathrm{eff}} \approx 2.0$ (Fig. \ref{fig:hsc_phys_props_gradient_age_reff_b}), resulting in an age difference between outer and inner clumps of $\sim 100\,\mathrm{Myr}$. 

\citet{Guo2018} found a larger age difference between the outer and inner clumps, ranging from $\sim 700\,\mathrm{Myr}$ for clumps in galaxies with $\log$($M_{\mathrm{gal}}/M_\odot$) $< 10.6$ at $z < 1.0$ to $\sim 250-300\,\mathrm{Myr}$ for clumps in galaxies with similar stellar mass but at higher redshift $1.0 \leq z < 2.0$. However, these age differences are difficult to compare directly to the results from our study. \citet{Guo2018} used different clump detection methods and the clump definition is based on a clump-galaxy UV-flux ratio. Also, the assumed SFH (exponentially declining or simple $\tau$ SFH) of the clumps differs from the assumptions made for our clump sample and the galactocentric distance is normalised by the estimated semi-major axis of the galaxy instead of the effective radius. %Furthermore, the stellar age gradients as a function of the radial distance from the galaxy centre can also be affected by the decreasing spatial resolution with increasing redshift due to blending of smaller clumps and/or contamination by disk stars. The inferred clump ages tend to slightly increase as a function of redshift (Fig. \ref{fig:hsc_discussion_form_evol_gradients_age_red}), but a similar trend is also observed from the simulated clumps which show only flat age gradients (see Table \ref{tab:hsc_phys_props_gradient_age}, for example). While some blending of clumps may still be present, the impact on the observed stellar age gradients is likely to be minor.

\begin{figure}
    \centering
    \subfloat[\centering Clump stellar mass distribution. \label{fig:hsc_discussion_form_evol_gradients_mass_a}]{{\includegraphics[width=1\columnwidth]{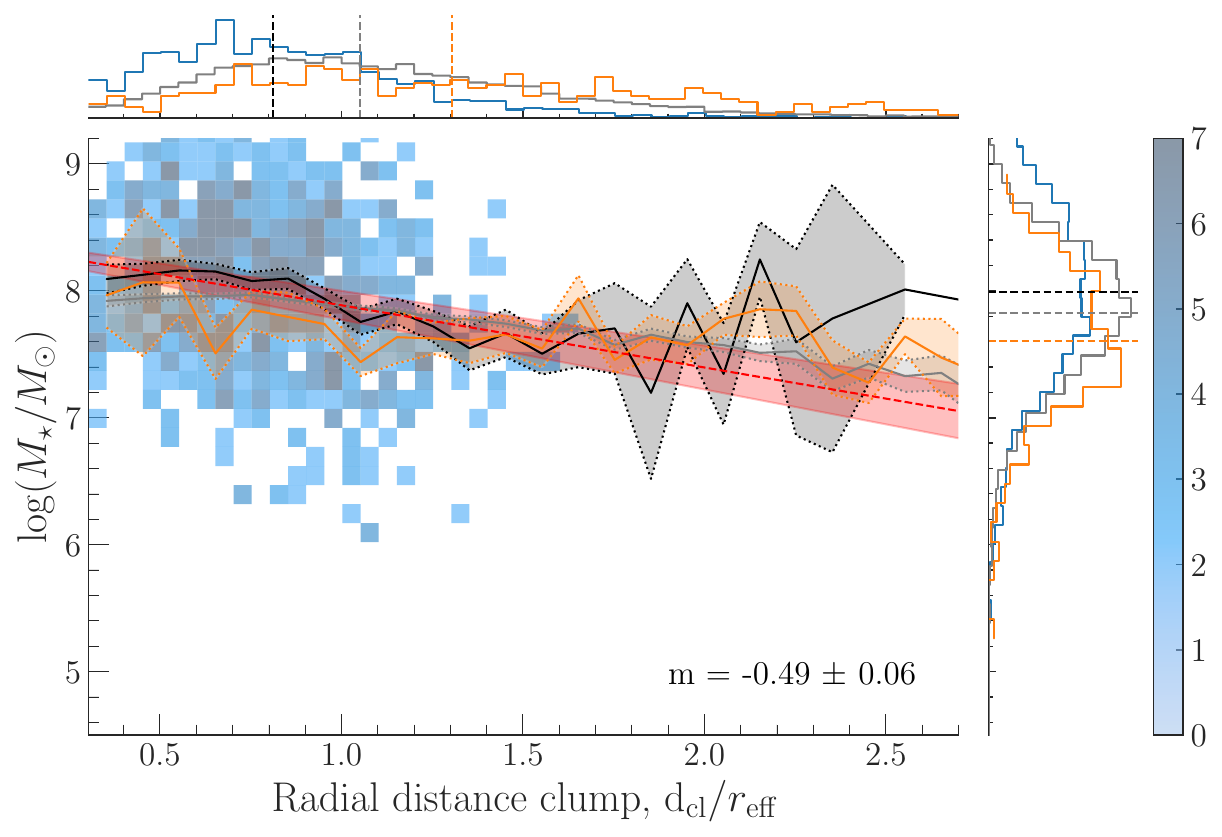} }}
    \\
    \subfloat[\centering Clump stellar age distribution. \label{fig:hsc_discussion_form_evol_gradients_age_a}]{{\includegraphics[width=1\columnwidth]{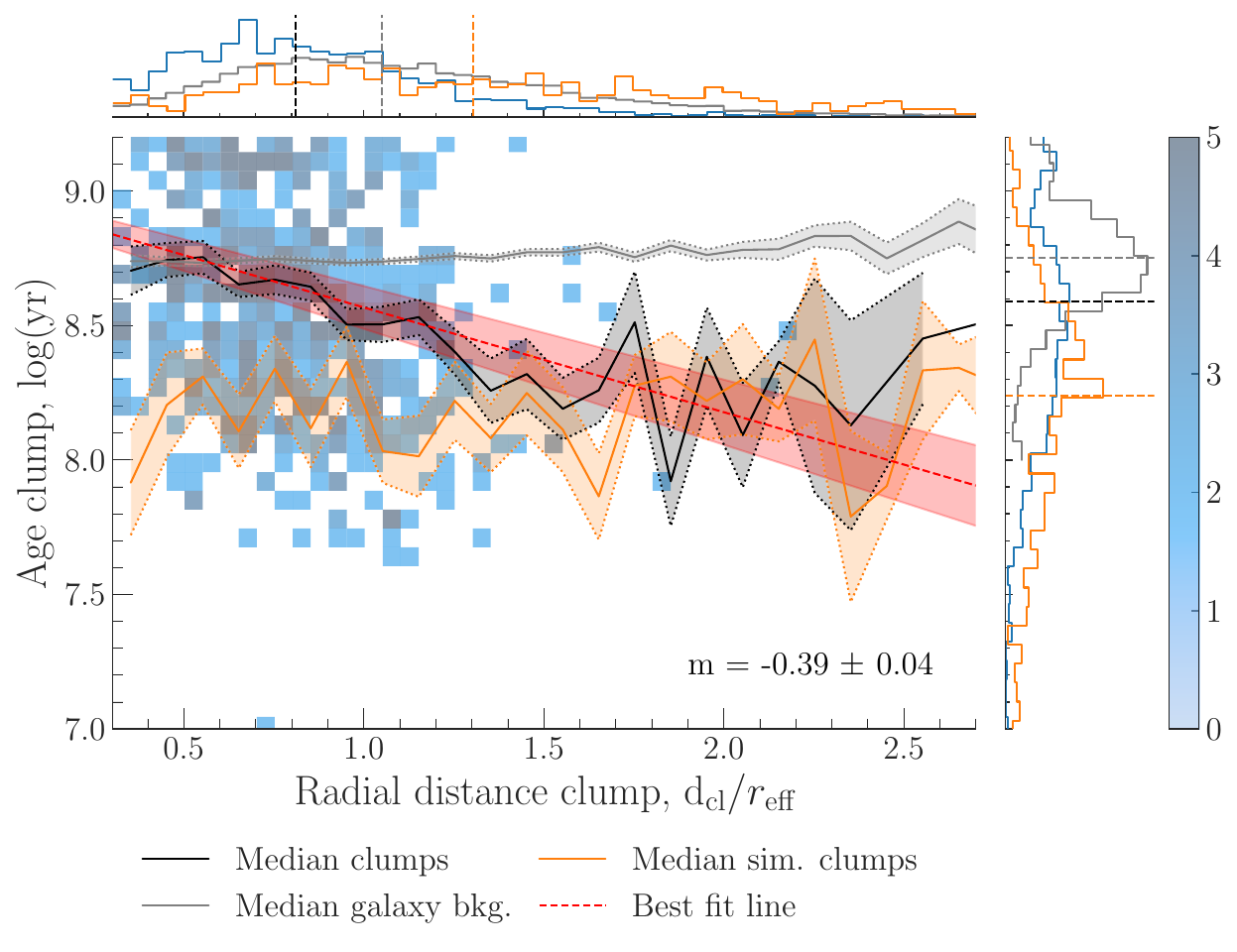} }}
    \caption[Clump stellar mass and age distribution as a function of the normalised galactocentric distance for different clump definitions.]{Similar to Fig. \ref{fig:hsc_phys_props_gradient_UR_reff} but showing the clump stellar mass (a) and age distributions (b) as a function of the normalised galactocentric distance for clumps defined by a \textit{u}-band clump-galaxy flux ratio $F_{u,\mathrm{cl}}/F_{u,\mathrm{gal}} \geq 0.08$ from the CLAUDS/HSC-SSP galaxy sample.}
    \label{fig:hsc_discussion_form_evol_gradients_mass_age}
\end{figure}

If we apply a clump definition that is based on a clump-galaxy \textit{u}-band flux ratio of $F_{u,\mathrm{cl}}/F_{u,\mathrm{gal}} \geq 0.08$ and that is comparable to GSFC definitions used by \citet{Guo2018} and \citet{Adams2022}, we still observe negative stellar mass and age gradients. The estimated stellar mass gradient is flatter ($m=-0.49$, Fig. \ref{fig:hsc_discussion_form_evol_gradients_mass_a}) and the estimated age gradient is steeper for a flux ratio-based definition of clumps ($m=-0.39$, Fig. \ref{fig:hsc_discussion_form_evol_gradients_age_a}) compared to that estimated for our full sample of clumps (stellar mass gradient: $m=-0.54$, age gradient: $m=-0.24$). With that steeper age gradient, the age difference between clumps located at a galactocentric distance of $\mathrm{d}_{\mathrm{cl}}/r_{\mathrm{eff}} \approx 0.3$ and those that are located at a distance of $\mathrm{d}_{\mathrm{cl}}/r_{\mathrm{eff}} \approx 2.0$ is $\sim 350\,\mathrm{Myr}$ between the inner and outer clumps. This value aligns well with the observations from \citet{Guo2018}.

Age gradients that are observed in the host galaxy could also be caused by the inside-out growth of the galactic disk. Observations of the Milky Way and nearby spiral galaxies show that the inner disks are older and have higher metallicities than the outer disk regions \citep[e.g.][]{Pilkington2012,Frankel2019}. We show in Figure \ref{fig:hsc_phys_props_clumps_hist_age} and \ref{fig:hsc_phys_props_gradient_age_reff} that the clumps evolve differently compared to their surrounding galactic disk environment and tend to be younger than the intra-clump regions. The observed age gradients of the clumps are also steeper than the age gradients of the galaxy background (Table \ref{tab:hsc_phys_props_gradient_age}), suggesting that the evolution of the clumps is dynamically separated from the evolution of the galactic disk. Hence, we argue that the observed radial age gradients are the result of the inward migration and cannot be explained by an inside-out growth of the galactic disk alone. 

% In contrast to a negative metallicity gradient that would be expected from an inside-out growth of the galactic disk, the inferred metallicities of our clumps appear to be independent of the radial distance from the galactic centre and the estimated slopes are not significantly different from a flat gradient. 
% We note, however, that the age-metallicity degeneracy is difficult to break with the limited photometric data and, although metallicity is accounted for during the SED fitting process, it is not regarded as a parameter of interest (Section \ref{sec:clump_sed}).

% -----------------------------------------------
% == Low-Redshift Analogues of GSFCs from high-Redshift Galaxies
\subsection{Low-redshift analogues of GSFCs from high-redshift galaxies}\label{sec:hsc_discussion_analogues}
\begin{figure}
    \centering
    \subfloat[\centering Distribution of stellar mass. \label{fig:hsc_discussion_form_evol_gradients_stats_comp_a}]{{\includegraphics[width=0.5\columnwidth]{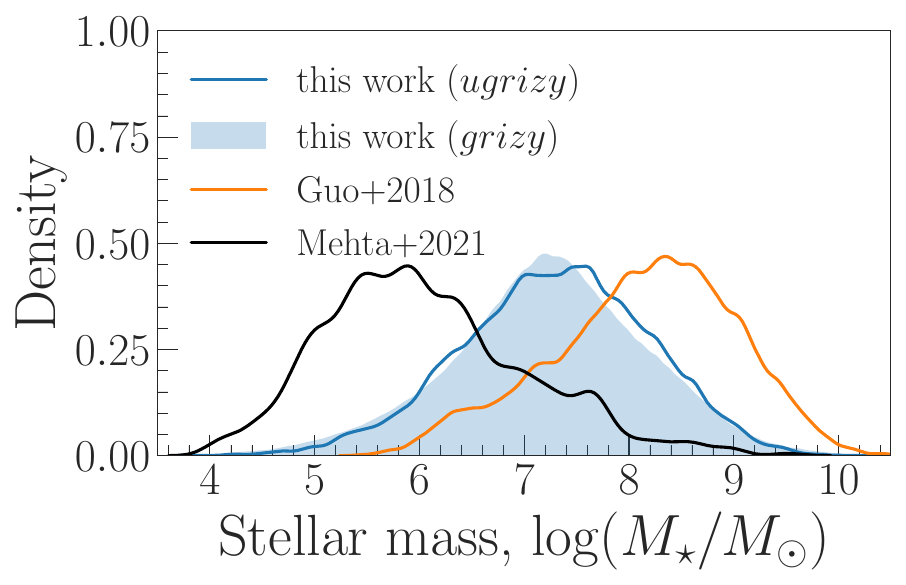} }}
    \subfloat[\centering Distribution of SFR. \label{fig:hsc_discussion_form_evol_gradients_stats_comp_b}]{{\includegraphics[width=0.5\columnwidth]{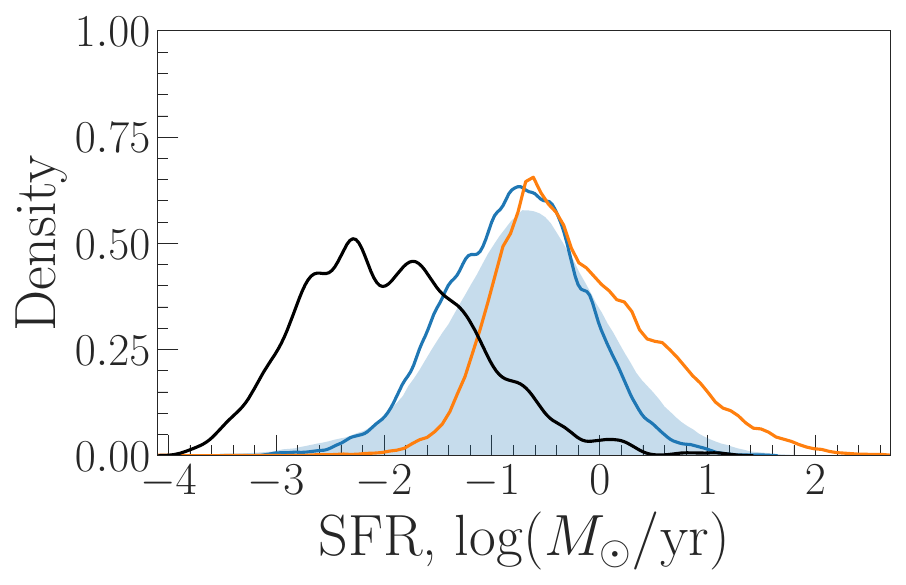} }}
    \\
    \subfloat[\centering Distribution of stellar age. \label{fig:hsc_discussion_form_evol_gradients_stats_comp_c}]{{\includegraphics[width=0.5\columnwidth]{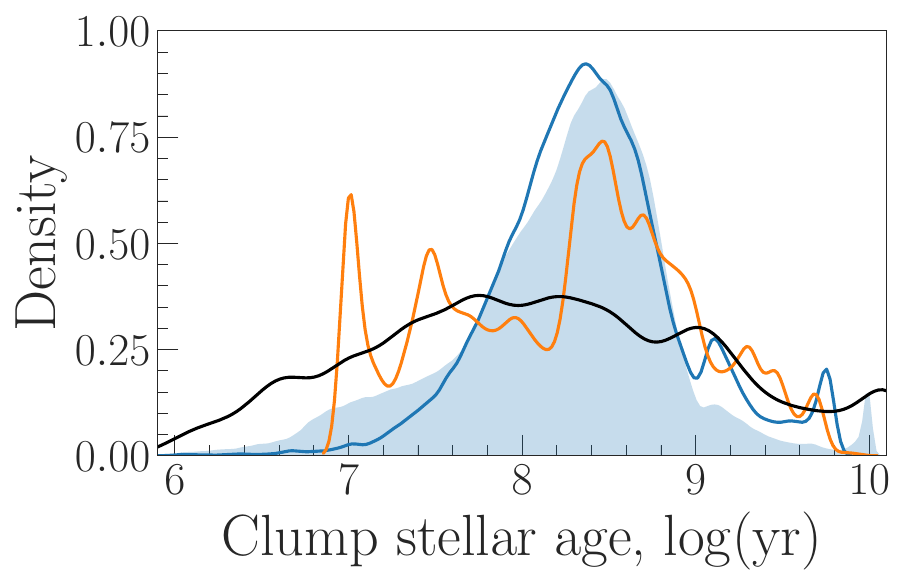} }}
    \subfloat[\centering Distribution of sSFR. \label{fig:hsc_discussion_form_evol_gradients_stats_comp_d}]{{\includegraphics[width=0.5\columnwidth]{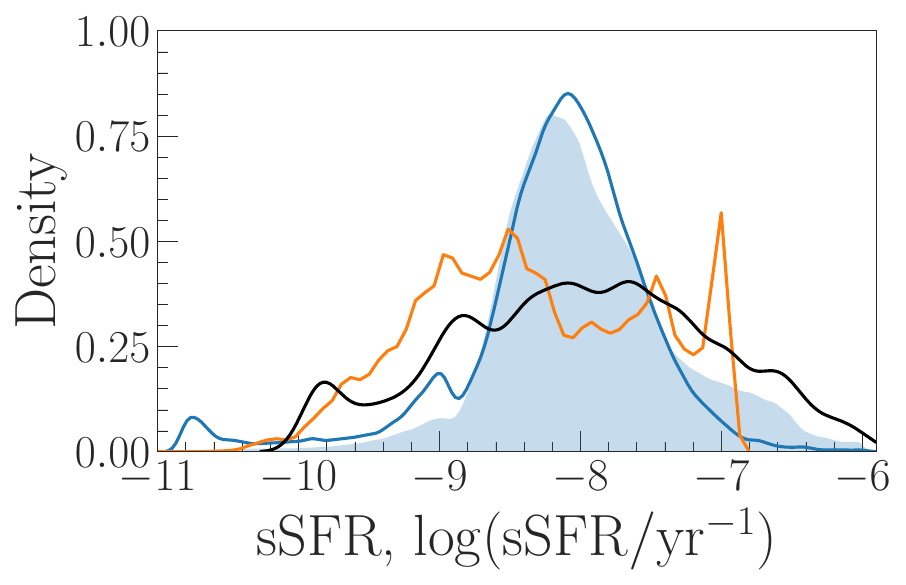} }}
    \caption[Distribution of clump stellar mass, age, SFR and sSFR from different studies.]{Distribution of clump stellar mass, age, SFR and sSFR from different studies. The distributions of the \textit{ugrizy} clumps from this study are shown in blue ($z\leq 0.32$), the \textit{grizy} clumps in orange, the clumps analysed by \citet[][$0.5\leq z \leq3.0$]{Guo2018} in grey and the clump sample used by \citet[][$z < 0.06$]{Mehta2021} in black.}
    \label{fig:hsc_discussion_form_evol_gradients_stats_comp}
\end{figure}

To determine whether the physical properties (Section \ref{sec:hsc_phys_props_clumps} and Fig. \ref{fig:hsc_phys_props_clumps_hist}) that we measured for our samples of low-redshift clumps with 6-filter band photometry and 5-filter band photometry are similar to those estimated for high-redshift clumps, we compare the distribution of the stellar masses, ages, SFRs and sSFRs of our clump sample to the clumps analysed by \citet{Guo2018}. These distributions are shown in Figure \ref{fig:hsc_discussion_form_evol_gradients_stats_comp}. For a comparison with estimated properties from a different sample of low-redshift clumps, we also plot the clump data from \citet{Mehta2021}. 

Figure \ref{fig:hsc_discussion_form_evol_gradients_stats_comp_a} shows that the distributions of the measured clump stellar mass are different for high- and low-redshift clumps. The low-redshift sample from \citet{Mehta2021} is less massive than our sample of low-redshift clumps. This difference is not caused by the different redshift range that both samples probe but probably from a combination of different clump detection methods, background subtraction methods and/or assumptions made for the SED fitting process. On the other hand, the high-redshift sample from \citet{Guo2018} is more massive than the \textit{ugrizy} and \textit{grizy} clump sample. This could indicate that either high-redshift clumps are relatively more massive as might be expected due to the higher gas fraction in those galaxies \citep[e.g.][]{Wang2022} or that the stellar masses of high-redshift clumps are overestimated due to a combination of sample incompleteness, background light contamination and limits in spatial resolution, as pointed out by \citet{HuertasCompany2020}. Moreover, more recent observations using JWST \citep[e.g.][]{Sok2026,Kalita2025} and involving gravitationally lensed clumpy galaxies \citep[e.g.][]{Mestric2022,Claeyssens2023,Messa2024,Claeyssens2025} also indicate that clump stellar masses are generally lower than previously observed. For example, \citet{Claeyssens2025} reported a median stellar mass of $10^{6.7}\,M_\odot$ for clumps detected in galaxies at $0.7<z<1.5$ and of $10^{7.1}\,M_\odot$ at $3.5<z<5.5$. These values are in better agreement to the clump stellar masses we observe from our sample (median stellar mass \textit{ugrizy} clumps: $10^{7.43}\,M_\odot$ and \textit{grizy} clumps: $10^{7.36}\,M_\odot$).

The distributions of the SFR (Fig. \ref{fig:hsc_discussion_form_evol_gradients_stats_comp_b}), stellar age (Fig. \ref{fig:hsc_discussion_form_evol_gradients_stats_comp_c}) and sSFR (Fig. \ref{fig:hsc_discussion_form_evol_gradients_stats_comp_d}) span similar property ranges for all three clump samples, although the SFR measurements from \citet{Mehta2021} peak at lower values due to the lower stellar mass estimates while having similar sSFRs in comparison with our sample. The distribution of SFR measurements from \citet{Guo2018} extends to higher SFRs due to their more massive clump sample but have comparable sSFRs to the other samples.

These results suggest that our observed low-redshift clumps are very similar to those observed at high-redshift and that at least a subset of low-redshift SFGs host clumps that are analogues to the clumps detected in clumpy galaxies at high redshift.

% ================================================
% = Conclusion
% ================================================
\section{Conclusions}\label{sec:conclusion}
In this work, we have presented the analysis of the physical properties of 711,307 clumps from a mass-complete sample of 211,247 galaxies at redshifts $z\leq0.32$. Of those, 14,358 clumps (in 4,162 galaxies) have 6-filter band photometry (\textit{ugrizy}) and 696,949 clumps (in 207,085 galaxies) 5-filter band photometry (\textit{grizy}).

The clumps in our sample were identified using a DL-based object detection model that was trained to detect objects which are similar to those identified visually by human beings. We carefully validated our detections and aperture photometry measurements of the clumps using a large sample of simulated clumps that were injected into the same galaxy images.

We analysed the physical clump properties (colours (\textit{u}-\textit{r}) and (\textit{g}-\textit{r}), stellar mass, age, SFR, sSFR, metallicity and dust attenuation), which we inferred through SED fitting using a Bayesian approach and nested sampling, and have outlined the robust estimation of the cSMF and the radial gradients that describe the variations of the physical properties of the clumps as a function of the distance from the galactic centre.

This is the first study to this date that has analysed such a large sample of star-forming clumps in low-redshift galaxies. Our large clump sample enabled us to conduct a thorough examination of the physical clump properties and to conduct multiple splits of the sample, which was previously difficult to achieve. We have provided robust estimates of the cSMF and the radial variations of the clump properties as a function of galactocentric distance, all in different host galaxy redshift and stellar mass bins. We have also explored the differences between the cSMFs for clump populations at different ages.

The key results are summarised in the following points:
\begin{enumerate}
    \item Using a clump definition based on a clump-galaxy stellar mass ratio of $\log$($M_{\mathrm{cl}}/M_{\mathrm{gal}}$) $\geq -6.6$, we observe distributions of the stellar mass, age, SFR and sSFR of the clumps are similar to the clump property distributions observed from gravitationally lensed and unlensed high-redshift clumps.
    \item The observed cSMF for clumps with estimated ages of $t_{\mathrm{age}} \leq 100\, \mathrm{Myr}$ follows a power-law with an estimated exponent of $\alpha=-1.976 \pm 0.016$. The cSMF of very young clumps is an estimate of the initial cSMF as the majority of clumps have not yet been destroyed by stellar feedback and is consistent with in situ clump formation through VDI processes. The slope of the cSMF generally flattens with increasing stellar mass of the host galaxy and becomes steeper for older clump populations. A clear trend of the power-law slopes with redshift cannot be observed. The change of the cSMF with age of the clumps indicates that the aggregated cSMF is not only determined by formation processes but also reflects the evolution of a coeval clump population within the host galaxy environment.
    \item The large number of clumps detected in this sample of low-redshift galaxies enables a robust estimation of the gradients that describe the variations of the physical properties of the clumps as a function of the distance from the galactic centre. The results show that star-forming clumps in low-redshift galaxies evolve (1) differently compared to the intra-clump regions of the galactic disk and (2) similarly to the GSFCs that are observed in high-redshift galaxies.
    \item The clumps from the mass-complete CLAUDS/HSC-SSP galaxy sample show a negative (\textit{u}-\textit{r}) colour gradient with a slope of $m=-0.162 \pm 0.016$, a negative (\textit{g}-\textit{r}) colour gradient with a slope of $m=-0.102 \pm 0.008$, a negative stellar mass gradient with a slope of $m=-0.539 \pm 0.016$ and a negative stellar age gradient with a slope of $m=-0.235 \pm 0.010$ as a function of the radial distance from the galaxy centre. The gradients are robust against uncertainties introduced by the background subtraction method, the redshift measurement method, the determination of the effective radius used for normalising the radial distance and changes to the assumed SFH of the clumps. The existence of the negative stellar mass and age gradients is not changed by applying different clump definitions.
    \item The data presented in this paper provides further support for an inward migration scenario for clumps that is predicted from theoretical and simulation studies. It can help to better constrain clump formation and evolution simulations by providing a sample of observed star-forming clumps of unprecedented size as, until now, constraints have only been derived from much smaller samples of galaxies with observed clumps.
\end{enumerate}

\section*{Acknowledgements}
% We would like to thank the anonymous referee for their valuable comments and insight, which improved the quality of this paper.

JJP acknowledges funding from the Science and Technology Facilities Council (STFC) Grant Code ST/X508640/1. HD and SS acknowledge funding via the ELSA project. ``ELSA: Euclid Legacy Science Advanced analysis tools'' (Grant Agreement no. 101135203) is funded by the European Union. Views and opinions expressed are however those of the author(s) only and do not necessarily reflect those of the European Union or Innovate UK. Neither the European Union nor the granting authority can be held responsible for them. UK participation is funded through the UK Horizon guarantee scheme under Innovate UK grant 10093177. LFF acknowledges partial support from NASA awards 80NSSC24K1277 and 80NSSC20M0057.

This research made use of the open-source Python scientific computing ecosystem, including \textsc{NumPy} \citep{Harris2020}, \textsc{Matplotlib} \citep{Hunter2007}, \textsc{seaborn} \citep{Waskom2021} and \textsc{Pandas} \citep{McKinney2010}. This research made use of \textsc{Astropy}, a community-developed core Python package for Astronomy \citep{astropy2022} and the \textsc{Photutils} Python package \citep{LarryBradley2025}.

The authors acknowledge the Digital Research Alliance of Canada (\url{https://alliancecan.ca/}) and the Minnesota Supercomputing Institute (MSI, \url{https://www.msi.umn.edu/}) at The University of Minnesota for providing high-performance computing (HPC) resources that have contributed to the research results reported within this paper.

%%%%%%%%%%%%%%%%%%%%%%%%%%%%%%%%%%%%%%%%%%%%%%%%%%
\section*{Data Availability}
The training data and Python code examples for the FRCNN models and the adjusted feature extraction backbone are available from \citet{Popp2025} and a public Github repository\footnote{\url{https://github.com/ou-astrophysics/Zoobot-for-image-segmentation-and-object-detection}}. The catalogue of star-forming clumps, including the measured photometry and estimated physical properties, is available from \citet{Popp2026}.



\bibliographystyle{mnras}
\bibliography{LIB_Clumpy_Galaxies} % if your bibtex file is called example.bib

%\begin{thebibliography}{99}
%\end{thebibliography}

%%%%%%%%%%%%%%%%%%%%%%%%%%%%%%%%%%%%%%%%%%%%%%%%%%

%%%%%%%%%%%%%%%%% APPENDICES %%%%%%%%%%%%%%%%%%%%%

\appendix
% -----------------------------------------------
% SED Fitting Parameters and Execution
\section{Prospector parameters and SED fitting execution}\label{sec:hsc_sed_fitting_exec}
SED fitting with \textsc{Prospector} was set up using dynamic nested sampling with \textsc{dynesty} \citep{Speagle2020}. We run the nested sampler with the number of initial live points (\texttt{nlive\_init}) set to 400 and iteratively adding an additional 100 live points with each batch (\texttt{nlive\_batch}) after a first estimation of the prior volume and importance weights has been established with the initial live points. The additional batches of live points (up to 10 batches, set through \texttt{nested\_maxbatch}) are placed in locations of the prior volume that have not been sampled densely enough. Generally, the live points were selected from the prior volume using multiple bounding ellipsoids to describe the iso-likelihood contours (\texttt{nested\_bound=`multi'}). The posterior was sampled with a random walk (\texttt{nested\_method=`rwalk'}) up to 48 steps away from each live point (\texttt{nested\_walks}).

\textsc{dynesty} allows users to define stopping criteria, so that sampling is stopped when a certain threshold is reached in any of them. These criteria contain hard stops defined by the maximum number of iterations (\texttt{nested\_maxiter}) and log-likelihood calls made during a single run (\texttt{nested\_maxcall}) or the maximum number of dynamic batches that is controlled by the parameter \texttt{nested\_maxbatch}. These stopping criteria were set to their default values and are listed with all other run parameters for the \textsc{Prospector} fits in Table \ref{tab:hsc_sed_fitting_exec}.

\begin{table}
	\centering
	\caption[Run parameters for SED fitting with dynamic nested sampling using \textsc{Prospector}.]{Run parameters for SED fitting with dynamic nested sampling using \textsc{Prospector}. For details see \citet{Johnson2021}.}
    \label{tab:hsc_sed_fitting_exec}
	\footnotesize
        \begin{tabular}{ll}
		\hline
		\textsc{Prospector} parameter         & Value \\
		\hline
        \texttt{nested\_bound}                & \texttt{`multi'} \\
        \texttt{nested\_method}               & \texttt{`rwalk'} \\
        \texttt{nested\_walks}                & \texttt{48} \\
        \texttt{nlive\_init}                  & \texttt{400} \\
        \texttt{nlive\_batch}                 & \texttt{100} \\
        \texttt{nested\_dlogz\_init}          & \texttt{0.05} \\
        \texttt{nested\_maxcall}              & \texttt{50000000} \\
        \texttt{nested\_maxiter}              & \texttt{1000000} \\
        \texttt{nested\_maxbatch}             & \texttt{10} \\
        \texttt{nested\_bootstrap}            & \texttt{0} \\
        \texttt{nested\_target\_n\_effective} & \texttt{3000} \\
        \hline
	\end{tabular}
\end{table}

In addition to the hard stops, two other stopping criteria can be defined that depend on the evidence and on the effective prior sample size (EPSS). We set the limit for the evidence to $\texttt{nested\_dlogz\_init}=0.05$, so that the fitting run stops if the fraction of the estimated remaining evidence falls below that value. The EPSS criterion is based on the minimum number of effective samples reached (\texttt{nested\_target\_n\_effective}) which we set to 3,000. The threshold of 3,000 was chosen after tests on a random sample of 1,000 real clump observations for which we run \textsc{Prospector} fits with \texttt{nested\_target\_n\_effective} set to 500, 1,000, 3,000, 5,000, 10,000 and 20,000. The SED model was constructed similarly to the model described in Table \ref{tab:hsc_sed_fitting_priors} and fitted to the measured \textit{ugrizy} photometry of the clump samples. In general, a higher effective sample threshold could improve the posterior quality as more informative samples are used to approximate the posterior. However, increasing the threshold also increases the computation time for each single fit which is shown in Table \ref{tab:hsc_sed_fitting_exec_ess} for a single CPU together with the average number of samples or iterations needed to reach a specific EPSS threshold. 

\begin{table}
	\centering
	\caption[Mean and standard deviation for the number of iterations and fitting times on a single CPU for different values of \texttt{nested\_target\_n\_effective}.]{Mean and standard deviation (SD) for the number of iterations and fitting times on a single CPU for different values of \texttt{nested\_target\_n\_effective} as stopping criterion.}
    \label{tab:hsc_sed_fitting_exec_ess}
	\footnotesize
        \begin{tabular}{rrrrr}
		\hline
		\multicolumn{1}{c}{Effective}   & \multicolumn{2}{c}{Iterations} & \multicolumn{2}{c}{Fitting time} \\
        \multicolumn{1}{c}{sample size} & \multicolumn{1}{c}{Mean} & \multicolumn{1}{c}{SD} & \multicolumn{1}{c}{Mean} & \multicolumn{1}{c}{SD} \\
		\hline
           500 &  2,278 &   198 &  4m 30s &  1m 49s \\
         1,000 &  3,012 &   301 &  5m 27s &  4m 32s \\
         3,000 &  4,363 &   346 &  8m 33s &  6m 12s \\
         5,000 &  8,129 &   851 & 11m 47s &  9m 05s \\
        10,000 & 15,301 & 1,206 & 19m 11s & 15m 12s \\
        20,000 & 28,553 & 2,589 & 34m 31s & 24m 11s \\
        \hline
	\end{tabular}
\end{table}

A reasonable trade-off between posterior quality and computational time is achieved by a threshold value of 3,000 for the parameter \texttt{nested\_target\_n\_effective}. Below a threshold of 3,000 the posterior quality decreased significantly and the resulting point estimates differed significantly from estimates that were obtained with a higher threshold for the required EPSS. The MAP estimates for the physical properties $\log(M_\star/M_\odot)$, $\log(t_{\mathrm{age}}/\mathrm{yr})$, $\log(Z_\star/Z_\odot)$ and $A_V$ differed by $\leq 0.1\,\mathrm{dex}$ between fits run with a EPSS thresholds of 3,000, 5,000, 10,000 and 20,000, but the differences of fits with EPSS thresholds of 500 and 1,000 differed by $\geq 0.5\,\mathrm{dex}$ from those with higher EPSS thresholds for the majority of the clump sample.

We run the fits on different high-performance computing (HPC) clusters provided by the Digital Research Alliance of Canada\footnote{\url{https://alliancecan.ca/}} and the Minnesota Supercomputing Institute\footnote{\url{https://msi.umn.edu/}}. Up to 20 nodes with 128 CPUs each were used in parallel on the HPC clusters.

For each fit, a result file was generated containing every sample obtained from the posterior distribution, the associated weights, log-likelihood and log-probability as well as the cumulative evidence at each iteration. The result files were processed to determine the different point estimates MAP, mode and median with their corresponding credible intervals from the posterior samples.

% -----------------------------------------------
% SED Fitting with 5-Band Photometry
\section{SED fitting with 5- vs. 6-filter band photometry}\label{sec:hsc_sed_eval_sed_GRIZY}
\begin{figure*}
    \centering
    \includegraphics[width=0.9\textwidth]{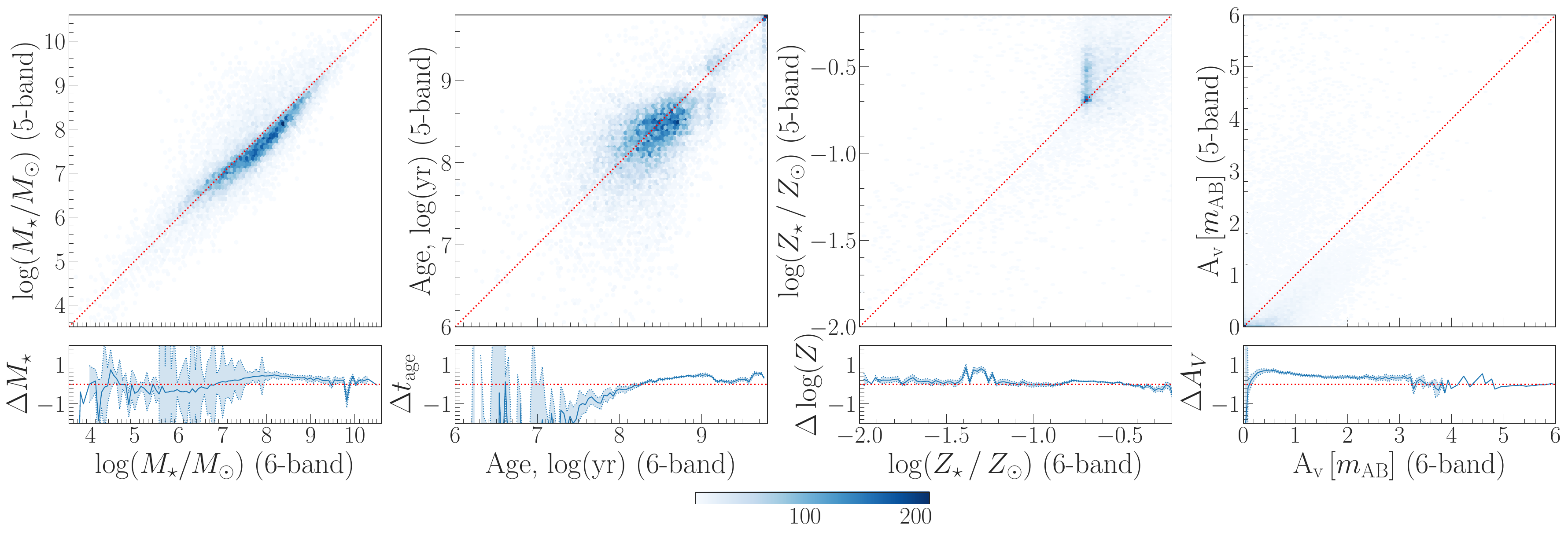}
    \caption[Comparison of the MAP estimates of the physical parameters from a model with observed 6-band (\textit{ugrizy}) and with observed 5-band photometry (\textit{grizy}).]{Comparison of the MAP estimates of the physical parameters stellar mass ($M_\star$), age ($t_{\mathrm{age}}$), metallicity ($Z_\star$) and dust attenuation ($A_V$) from a model with observed 6-band (\textit{ugrizy}) and with observed 5-band photometry (\textit{grizy}). The lower plot in each row shows the median of the normalised differences and the 16th and 84th percentiles as shaded areas. The differences for the each parameter are calculated as $\Delta \boldsymbol{\theta} = (\widehat{\boldsymbol{\theta}}_{\mathrm{MAP},\,6\mathrm{-band}}-\widehat{\boldsymbol{\theta}}_{\mathrm{MAP},\,5\mathrm{-band}})/\widehat{\boldsymbol{\theta}}_{\mathrm{MAP},\,6\mathrm{-band}}$. The red dotted lines mark the one-to-one relation and $\Delta \boldsymbol{\theta} = 0$ for visual comparison.}
    \label{fig:hsc_sed_eval_sed_GRIZY_comp}
\end{figure*}

We fitted SED models with a constant SFH and the same prior distributions we have described in Section \ref{sec:clump_sed} to our sample of clumps from the CLAUDS/HSC-SSP galaxy sample but omitted the \textit{u}-band photometry data points. The MAP estimates $\widehat{\boldsymbol{\theta}}_{\mathrm{MAP}}$ from both posterior distributions, resulting from the fits on the \textit{grizy} and \textit{ugrizy} datasets, are compared in Figure \ref{fig:hsc_sed_eval_sed_GRIZY_comp}. The plot also shows the normalised differences $\Delta \boldsymbol{\theta} = (\widehat{\boldsymbol{\theta}}_{\mathrm{MAP},\,6\mathrm{-band}}-\widehat{\boldsymbol{\theta}}_{\mathrm{MAP},\,5\mathrm{-band}})/\widehat{\boldsymbol{\theta}}_{\mathrm{MAP},\,6\mathrm{-band}}$ for each of the physical properties stellar mass ($M_\star$), age ($t_{\mathrm{age}}$), metallicity ($Z_\star$) and dust attenuation ($A_V$). 

The medians of the normalised differences are between $-0.5 < \Delta \boldsymbol{\theta} < 0.5$ over most of the ranges of the estimated parameters. However, estimated ages differ considerably, especially for ages that are estimated to be $<10^8\,\mathrm{yr}$ using the \textit{ugrizy} data. The \textit{grizy} model generally estimates higher ages for those clump candidates. At very low $A_V$ values estimated using the \textit{ugrizy} data, the \textit{grizy} model gives a stronger dust attenuation estimate, while it tends to underestimate $A_V$ compared to the 6-channel model for the range of $A_V \gtrsim 0.1\,m_{\mathrm{AB}}$. The stellar masses for most clump candidates are underestimated by the 5-channel model compared to the 6-channel model for the upper half of the mass range ($\gtrsim 10^7\,M_\odot$) but overestimated for the lower half. Estimates from both models for metallicity are very similar for the majority of clump candidates. 

The MAP estimates from the \textit{grizy} model differ from those of the \textit{ugrizy} model for the same clump candidates. Although Figure \ref{fig:hsc_sed_eval_sed_GRIZY_comp} implies that inferred estimates of the stellar mass, metallicity and dust attenuation from five photometry data points are comparable to the estimates from six data points, the age estimates from both models differ significantly and appear less reliable for the \textit{grizy} data for ages of $t_{\mathrm{age}} \lesssim 10^8\,\mathrm{yr}$. However, the $95\%$ credible intervals around the MAP estimates from both models do overlap in more than $90\%$ of all cases. From the 19,764 clump candidates that have measured magnitudes brighter than the detection limit in all filter bands, the credible intervals (CIs) inferred from the 5-channel fits overlap with the CIs from the 6-channel fits for 17,805 ($90.1\%$) clump candidates in all four physical parameters. Taking the CIs of $\widehat{\theta}_{t_{\mathrm{age}}, \mathrm{MAP}}$ alone, there are 19,011 ($96.2\%$) detected clumps with overlapping CIs from both models. The least overlap is seen for dust attenuation $A_V$ (18,187 or $92.0\%$). The CIs for stellar mass and metallicity both overlap for $>99\%$ objects.

% -----------------------------------------------
% SED Fitting of the underlying Galaxy Light
\section{SED fitting of the underlying galaxy light}\label{sec:hsc_sed_eval_sed_background}
We assume that the clumps are located in an inter-galactic environment that is older than the clumps and has long since reached its star-formation peak. The SFH is assumed to be an exponentially declining or simple $\tau$ model which has been successfully used to describe typical star-forming galaxies \citep[e.g.][]{Brinchmann2004}. The age prior distribution was set to $\log(t_{\mathrm{age}}) = \mathcal{U}(8.00, 10.14)$ to allow for stellar population ages of up to $13.8\,\mathrm{Gyr}$. Metallicity is fixed to solar metallicity following the results from \citet{Brinchmann2004} for nearby SFGs. Dust attenuation is expected to be lower than in actively star-forming regions so the prior for $A_V = \mathcal{U}(0.0, 2.0)$. 

The prior distributions are listed in Table \ref{tab:hsc_sed_eval_sed_background} and all other parameters and settings for the SED fitting procedures are the same as for the clump candidates (Section \ref{sec:clump_sed} and Appendix \ref{sec:hsc_sed_fitting_exec}).

\begin{table}
	\centering
	\caption[Prior distributions of the SED model with an exponentially declining (simple $\tau$) SFH for the diffuse galactic background.]{Prior distributions of the SED model with an exponentially declining (simple $\tau$) SFH for the diffuse galactic background. Stellar and gas metallicity are fixed to solar metallicity and redshift is fixed to the redshift of the host galaxy.}
    \label{tab:hsc_sed_eval_sed_background}
	\footnotesize
        \begin{tabular}{lllr}
		\hline
		Physical parameter & Units & Distribution & Prior range \\
		\hline
        $\tau$             & Gyr                     & log-uniform & $[0.1, 10]$ \\
        $M_\star$          & $M_\odot$               & log-uniform & $[10^2, 10^{11}]$ \\
        $t_{\mathrm{age}}$ & Gyr                     & uniform     & $[0.1, 13.8]$ \\
        $Z_\star$          & $\log(Z_\star/Z_\odot)$ & (fixed)     & $0.0$ \\
        $A_V$              & $m_{\mathrm{AB}}$       & uniform     & $[0.0, 2.0]$ \\
        Ionisation         & $\log(U)$               & (fixed)     & $-2.0$ \\
        $Z_{\mathrm{gas}}$ & $\log(Z_{\mathrm{gas}}/Z_\odot)$ & (fixed) & $0.0$ \\
        Redshift           & $z$                     & (fixed)     & $[0.005, 0.5]$ \\
        \hline
	\end{tabular}
\end{table}

% -----------------------------------------------
% Values and uncertainties of the cSMF
\section{Estimated slopes of the cSMF}\label{sec:cSMF_tables}
Assuming a scale-free relation \citep[see][for example]{Guszejnov2018} and following the notation recommended by \citet{Hopkins2018}, an empirical stellar mass function can be approximated by a power-law model of the form:
\begin{equation}\label{eq:hsc_phys_props_cSMF_nonlog}
    \frac{\dd N}{\dd M_\star} \propto M_\star^{\alpha},
\end{equation}
where $\dd N/\dd M_\star$ is the number count of clumps in the stellar mass bin $\dd M_\star$ and $\alpha$ a dimensionless exponent. The cSMF can also be expressed per logarithmic stellar mass interval and written in linear form as:
\begin{equation}\label{eq:hsc_phys_props_cSMF_linear}
    \log \left(\frac{\dd N}{\dd \log(M_\star)}\right) = \Gamma \log(M_\star)+c,
\end{equation}
where $\Gamma = \alpha + 1$ and $c$ is an empirically determined constant. For consistency with other publications, we will report the power-law exponent $\alpha = \Gamma -1$ that is used in Equation \ref{eq:hsc_phys_props_cSMF_nonlog} and also include the sign of the exponent instead of reporting only the absolute value to avoid ambiguity.

Table \ref{tab:hsc_phys_props_cSMF_clump_age} lists the estimated power-law exponents $\alpha$ for different subsamples of the mass-complete sample of the CLAUDS/HSC-SSP galaxy sample with 6-filter band photometry and for different subsamples of clumps based on their estimated ages.

\begin{table*}
	\centering
	\caption[Measured slopes $\alpha$ of the best power-law fit to the cSMF for different samples.]{Measured slopes $\alpha$ of the best power-law fit to the cSMF for different samples. The linear models (see Equation \ref{eq:hsc_phys_props_cSMF_linear}) were fitted to the massive ends of the cSMF and the fitted values for $\alpha = \Gamma -1$ are listed with their standard errors and the 95\% confidence interval for $\alpha$.}
    \label{tab:hsc_phys_props_cSMF_clump_age}
	\footnotesize
        \begin{tabular}{llrrrr}
		\hline
		\multicolumn{1}{c}{Redshift} & \multicolumn{1}{c}{Galaxy mass} & \multicolumn{1}{c}{Galaxies} & \multicolumn{1}{c}{Clumps} & \multicolumn{1}{c}{Slope $\alpha$} & \multicolumn{1}{c}{95\% conf. interval} \\
        \multicolumn{1}{c}{$z$} & \multicolumn{1}{c}{$\log(M_{\mathrm{gal}}/M_\odot)$} & & & & \\
		\hline
        \multicolumn{6}{l}{\textbf{All clumps}} \\
        \hline
        $\leq 0.32$ & $\geq 9.0$              & 4,162 & 14,358 & $-2.297 \pm 0.004$  & $[-2.305, -2.288]$ \\ 
        $0.0 \leq z \leq 0.1$ & $[9.0, 9.9)$  &   251 &  1,167 & $-2.534 \pm 0.055$  & $[-2.654, -2.414]$ \\
        $0.0 \leq z \leq 0.1$ & $[9.8, 10.6)$ &   170 &  1,266 & $-2.358 \pm 0.077$  & $[-2.526, -2.190]$ \\
        $0.0 \leq z \leq 0.1$ & $\geq 10.6$   &    88 &    539 & $-1.692 \pm 0.021$  & $[-1.736, -1.649]$ \\
        $0.1 < z \leq 0.2$ & $[9.0, 9.9)$     &   671 &  1,864 & $-2.818 \pm 0.068$  & $[-2.966, -2.669]$ \\
        $0.1 < z \leq 0.2$ & $[9.8, 10.6)$    &   866 &  3,442 & $-2.610 \pm 0.062$  & $[-2.744, -2.477]$ \\
        $0.1 < z \leq 0.2$ & $\geq 10.6$      &   396 &  1,572 & $-2.220 \pm 0.048$  & $[-2.323, -2.116]$ \\
        $z > 0.2$ & $[9.0, 9.9)$              &   423 &    862 & $-2.776 \pm 0.087$  & $[-2.971, -2.582]$ \\
        $z > 0.2$ & $[9.8, 10.6)$             & 1,375 &  3,575 & $-2.530 \pm 0.045$  & $[-2.626, -2.435]$ \\
        $z > 0.2$ & $\geq 10.6$               & 1,003 &  2,230 & $-2.273 \pm 0.034$  & $[-2.345, -2.201]$ \\
        \hline
        \multicolumn{6}{l}{\textbf{Clumps in SFGs}} \\
        \hline
        $\leq 0.32$ & $\geq 9.0$              & 3,392 & 12,514 & $-2.370 \pm 0.005$  & $[-2.381, -2.360]$ \\
        \hline
        \multicolumn{6}{l}{\textbf{Clumps in non-SFGs}} \\
        \hline
        $\leq 0.32$ & $\geq 9.0$              &   770 &  1,844 & $-2.093 \pm 0.008$  & $[-2.109, -2.078]$ \\
		\hline
        \multicolumn{6}{l}{\textbf{Clump stellar age:} $t_{\mathrm{age}}\leq 100\,\mathrm{Myr}$} \\
        \hline
        $\leq 0.32$ & $\geq 9.0$              & 1,399 & 2,972 & $-1.976 \pm 0.016$  & $[-2.008, -1.943]$ \\
        $0.0 \leq z \leq 0.1$ & $[9.0, 9.9)$  & 140 & 485 & $-2.315 \pm 0.091$  & $[-2.514, -2.116]$ \\
        $0.0 \leq z \leq 0.1$ & $[9.8, 10.6)$ & 88 & 466 & $-1.993 \pm 0.032$  & $[-2.061, -1.925]$ \\
        $0.0 \leq z \leq 0.1$ & $\geq 10.6$   & 43 & 183 & $-1.721 \pm 0.051$  & $[-1.829, -1.612]$ \\
        $0.1 < z \leq 0.2$ & $[9.0, 9.9)$     & 218 & 339 & $-2.506 \pm 0.106$  & $[-2.743, -2.269]$ \\
        $0.1 < z \leq 0.2$ & $[9.8, 10.6)$    & 288 & 538 & $-2.148 \pm 0.047$  & $[-2.249, -2.047]$ \\
        $0.1 < z \leq 0.2$ & $\geq 10.6$      & 129 & 243 & $-1.950 \pm 0.048$  & $[-2.054, -1.846]$ \\
        $z > 0.2$ & $[9.0, 9.9)$              & 122 & 167 & $-2.260 \pm 0.117$  & $[-2.530, -1.989]$ \\
        $z > 0.2$ & $[9.8, 10.6)$             & 401 & 581 & $-2.061 \pm 0.041$  & $[-2.148, -1.975]$ \\
        $z > 0.2$ & $\geq 10.6$               & 249 & 351 & $-1.723 \pm 0.023$  & $[-1.772, -1.675]$ \\
		\hline
        \multicolumn{6}{l}{\textbf{Clump stellar age:} $100 < t_{\mathrm{age}}\leq 300\,\mathrm{Myr}$} \\
        \hline
        $\leq 0.32$ & $\geq 9.0$              & 2,619 & 5,570 & $-2.209 \pm 0.018$  & $[-2.246, -2.173]$ \\
        $0.0 \leq z \leq 0.1$ & $[9.0, 9.9)$  & 173 & 397 & $-2.527 \pm 0.065$  & $[-2.672, -2.381]$ \\
        $0.0 \leq z \leq 0.1$ & $[9.8, 10.6)$ & 113 & 374 & $-1.956 \pm 0.030$  & $[-2.020, -1.892]$ \\
        $0.0 \leq z \leq 0.1$ & $\geq 10.6$   & 59 & 197 & $-1.941 \pm 0.047$  & $[-2.043, -1.839]$ \\
        $0.1 < z \leq 0.2$ & $[9.0, 9.9)$     & 467 & 900 & $-2.948 \pm 0.103$  & $[-3.177, -2.719]$ \\
        $0.1 < z \leq 0.2$ & $[9.8, 10.6)$    & 531 & 1,252 & $-2.325 \pm 0.063$  & $[-2.459, -2.190]$ \\
        $0.1 < z \leq 0.2$ & $\geq 10.6$      & 213 & 540 & $-1.900 \pm 0.027$  & $[-1.957, -1.843]$ \\
        $z > 0.2$ & $[9.0, 9.9)$              & 280 & 438 & $-2.687 \pm 0.125$  & $[-2.975, -2.399]$ \\
        $z > 0.2$ & $[9.8, 10.6)$             & 862 & 1,545 & $-2.459 \pm 0.061$  & $[-2.590, -2.327]$ \\
        $z > 0.2$ & $\geq 10.6$               & 527 & 868 & $-1.951 \pm 0.031$  & $[-2.017, -1.886]$ \\
		\hline
        \multicolumn{6}{l}{\textbf{Clump stellar age:} $t_{\mathrm{age}}>300\,\mathrm{Myr}$} \\
        \hline
        $\leq 0.32$ & $\geq 9.0$              & 2,837 & 5,816 & $-2.240 \pm 0.026$  & $[-2.293, -2.187]$ \\
        $0.0 \leq z \leq 0.1$ & $[9.0, 9.9)$  & 135 & 285 & $-2.434 \pm 0.094$  & $[-2.643, -2.225]$ \\
        $0.0 \leq z \leq 0.1$ & $[9.8, 10.6)$ & 132 & 426 & $-2.323 \pm 0.102$  & $[-2.544, -2.101]$ \\
        $0.0 \leq z \leq 0.1$ & $\geq 10.6$   & 62 & 159 & $-1.707 \pm 0.044$  & $[-1.802, -1.612]$ \\
        $0.1 < z \leq 0.2$ & $[9.0, 9.9)$     & 377 & 625 & $-2.784 \pm 0.123$  & $[-3.058, -2.509]$ \\
        $0.1 < z \leq 0.2$ & $[9.8, 10.6)$    & 710 & 1,652 & $-2.540 \pm 0.103$  & $[-2.760, -2.320]$ \\
        $0.1 < z \leq 0.2$ & $\geq 10.6$      & 317 & 789 & $-2.134 \pm 0.050$  & $[-2.239, -2.029]$ \\
        $z > 0.2$ & $[9.0, 9.9)$              & 188 & 257 & $-2.545 \pm 0.177$  & $[-2.953, -2.138]$ \\
        $z > 0.2$ & $[9.8, 10.6)$             & 869 & 1,449 & $-2.427 \pm 0.056$  & $[-2.547, -2.307]$ \\
        $z > 0.2$ & $\geq 10.6$               & 650 & 1,011 & $-2.161 \pm 0.041$  & $[-2.248, -2.074]$ \\
		\hline
	\end{tabular}
\end{table*}
\section{Values and uncertainties of the observed radial gradients}\label{sec:grad_tables}
We provide a full list of our estimated (\textit{u}-\textit{r}) and (\textit{g}-\textit{r}) colour gradients in Table \ref{tab:hsc_phys_props_gradient_UR} and Table \ref{tab:hsc_phys_props_gradient_GR}, our estimated stellar mass gradients in Table \ref{tab:hsc_phys_props_gradient_mass} and our estimated age gradients of the clumps in Table \ref{tab:hsc_phys_props_gradient_age}. The slopes of the gradients were estimated using linear models that were fitted to the distributions of the clump physical properties as a function of the radial distance. The tables list the gradient  with its standard error, the 95\% confidence interval and the $p$-value from the F-test for statistical significance. Furthermore, we added the gradient estimates for the diffuse galaxy background at the clumps' positions and for our sample of simulated clumps. These values are used to compare the measurements of the real clumps to the randomly distributed set of simulated clumps and to the galactic environment the clumps are embedded in (see main text).

\begin{table*}
	\centering
	\caption[Values and uncertainties of the radial (\textit{u}-\textit{r}) colour gradient $m$.]{Values and uncertainties of the radial (\textit{u}-\textit{r}) colour gradients $m$ for the total sample and binned by redshift and host galaxy mass. Linear models of the form $(u-r) = m (\mathrm{d}_{\mathrm{cl}}/r_{\mathrm{eff}})+c$ were fitted to the distributions of the (\textit{u}-\textit{r}) colours as a function of the normalised radial distance $\mathrm{d}_{\mathrm{cl}}/r_{\mathrm{eff}}$ for the clumps, the underlying galaxy background and the simulated clumps. Only clumps from the mass-complete galaxy sample ($\log(M_{\mathrm{gal}}/M_\odot) \geq 9.0,\,z \leq 0.32$) that lie between $0.3 \leq \mathrm{d}_{\mathrm{cl}}/r_{\mathrm{eff}} \leq 2.7$ were considered. The table lists the gradient $m$ with its standard error, the constant $c$ with its standard error, the 95\% confidence interval for $m$ and the $p$-value from the F-test for statistical significance. The stellar mass bins for the galaxies with redshift $z\leq 0.32$ are: $9.0 \leq \log(M_{\mathrm{gal}}/M_\odot) < 9.8$ (low-mass), $9.8 \leq \log(M_{\mathrm{gal}}/M_\odot) < 10.6$ (medium-mass) and $\log(M_{\mathrm{gal}}/M_\odot) \geq 10.6$ (high-mass).}
    \label{tab:hsc_phys_props_gradient_UR}
	\footnotesize
        \begin{tabular}{lrrrrr}
		\hline
		Sample & \multicolumn{1}{l}{Count} & \multicolumn{1}{l}{Gradient $m$} & \multicolumn{1}{l}{Constant $c$} & \multicolumn{1}{l}{95\% conf. interval}  & \multicolumn{1}{l}{$p$-value} \\
		\hline
        \multicolumn{5}{l}{\textbf{\textit{ugrizy} clumps}} \\
        \hline
        Total & 14,249 & $-0.162 \pm 0.016$ & $1.421 \pm 0.019$ & $[-0.193, -0.132]$ & $6.96 \times 10^{-25}$ \\
        \textit{thereof:} $z \leq 0.1$ & 2,897 & $-0.189 \pm 0.032$ & $1.170 \pm 0.039$ & $[-0.253, -0.126]$ & $5.05 \times 10^{-09}$ \\
        \textit{thereof:} $0.1 < z \leq 0.2$ & 6,832 & $-0.171 \pm 0.023$ & $1.456 \pm 0.027$ & $[-0.215, -0.127]$ & $3.77 \times 10^{-14}$ \\
        \textit{thereof:} $0.2 < z \leq 0.32$ & 4,520 & $-0.123 \pm 0.028$ & $1.520 \pm 0.033$ & $[-0.179, -0.067]$ & $1.55 \times 10^{-05}$ \\
        \textit{thereof:} low-mass gal. & 3,650 & $-0.181 \pm 0.028$ & $1.168 \pm 0.029$ & $[-0.236, -0.126]$ & $1.17 \times 10^{-10}$ \\
        \textit{thereof:} med.-mass gal. & 6,960 & $-0.203 \pm 0.022$ & $1.473 \pm 0.027$ & $[-0.247, -0.160]$ & $8.12 \times 10^{-20}$ \\
        \textit{thereof:} high-mass gal. & 3,639 & $-0.374 \pm 0.033$ & $1.939 \pm 0.043$ & $[-0.438, -0.309]$ & $1.69 \times 10^{-29}$ \\
        \textit{thereof:} in SFGs & 12,414 & $-0.225 \pm 0.016$ & $1.406 \pm 0.019$ & $[-0.257, -0.193]$ & $1.84 \times 10^{-42}$ \\
        \textit{thereof:} in non-SFGs & 1,835 & $-0.343 \pm 0.044$ & $2.216 \pm 0.061$ & $[-0.429, -0.257]$ & $8.86 \times 10^{-15}$ \\
		\hline
        \multicolumn{5}{l}{\textbf{Galaxy background}} \\
        \hline
        Total & 14,057 & $0.075 \pm 0.010$ & $1.647 \pm 0.012$ & $[0.056, 0.094]$ & $1.8 \times 10^{-14}$ \\
        \textit{thereof:} $z \leq 0.1$ & 2,781 & $-0.034 \pm 0.020$ & $1.610 \pm 0.025$ & $[-0.074, 0.006]$ & $0.095$ \\
        \textit{thereof:} $0.1 < z \leq 0.2$ & 6,773 & $0.105 \pm 0.013$ & $1.623 \pm 0.016$ & $[0.079, 0.131]$ & $5.12 \times 10^{-15}$ \\
        \textit{thereof:} $0.2 < z \leq 0.32$ & 4,503 & $0.112 \pm 0.019$ & $1.690 \pm 0.022$ & $[0.076, 0.148]$ & $1.58 \times 10^{-09}$ \\
        \textit{thereof:} low-mass gal. & 3,461 & $0.021 \pm 0.018$ & $1.498 \pm 0.019$ & $[-0.014, 0.056]$ & $0.245$ \\
        \textit{thereof:} med.-mass gal. & 6,522 & $0.058 \pm 0.014$ & $1.643 \pm 0.016$ & $[0.031, 0.085]$ & $2.24 \times 10^{-05}$ \\
        \textit{thereof:} high-mass gal. & 4,074 & $-0.052 \pm 0.019$ & $2.009 \pm 0.024$ & $[-0.088, -0.016]$ & $0.005$ \\
		\hline
        \multicolumn{5}{l}{\textbf{Simulated clumps}} \\
        \hline
        Total & 1,586 & $-0.064 \pm 0.029$ & $1.190 \pm 0.045$ & $[-0.121, -0.007]$ & $0.029$ \\
        \textit{thereof:} $z \leq 0.1$ & 361 & $0.054 \pm 0.059$ & $0.782 \pm 0.095$ & $[-0.063, 0.171]$ & $0.363$ \\
        \textit{thereof:} $0.1 < z \leq 0.2$ & 700 & $-0.077 \pm 0.043$ & $1.169 \pm 0.068$ & $[-0.162, 0.008]$ & $0.076$ \\
        \textit{thereof:} $0.2 < z \leq 0.32$ & 525 & $-0.076 \pm 0.049$ & $1.420 \pm 0.072$ & $[-0.171, 0.019]$ & $0.118$ \\
        \textit{thereof:} low-mass gal. & 401 & $0.037 \pm 0.061$ & $0.965 \pm 0.082$ & $[-0.082, 0.157]$ & $0.541$ \\
        \textit{thereof:} med.-mass gal. & 539 & $-0.024 \pm 0.051$ & $1.093 \pm 0.080$ & $[-0.125, 0.077]$ & $0.636$ \\
        \textit{thereof:} high-mass gal. & 646 & $-0.202 \pm 0.045$ & $1.498 \pm 0.075$ & $[-0.291, -0.113]$ & $1.03 \times 10^{-05}$ \\
		\hline
	\end{tabular}
\end{table*}

\begin{table*}
	\centering
	\caption[Values and uncertainties of the radial (\textit{g}-\textit{r}) colour gradient $m$.]{Similar to Table \ref{tab:hsc_phys_props_gradient_UR} but showing the gradient values $m$ of the linear models $(g-r) = m (\mathrm{d}_{\mathrm{cl}}/r_{\mathrm{eff}})+c$ fitted to the distributions of the (\textit{g}-\textit{r}) colours as a function of the normalised radial distance.}
    \label{tab:hsc_phys_props_gradient_GR}
	\footnotesize
        \begin{tabular}{lrrrrr}
		\hline
		Sample & \multicolumn{1}{l}{Count} & \multicolumn{1}{l}{Gradient $m$} & \multicolumn{1}{l}{Constant $c$} & \multicolumn{1}{l}{95\% conf. interval}  & \multicolumn{1}{l}{$p$-value} \\
		\hline
        \multicolumn{5}{l}{\textbf{\textit{ugrizy} clumps}} \\
        \hline
        Total & 14,249 & $-0.102 \pm 0.008$ & $0.690 \pm 0.010$ & $[-0.118, -0.086]$ & $1.98 \times 10^{-35}$ \\
        \textit{thereof:} $z \leq 0.1$ & 2,897 & $-0.104 \pm 0.017$ & $0.480 \pm 0.020$ & $[-0.137, -0.071]$ & $8.23 \times 10^{-10}$ \\
        \textit{thereof:} $0.1 < z \leq 0.2$ & 6,832 & $-0.088 \pm 0.011$ & $0.661 \pm 0.014$ & $[-0.110, -0.066]$ & $1.01 \times 10^{-14}$ \\
        \textit{thereof:} $0.2 < z \leq 0.32$ & 4,520 & $-0.121 \pm 0.014$ & $0.867 \pm 0.017$ & $[-0.149, -0.093]$ & $3.41 \times 10^{-17}$ \\
        \textit{thereof:} low-mass gal. & 3,650 & $-0.098 \pm 0.015$ & $0.498 \pm 0.016$ & $[-0.127, -0.069]$ & $5.75 \times 10^{-11}$ \\
        \textit{thereof:} med.-mass gal. & 6,960 & $-0.154 \pm 0.011$ & $0.765 \pm 0.014$ & $[-0.176, -0.132]$ & $5.99 \times 10^{-42}$ \\
        \textit{thereof:} high-mass gal. & 3,639 & $-0.213 \pm 0.017$ & $0.982 \pm 0.022$ & $[-0.246, -0.180]$ & $5.06 \times 10^{-36}$ \\
		\hline
        \multicolumn{5}{l}{\textbf{\textit{grizy} clumps}} \\
        \hline
        Total & 661,691 & $-0.080 \pm 0.002$ & $0.497 \pm 0.002$ & $[-0.084, -0.077]$ & $0.000$ \\
        \textit{thereof:} $z \leq 0.1$ & 161,340 & $-0.035 \pm 0.003$ & $0.236 \pm 0.004$ & $[-0.042, -0.029]$ & $1.12 \times 10^{-26}$ \\
        \textit{thereof:} $0.1 < z \leq 0.2$ & 316,519 & $-0.064 \pm 0.003$ & $0.487 \pm 0.003$ & $[-0.069, -0.059]$ & $6.12 \times 10^{-131}$ \\
        \textit{thereof:} $z > 0.2$ & 183,832 & $-0.120 \pm 0.004$ & $0.709 \pm 0.004$ & $[-0.127, -0.112]$ & $4.09 \times 10^{-234}$ \\
        \textit{thereof:} low-mass gal. & 171,384 & $-0.063 \pm 0.004$ & $0.285 \pm 0.004$ & $[-0.070, -0.056]$ & $3.42 \times 10^{-71}$ \\
        \textit{thereof:} med.-mass gal. & 324,122 & $-0.121 \pm 0.003$ & $0.556 \pm 0.003$ & $[-0.126, -0.116]$ & $0.000$ \\
        \textit{thereof:} high-mass gal. & 166,185 & $-0.199 \pm 0.003$ & $0.813 \pm 0.005$ & $[-0.206, -0.193]$ & $0.000$ \\
		\hline
        \multicolumn{5}{l}{\textbf{All clumps}} \\
        \hline
        Total & 675,940 & $-0.081 \pm 0.002$ & $0.501 \pm 0.002$ & $[-0.084, -0.077]$ & $0.000$ \\
        \textit{thereof:} $z \leq 0.1$ & 164,237 & $-0.037 \pm 0.003$ & $0.241 \pm 0.004$ & $[-0.043, -0.030]$ & $2.63 \times 10^{-29}$ \\
        \textit{thereof:} $0.1 < z \leq 0.2$ & 323,351 & $-0.064 \pm 0.003$ & $0.491 \pm 0.003$ & $[-0.069, -0.059]$ & $1.91 \times 10^{-136}$ \\
        \textit{thereof:} $z > 0.2$ & 188,352 & $-0.119 \pm 0.004$ & $0.712 \pm 0.004$ & $[-0.126, -0.112]$ & $8.75 \times 10^{-241}$ \\
        \textit{thereof:} low-mass gal. & 175,034 & $-0.064 \pm 0.003$ & $0.290 \pm 0.004$ & $[-0.071, -0.057]$ & $1.8 \times 10^{-75}$ \\
        \textit{thereof:} med.-mass gal. & 331,082 & $-0.122 \pm 0.003$ & $0.560 \pm 0.003$ & $[-0.127, -0.116]$ & $0.000$ \\
        \textit{thereof:} high-mass gal. & 169,824 & $-0.200 \pm 0.003$ & $0.816 \pm 0.004$ & $[-0.206, -0.193]$ & $0.000$ \\
		\hline
        \multicolumn{5}{l}{\textbf{Galaxy background}} \\
        \hline
        Total & 14,057 & $-0.002 \pm 0.006$ & $0.592 \pm 0.007$ & $[-0.013, 0.009]$ & $0.755$ \\
        \textit{thereof:} $z \leq 0.1$ & 2,781 & $-0.010 \pm 0.010$ & $0.488 \pm 0.012$ & $[-0.030, 0.009]$ & $0.306$ \\
        \textit{thereof:} $0.1 < z \leq 0.2$ & 6,773 & $0.007 \pm 0.008$ & $0.553 \pm 0.009$ & $[-0.008, 0.022]$ & $0.380$ \\
        \textit{thereof:} $0.2 < z \leq 0.32$ & 4,503 & $-0.006 \pm 0.011$ & $0.710 \pm 0.013$ & $[-0.027, 0.016]$ & $0.610$ \\
        \textit{thereof:} low-mass gal. & 3,461 & $-0.004 \pm 0.011$ & $0.453 \pm 0.011$ & $[-0.024, 0.017]$ & $0.736$ \\
        \textit{thereof:} med.-mass gal. & 6,522 & $-0.038 \pm 0.008$ & $0.608 \pm 0.009$ & $[-0.053, -0.022]$ & $1.65 \times 10^{-06}$ \\
        \textit{thereof:} high-mass gal. & 4,074 & $-0.085 \pm 0.010$ & $0.850 \pm 0.013$ & $[-0.104, -0.066]$ & $8.87 \times 10^{-18}$ \\
		\hline
        \multicolumn{5}{l}{\textbf{Simulated clumps}} \\
        \hline
        Total & 1,586 & $-0.043 \pm 0.018$ & $0.564 \pm 0.028$ & $[-0.078, -0.008]$ & $0.017$ \\
        \textit{thereof:} $z \leq 0.1$ & 361 & $0.018 \pm 0.038$ & $0.227 \pm 0.061$ & $[-0.056, 0.093]$ & $0.633$ \\
        \textit{thereof:} $0.1 < z \leq 0.2$ & 700 & $-0.041 \pm 0.024$ & $0.539 \pm 0.038$ & $[-0.088, 0.006]$ & $0.089$ \\
        \textit{thereof:} $0.2 < z \leq 0.32$ & 525 & $-0.033 \pm 0.027$ & $0.747 \pm 0.040$ & $[-0.086, 0.021]$ & $0.230$ \\
        \textit{thereof:} low-mass gal. & 401 & $-0.006 \pm 0.037$ & $0.421 \pm 0.050$ & $[-0.078, 0.067]$ & $0.881$ \\
        \textit{thereof:} med.-mass gal. & 539 & $-0.030 \pm 0.032$ & $0.513 \pm 0.049$ & $[-0.093, 0.033]$ & $0.347$ \\
        \textit{thereof:} high-mass gal. & 646 & $-0.130 \pm 0.027$ & $0.785 \pm 0.045$ & $[-0.182, -0.077]$ & $1.76 \times 10^{-06}$ \\
        \hline
	\end{tabular}
\end{table*}

\begin{table*}
	\centering
	\caption[Values and uncertainties of the radial clump stellar mass gradient $m$.]{Similar to Table \ref{tab:hsc_phys_props_gradient_UR} but showing the gradient values $m$ of the linear models $\log(M_{\mathrm{cl}}/M_\odot) = m (\mathrm{d}_{\mathrm{cl}}/r_{\mathrm{eff}})+c$ fitted to the distributions of the clump stellar mass as a function of the normalised radial distance.}
    \label{tab:hsc_phys_props_gradient_mass}
	\footnotesize
        \begin{tabular}{lrrrrr}
		\hline
		Sample & \multicolumn{1}{l}{Count} & \multicolumn{1}{l}{Gradient $m$} & \multicolumn{1}{l}{Constant $c$} & \multicolumn{1}{l}{95\% conf. interval}  & \multicolumn{1}{l}{$p$-value} \\
		\hline
        \multicolumn{5}{l}{\textbf{\textit{ugrizy} clumps}} \\
        \hline
        Total & 14,249 & $-0.539 \pm 0.016$ & $7.903 \pm 0.019$ & $[-0.571, -0.507]$ & $9.49 \times 10^{-233}$ \\
        \textit{thereof:} $z \leq 0.1$ & 2,897 & $-0.539 \pm 0.034$ & $7.101 \pm 0.041$ & $[-0.606, -0.473]$ & $4.62 \times 10^{-55}$ \\
        \textit{thereof:} $0.1 < z \leq 0.2$ & 6,832 & $-0.543 \pm 0.020$ & $7.927 \pm 0.023$ & $[-0.581, -0.504]$ & $1.74 \times 10^{-160}$ \\
        \textit{thereof:} $z > 0.2$ & 4,520 & $-0.522 \pm 0.023$ & $8.369 \pm 0.027$ & $[-0.567, -0.478]$ & $3.01 \times 10^{-109}$ \\
        \textit{thereof:} low-mass gal. & 3,650 & $-0.449 \pm 0.032$ & $7.303 \pm 0.033$ & $[-0.511, -0.387]$ & $6.43 \times 10^{-45}$ \\
        \textit{thereof:} med.-mass gal. & 6,960 & $-0.696 \pm 0.022$ & $8.157 \pm 0.026$ & $[-0.738, -0.653]$ & $7.88 \times 10^{-210}$ \\
        \textit{thereof:} high-mass gal. & 3,639 & $-0.840 \pm 0.029$ & $8.631 \pm 0.038$ & $[-0.898, -0.783]$ & $2.28 \times 10^{-161}$ \\
        \textit{thereof:} in SFGs & 12,414 & $-0.573 \pm 0.017$ & $7.870 \pm 0.020$ & $[-0.607, -0.539]$ & $1.94 \times 10^{-231}$ \\
        \textit{thereof:} in non-SFGs & 1,835 & $-0.783 \pm 0.044$ & $8.683 \pm 0.061$ & $[-0.870, -0.696]$ & $2.28 \times 10^{-64}$ \\
		\hline
        \multicolumn{5}{l}{\textbf{\textit{grizy} clumps}} \\
        \hline
        Total & 661,691 & $-0.589 \pm 0.002$ & $7.901 \pm 0.003$ & $[-0.594, -0.584]$ & $0.000$ \\
        \textit{thereof:} $z \leq 0.1$ & 161,340 & $-0.546 \pm 0.004$ & $7.093 \pm 0.005$ & $[-0.555, -0.538]$ & $0.000$ \\
        \textit{thereof:} $0.1 < z \leq 0.2$ & 316,519 & $-0.546 \pm 0.003$ & $7.955 \pm 0.004$ & $[-0.551, -0.540]$ & $0.000$ \\
        \textit{thereof:} $z > 0.2$ & 183,832 & $-0.565 \pm 0.004$ & $8.366 \pm 0.005$ & $[-0.572, -0.557]$ & $0.000$ \\
        \textit{thereof:} low-mass gal. & 171,384 & $-0.459 \pm 0.005$ & $7.337 \pm 0.005$ & $[-0.468, -0.450]$ & $0.000$ \\
        \textit{thereof:} med.-mass gal. & 324,122 & $-0.685 \pm 0.003$ & $8.063 \pm 0.004$ & $[-0.691, -0.678]$ & $0.000$ \\
        \textit{thereof:} high-mass gal. & 166,185 & $-0.894 \pm 0.005$ & $8.605 \pm 0.006$ & $[-0.903, -0.885]$ & $0.000$ \\
        \textit{thereof:} in SFGs & 579,405 & $-0.588 \pm 0.003$ & $7.847 \pm 0.003$ & $[-0.594, -0.583]$ & $0.000$ \\
        \textit{thereof:} in non-SFGs & 82,286 & $-0.925 \pm 0.007$ & $8.715 \pm 0.009$ & $[-0.938, -0.912]$ & $0.000$ \\
		\hline
        \multicolumn{5}{l}{\textbf{All clumps}} \\
        \hline
        Total & 675,940 & $-0.588 \pm 0.002$ & $7.901 \pm 0.003$ & $[-0.592, -0.583]$ & $0.000$ \\
        \textit{thereof:} $z \leq 0.1$ & 164,237 & $-0.546 \pm 0.004$ & $7.094 \pm 0.005$ & $[-0.555, -0.538]$ & $0.000$ \\
        \textit{thereof:} $0.1 < z \leq 0.2$ & 323,351 & $-0.545 \pm 0.003$ & $7.954 \pm 0.003$ & $[-0.551, -0.540]$ & $0.000$ \\
        \textit{thereof:} $z > 0.2$ & 188,352 & $-0.563 \pm 0.004$ & $8.366 \pm 0.004$ & $[-0.571, -0.556]$ & $0.000$ \\
        \textit{thereof:} low-mass gal. & 175,034 & $-0.458 \pm 0.005$ & $7.337 \pm 0.005$ & $[-0.467, -0.450]$ & $0.000$ \\
        \textit{thereof:} med.-mass gal. & 331,082 & $-0.685 \pm 0.003$ & $8.065 \pm 0.004$ & $[-0.691, -0.678]$ & $0.000$ \\
        \textit{thereof:} high-mass gal. & 169,824 & $-0.893 \pm 0.005$ & $8.606 \pm 0.006$ & $[-0.901, -0.884]$ & $0.000$ \\
        \textit{thereof:} in SFGs & 591,819 & $-0.588 \pm 0.003$ & $7.847 \pm 0.003$ & $[-0.593, -0.583]$ & $0.000$ \\
        \textit{thereof:} in non-SFGs & 84,121 & $-0.922 \pm 0.007$ & $8.714 \pm 0.009$ & $[-0.935, -0.909]$ & $0.000$ \\
		\hline
        \multicolumn{5}{l}{\textbf{Galaxy background}} \\
        \hline
        Total & 14,057 & $-0.329 \pm 0.014$ & $8.107 \pm 0.017$ & $[-0.357, -0.301]$ & $1.81 \times 10^{-114}$ \\
        \textit{thereof:} $z \leq 0.1$ & 2,781 & $-0.378 \pm 0.030$ & $7.613 \pm 0.037$ & $[-0.436, -0.320]$ & $1.19 \times 10^{-35}$ \\
        \textit{thereof:} $0.1 < z \leq 0.2$ & 6,773 & $-0.286 \pm 0.015$ & $8.083 \pm 0.019$ & $[-0.316, -0.257]$ & $3.65 \times 10^{-77}$ \\
        \textit{thereof:} $0.2 < z \leq 0.32$ & 4,503 & $-0.302 \pm 0.019$ & $8.465 \pm 0.023$ & $[-0.339, -0.264]$ & $1.89 \times 10^{-54}$ \\
        \textit{thereof:} low-mass gal. & 3,461 & $-0.497 \pm 0.031$ & $7.753 \pm 0.034$ & $[-0.558, -0.435]$ & $8.80 \times 10^{-54}$ \\
        \textit{thereof:} med.-mass gal. & 6,522 & $-0.453 \pm 0.017$ & $8.258 \pm 0.021$ & $[-0.486, -0.420]$ & $4.96 \times 10^{-143}$ \\
        \textit{thereof:} high-mass gal. & 4,074 & $-0.497 \pm 0.018$ & $8.703 \pm 0.024$ & $[-0.532, -0.462]$ & $3.26 \times 10^{-148}$ \\
		\hline
        \multicolumn{5}{l}{\textbf{Simulated clumps}} \\
        \hline
        Total & 1,586 & $-0.201 \pm 0.035$ & $7.554 \pm 0.054$ & $[-0.269, -0.133]$ & $9.72 \times 10^{-09}$ \\
        \textit{thereof:} $z \leq 0.1$ & 361 & $-0.090 \pm 0.058$ & $6.435 \pm 0.092$ & $[-0.204, 0.024]$ & $0.121$ \\
        \textit{thereof:} $0.1 < z \leq 0.2$ & 700 & $-0.168 \pm 0.037$ & $7.562 \pm 0.058$ & $[-0.241, -0.095]$ & $6.56 \times 10^{-06}$ \\
        \textit{thereof:} $z > 0.2$ & 525 & $-0.144 \pm 0.036$ & $8.057 \pm 0.053$ & $[-0.215, -0.073]$ & $7.09 \times 10^{-05}$ \\
        \textit{thereof:} low-mass gal. & 401 & $-0.192 \pm 0.080$ & $7.150 \pm 0.108$ & $[-0.350, -0.034]$ & $0.017$ \\
        \textit{thereof:} med.-mass gal. & 539 & $-0.330 \pm 0.058$ & $7.712 \pm 0.090$ & $[-0.444, -0.216]$ & $2.28 \times 10^{-08}$ \\
        \textit{thereof:} high-mass gal. & 646 & $-0.332 \pm 0.046$ & $8.025 \pm 0.077$ & $[-0.424, -0.241]$ & $2.34 \times 10^{-12}$ \\
        \hline
	\end{tabular}
\end{table*}

\begin{table*}
	\centering
	\caption[Values and uncertainties of the radial clump stellar age gradient $m$.]{Similar to Table \ref{tab:hsc_phys_props_gradient_UR} but showing the gradient values $m$ of the linear models $\log(t_{\mathrm{age}}/\mathrm{yr}) = m (\mathrm{d}_{\mathrm{cl}}/r_{\mathrm{eff}})+c$ fitted to the distributions of the clump stellar age as a function of the normalised radial distance.}
    \label{tab:hsc_phys_props_gradient_age}
	\footnotesize
        \begin{tabular}{lrrrrr}
		\hline
		Sample & \multicolumn{1}{l}{Count} & \multicolumn{1}{l}{Gradient $m$} & \multicolumn{1}{l}{Constant $c$} & \multicolumn{1}{l}{95\% conf. interval}  & \multicolumn{1}{l}{$p$-value} \\
		\hline
        \multicolumn{5}{l}{\textbf{\textit{ugrizy} clumps}} \\
        \hline
        Total & 14,249 & $-0.235 \pm 0.010$ & $8.652 \pm 0.012$ & $[-0.255, -0.216]$ & $2.1 \times 10^{-124}$ \\
        \textit{thereof:} $z \leq 0.1$ & 2,897 & $-0.258 \pm 0.025$ & $8.539 \pm 0.030$ & $[-0.306, -0.210]$ & $2.11 \times 10^{-25}$ \\
        \textit{thereof:} $0.1 < z \leq 0.2$ & 6,832 & $-0.243 \pm 0.013$ & $8.720 \pm 0.016$ & $[-0.270, -0.217]$ & $1.96 \times 10^{-71}$ \\
        \textit{thereof:} $0.2 < z \leq 0.32$ & 4,520 & $-0.207 \pm 0.016$ & $8.620 \pm 0.019$ & $[-0.239, -0.175]$ & $1.39 \times 10^{-36}$ \\
        \textit{thereof:} low-mass gal. & 3,650 & $-0.181 \pm 0.020$ & $8.446 \pm 0.021$ & $[-0.219, -0.142]$ & $1.09 \times 10^{-19}$ \\
        \textit{thereof:} med.-mass gal. & 6,960 & $-0.299 \pm 0.014$ & $8.760 \pm 0.017$ & $[-0.327, -0.270]$ & $1.12 \times 10^{-93}$ \\
        \textit{thereof:} high-mass gal. & 3,639 & $-0.306 \pm 0.019$ & $8.819 \pm 0.024$ & $[-0.343, -0.269]$ & $8.93 \times 10^{-58}$ \\
        \textit{thereof:} in SFGs & 12,414 & $-0.247 \pm 0.011$ & $8.644 \pm 0.012$ & $[-0.268, -0.226]$ & $7.69 \times 10^{-116}$ \\
        \textit{thereof:} in non-SFGs & 1,835 & $-0.302 \pm 0.026$ & $8.881 \pm 0.036$ & $[-0.354, -0.250]$ & $1.63 \times 10^{-29}$ \\
		\hline
        \multicolumn{5}{l}{\textbf{\textit{grizy} clumps}} \\
        \hline
        Total & 661,691 & $-0.206 \pm 0.002$ & $8.465 \pm 0.002$ & $[-0.209, -0.203]$ & $0.000$ \\
        \textit{thereof:} $z \leq 0.1$ & 161,340 & $-0.193 \pm 0.003$ & $8.328 \pm 0.004$ & $[-0.200, -0.187]$ & $0.000$ \\
        \textit{thereof:} $0.1 < z \leq 0.2$ & 316,519 & $-0.196 \pm 0.002$ & $8.486 \pm 0.003$ & $[-0.200, -0.191]$ & $0.000$ \\
        \textit{thereof:} $z > 0.2$ & 183,832 & $-0.221 \pm 0.003$ & $8.533 \pm 0.004$ & $[-0.227, -0.214]$ & $0.000$ \\
        \textit{thereof:} low-mass gal. & 171,384 & $-0.112 \pm 0.003$ & $8.297 \pm 0.003$ & $[-0.118, -0.106]$ & $1.72 \times 10^{-269}$ \\
        \textit{thereof:} med.-mass gal. & 324,122 & $-0.208 \pm 0.002$ & $8.482 \pm 0.003$ & $[-0.213, -0.203]$ & $0.000$ \\
        \textit{thereof:} high-mass gal. & 166,185 & $-0.331 \pm 0.003$ & $8.664 \pm 0.004$ & $[-0.337, -0.324]$ & $0.000$ \\
        \textit{thereof:} in SFGs & 579,405 & $-0.185 \pm 0.002$ & $8.430 \pm 0.002$ & $[-0.189, -0.182]$ & $0.000$ \\
        \textit{thereof:} in non-SFGs & 82,286 & $-0.390 \pm 0.005$ & $8.797 \pm 0.007$ & $[-0.399, -0.380]$ & $0.000$ \\
		\hline
        \multicolumn{5}{l}{\textbf{All clumps}} \\
        \hline
        Total & 675,940 & $-0.206 \pm 0.002$ & $8.469 \pm 0.002$ & $[-0.210, -0.203]$ & $0.000$ \\
        \textit{thereof:} $z \leq 0.1$ & 164,237 & $-0.195 \pm 0.003$ & $8.332 \pm 0.004$ & $[-0.201, -0.188]$ & $0.000$ \\
        \textit{thereof:} $0.1 < z \leq 0.2$ & 323,351 & $-0.196 \pm 0.002$ & $8.490 \pm 0.003$ & $[-0.201, -0.192]$ & $0.000$ \\
        \textit{thereof:} $z > 0.2$ & 188,352 & $-0.220 \pm 0.003$ & $8.535 \pm 0.004$ & $[-0.226, -0.214]$ & $0.000$ \\
        \textit{thereof:} low-mass gal. & 175,034 & $-0.114 \pm 0.003$ & $8.300 \pm 0.003$ & $[-0.120, -0.107]$ & $5.88 \times 10^{-283}$ \\
        \textit{thereof:} med.-mass gal. & 331,082 & $-0.210 \pm 0.002$ & $8.487 \pm 0.003$ & $[-0.214, -0.205]$ & $0.000$ \\
        \textit{thereof:} high-mass gal. & 169,824 & $-0.330 \pm 0.003$ & $8.668 \pm 0.004$ & $[-0.336, -0.323]$ & $0.000$ \\
        \textit{thereof:} in SFGs & 591,819 & $-0.186 \pm 0.002$ & $8.434 \pm 0.002$ & $[-0.190, -0.183]$ & $0.000$ \\
        \textit{thereof:} in non-SFGs & 84,121 & $-0.388 \pm 0.005$ & $8.799 \pm 0.006$ & $[-0.397, -0.379]$ & $0.000$ \\
		\hline
        \multicolumn{5}{l}{\textbf{Galaxy background}} \\
        \hline
        Total & 14,057 & $0.012 \pm 0.005$ & $8.749 \pm 0.006$ & $[0.002, 0.023]$ & $0.016$ \\
        \textit{thereof:} $z \leq 0.1$ & 2,781 & $-0.041 \pm 0.013$ & $8.777 \pm 0.015$ & $[-0.066, -0.016]$ & $0.001$ \\
        \textit{thereof:} $0.1 < z \leq 0.2$ & 6,773 & $0.032 \pm 0.007$ & $8.742 \pm 0.008$ & $[0.018, 0.046]$ & $4.36 \times 10^{-06}$ \\
        \textit{thereof:} $0.2 < z \leq 0.32$ & 4,503 & $0.021 \pm 0.009$ & $8.734 \pm 0.011$ & $[0.003, 0.039]$ & $0.022$ \\
        \textit{thereof:} low-mass gal. & 3,461 & $-0.007 \pm 0.010$ & $8.674 \pm 0.011$ & $[-0.027, 0.012]$ & $0.450$ \\
        \textit{thereof:} med.-mass gal. & 6,522 & $-0.001 \pm 0.007$ & $8.754 \pm 0.009$ & $[-0.015, 0.013]$ & $0.939$ \\
        \textit{thereof:} high-mass gal. & 4,074 & $-0.041 \pm 0.010$ & $8.908 \pm 0.013$ & $[-0.061, -0.022]$ & $3.04 \times 10^{-05}$ \\
		\hline
        \multicolumn{5}{l}{\textbf{Simulated clumps}} \\
        \hline
        Total & 1,586 & $-0.017 \pm 0.025$ & $8.293 \pm 0.039$ & $[-0.066, 0.032]$ & $0.497$ \\
        \textit{thereof:} $z \leq 0.1$ & 361 & $0.005 \pm 0.054$ & $8.049 \pm 0.086$ & $[-0.101, 0.110]$ & $0.928$ \\
        \textit{thereof:} $0.1 < z \leq 0.2$ & 700 & $-0.058 \pm 0.033$ & $8.468 \pm 0.052$ & $[-0.124, 0.007]$ & $0.080$ \\
        \textit{thereof:} $0.2 < z \leq 0.32$ & 525 & $0.026 \pm 0.047$ & $8.227 \pm 0.070$ & $[-0.066, 0.118]$ & $0.582$ \\
        \textit{thereof:} low-mass gal. & 401 & $-0.037 \pm 0.057$ & $8.280 \pm 0.077$ & $[-0.149, 0.075]$ & $0.518$ \\
        \textit{thereof:} med.-mass gal. & 539 & $-0.054 \pm 0.044$ & $8.365 \pm 0.068$ & $[-0.141, 0.033]$ & $0.223$ \\
        \textit{thereof:} high-mass gal. & 646 & $0.003 \pm 0.038$ & $8.270 \pm 0.063$ & $[-0.072, 0.077]$ & $0.947$ \\
        \hline
	\end{tabular}
\end{table*}

% -----------------------------------------------
% Testing the robustness of the stellar mass and age gradients
\section{Testing the robustness of the stellar mass and age gradients}\label{sec:hsc_phys_props_gradient_robust}
In addition to the incompleteness affecting low-mass clumps, the observed gradients may also be influenced by the choice of method for measuring the effective radius that is used to normalize the radial distances and the various assumptions made during the SED fitting process to infer the stellar masses and ages of the clumps.

We first tested whether the gradients differ for galaxies with spectroscopic and photometric redshift. Galaxies for which redshift has been determined spectroscopically are generally at lower redshift and require observations of brighter objects than are required for a photometric redshift measurement. In the galaxy sample used in this study, the median redshift of galaxies with spectroscopic redshift is $\hat{z} = 0.11$, whereas the galaxies with photometric redshifts have a median redshift of $\hat{z} = 0.21$. The median apparent \textit{g}-band magnitude for spec-z (photo-z) galaxies is $\hat{g}=19.15\,m_{\mathrm{AB}}$ ($\hat{g}=20.81\,m_{\mathrm{AB}}$) and the median stellar mass is $\log$($M_{\mathrm{gal}}/M_\odot$) $=11.02$ ($\log$($M_{\mathrm{gal}}/M_\odot$) $=10.40$). %Recall also that the uncertainty for spectroscopic redshifts is usually smaller than for photometric redshifts and the SED fits are expected to recover a stellar mass and age of the clumps that is likely closer to the true value if the redshift measurement is more accurate.

The median stellar masses of the clumps as a function of the radial distance does not show a noticeably different trend compared to the baseline samples from the CLAUDS/HSC-SSP galaxies or the HSC-SSP galaxies (Table \ref{tab:hsc_phys_props_gradient_mass_variations}). However, the estimated gradients differ. The slopes estimated from the clumps in photo-z galaxies are flatter than the slope estimated from spec-z galaxies. This is expected as the spec-z galaxies are at lower redshift and tend to be brighter so that more low-mass clumps are detected in those galaxies. This shifts the median to lower stellar masses resulting in a less top-heavy distribution. Similarly, the stellar mass gradient estimated from the simulated clumps located in spec-z galaxies is not significantly different from $m=0.0$. This is due to the higher completeness of detected simulated low-mass clumps. 

The stellar mass gradients of the real clumps are still steeper than the gradients of the simulated clumps irrespective of the method of redshift estimation of the host galaxies. The redshift measurement method does not have a direct effect on the gradient measurement per se but it is a proxy for the degree of incompleteness affecting the low-mass clumps in the samples. 

As a second test, we compared the effect of different measurements of the effective or half-light radius that is used to normalise the distance of the clumps to the galaxy centre. We compared the half-light radii from the HSC-SSP PDR3 catalogue, the SDSS DR18 catalogue and the radii that we have remeasured for this study and that have been primarily used for the analysis of the clump property radial gradients. All real clump gradients are steeper than those measured from the simulated clumps using the same radius estimation method (Table \ref{tab:hsc_phys_props_gradient_mass_variations}), indicating that the choice of a specific half-light radius cannot explain the observed negative stellar mass gradients of the clumps.

While the gradient estimated with the SDSS half-light radius is close to the one estimated with our remeasured radius, the gradient using the HSC radius is less steep for the clumps from the CLAUDS/HSC-SSP and the HSC-SSP galaxy samples (Table \ref{tab:hsc_phys_props_gradient_mass_variations}). Furthermore, the HSC half-light radius is not available for $\sim20\%$ of the clump sample. 

As the HSC radius is significantly smaller than the other two radii measurements, the whole distribution of stellar masses as a function of the radial distance is shifted towards higher \textit{normalised} distances. This results in a median stellar mass that is higher in each normalised distance bin than the medians from the other radius estimation methods in the same normalised distance bin but also leads to a flatter gradient over the observed distance range.

The assumed SFH that is used during the SED fitting process can strongly affect the rate with which the clump stellar mass is formed. A constant SFH with its constant SFR would lead to a different stellar mass formed at time $t_1 = t_0 + t_{\mathrm{age}}$ than a bursty SFH or an exponential model for the SFH, i.e. a delayed $\tau$ SFH, given the same age $t_{\mathrm{age}}$ of the stellar population. Setting the metallicity during SED fitting to a fixed value, i.e. solar metallicity, can also affect the inferred stellar masses of the clumps and has an even more direct effect on the stellar age estimates. 

\begin{figure}
    \centering
    \includegraphics[width=0.8\columnwidth]{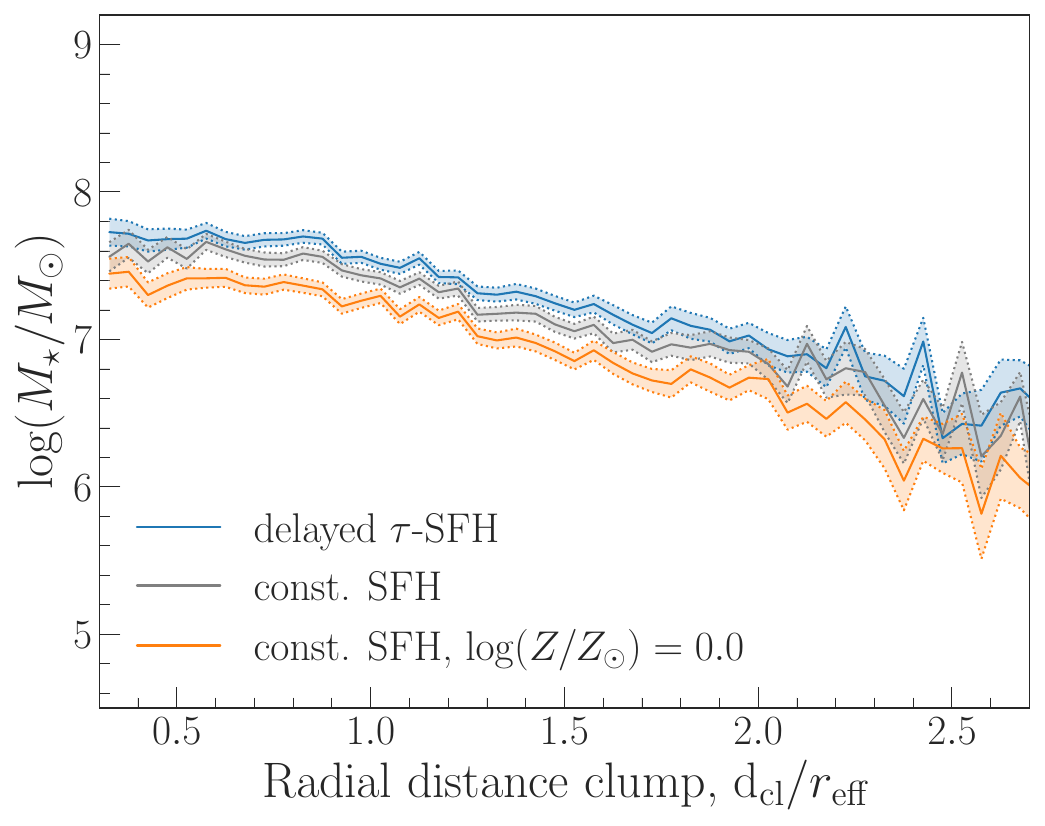}
    \caption[Median of the clump stellar mass distribution as a function of the normalised galactocentric distance for different SFH models.]{Median of the clump stellar mass distribution as a function of the normalised galactocentric distance for different SFH models. The stellar mass of the clumps was inferred with SED models using a constant SFH (grey), a delayed $\tau$ SFH (blue) or a constant SFH model with fixed metallicity (orange). Shaded areas show the standard error of the median.}
    \label{fig:hsc_phys_props_gradient_mass_SFH}
\end{figure}

In Figure \ref{fig:hsc_phys_props_gradient_mass_SFH}, we plot the median stellar mass as a function of the radial distance for the alternative models in comparison with the primary SED model with constant SFH and free metallicity. This lets us test whether a SED model with delayed $\tau$ SFH or a constant SFH but with fixed metallicity would result in different stellar mass gradients. The median stellar masses are similar and decline with similar slopes with increasing distance from the galactic centre for all three SED models. The estimated gradients (Table \ref{tab:hsc_phys_props_gradient_mass_variations}) are almost identical for the models with constant SFH regardless whether the metallicity is set fixed or inferred during the SED fitting process. The slope estimated for the SED model with delayed $\tau$ SFH is slightly less steep and significantly different from the other slopes (Welch's T-test: $p\text{-value}= 3.18 \times 10^{-5}$ compared with the primary SED model and $p\text{-value}= 1.13 \times 10^{-5}$ in comparison with the model with constant SFH and fixed metallicity). However, we note that the alternative (non-constant SFH) SED modelling assumptions would still result in clumps that are more massive closer to the galactic centre than further away. 

\begin{table*}
	\centering
	\caption[Values and uncertainties of the radial clump stellar mass gradient $m$ for different SED model assumptions and galaxy parameters.]{Similar to Table \ref{tab:hsc_phys_props_gradient_mass} but showing the gradient values $m$ of the linear models $\log(M_{\mathrm{cl}}/M_\odot) = m (\mathrm{d}_{\mathrm{cl}}/r_{\mathrm{eff}})+c$ fitted to the distributions of the clump stellar mass as a function of the normalised radial distance calculated for different SED model assumptions and galaxy parameters. The values for the baseline samples are repeated from Table \ref{tab:hsc_phys_props_gradient_mass}.}
    \label{tab:hsc_phys_props_gradient_mass_variations}
	\footnotesize
        \begin{tabular}{lrrrrr}
		\hline
		Sample & \multicolumn{1}{l}{Count} & \multicolumn{1}{l}{Gradient $m$} & \multicolumn{1}{l}{Constant $c$} & \multicolumn{1}{l}{95\% conf. interval}  & \multicolumn{1}{l}{$p$-value} \\
		\hline
        \multicolumn{5}{l}{\textbf{\textit{ugrizy} clumps}} \\
        \hline
        Baseline & 14,249 & $-0.539 \pm 0.016$ & $7.903 \pm 0.019$ & $[-0.571, -0.507]$ & $9.49 \times 10^{-233}$ \\
        Spectroscopic redshift & 3,933 & $-0.584 \pm 0.032$ & $7.571 \pm 0.041$ & $[-0.647, -0.522]$ & $1.18 \times 10^{-71}$ \\
        Photometric redshift & 10,316 & $-0.415 \pm 0.018$ & $7.917 \pm 0.020$ & $[-0.450, -0.380]$ & $5.13 \times 10^{-118}$ \\
        Half-light radius (HSC) & 11,559 & $-0.276 \pm 0.014$ & $7.825 \pm 0.021$ & $[-0.304, -0.248]$ & $5.50 \times 10^{-84}$ \\
        Half-light radius (SDSS) & 14,115 & $-0.440 \pm 0.016$ & $7.795 \pm 0.020$ & $[-0.472, -0.408]$ & $1.07 \times 10^{-154}$ \\
        delayed $\tau$-SFH & 14,907 & $-0.460 \pm 0.016$ & $7.863 \pm 0.019$ & $[-0.490, -0.429]$ & $1.07 \times 10^{-184}$ \\
        const. SFH, $\log(Z/Z_\odot)=0.0$ & 14,550 & $-0.540 \pm 0.018$ & $7.664 \pm 0.022$ & $[-0.576, -0.504]$ & $2.73 \times 10^{-182}$ \\
		\hline
        \multicolumn{5}{l}{\textbf{\textit{grizy} clumps}} \\
        \hline
        Baseline & 655,922 & $-0.533 \pm 0.003$ & $7.712 \pm 0.003$ & $[-0.539, -0.527]$ & $0.000$ \\
        Spectroscopic redshift & 189,772 & $-0.604 \pm 0.005$ & $7.483 \pm 0.006$ & $[-0.614, -0.595]$ & $0.000$ \\
        Photometric redshift & 466,220 & $-0.422 \pm 0.004$ & $7.719 \pm 0.004$ & $[-0.429, -0.415]$ & $0.000$ \\
        Half-light radius (HSC) & 538,070 & $-0.283 \pm 0.003$ & $7.643 \pm 0.004$ & $[-0.288, -0.278]$ & $0.000$ \\
        Half-light radius (SDSS) & 649,616 & $-0.418 \pm 0.003$ & $7.593 \pm 0.004$ & $[-0.424, -0.412]$ & $0.000$ \\
		\hline
        \multicolumn{5}{l}{\textbf{Simulated clumps}} \\
        \hline
        Baseline & 1,586 & $-0.201 \pm 0.035$ & $7.554 \pm 0.054$ & $[-0.269, -0.133]$ & $9.72 \times 10^{-09}$ \\
        Spectroscopic redshift & 572 & $-0.123 \pm 0.064$ & $7.037 \pm 0.105$ & $[-0.249, 0.002]$ & $0.055$ \\
        Photometric redshift & 1,014 & $-0.148 \pm 0.035$ & $7.705 \pm 0.052$ & $[-0.216, -0.081]$ & $1.95 \times 10^{-05}$ \\
        Half-light radius (HSC) & 915 & $-0.139 \pm 0.044$ & $7.638 \pm 0.078$ & $[-0.225, -0.053]$ & $0.002$ \\
        Half-light radius (SDSS) & 1,374 & $-0.059 \pm 0.038$ & $7.352 \pm 0.060$ & $[-0.133, 0.015]$ & $0.119$ \\
        \hline
	\end{tabular}
\end{table*}

The observed negative age gradients are robust against different choices of the radial distance measure, i.e. the different half-light radii, and are not biased by selecting galaxies with spectroscopic redshift measurements instead of galaxies with photometric redshift measurements. The gradient values vary between the different choices but are still negative and significant, while similar tests using the simulated clumps result in gradients that are not significantly different from zero (Table \ref{tab:hsc_phys_props_gradient_age_variations}).

\begin{figure}
    \centering
    \includegraphics[width=0.8\columnwidth]{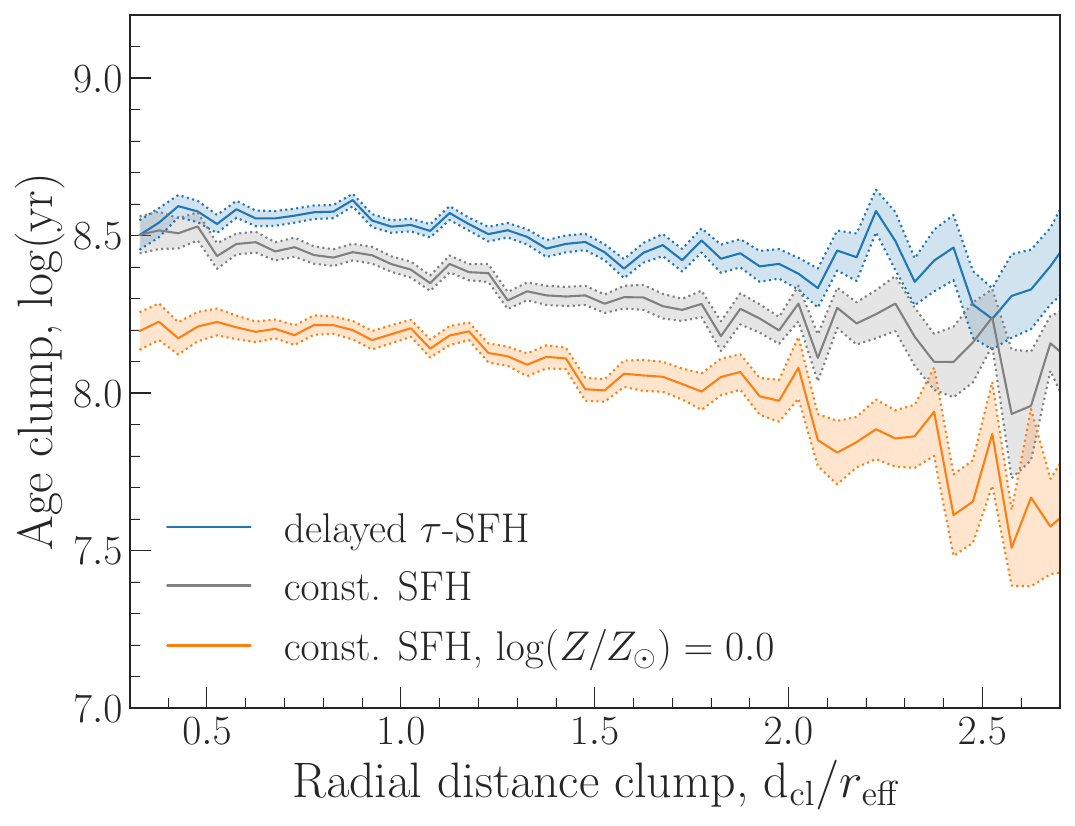}
    \caption[Median of the clump stellar age distribution as a function of the normalised galactocentric distance for different SFH models.]{Median of the clump stellar age distribution as a function of the normalised galactocentric distance for different SFH models. The stellar age of the clumps was inferred with SED models using a constant SFH (grey), a delayed $\tau$ SFH (blue) or a constant SFH model with fixed metallicity (orange). Shaded areas show the standard error of the median.}
    \label{fig:hsc_phys_props_gradient_age_SFH}
\end{figure}

However, the choice of the SFH model can impact the inferred clump ages and the resulting age gradient. In Figure \ref{fig:hsc_phys_props_gradient_age_SFH}, we plot the median stellar age of the clumps from the CLAUDS/HSC-SSP galaxy sample as a function of their distance from the galactic centre for the three SFHs tested during the SED fitting process. The stellar ages and their distribution versus radial distance bin differ significantly between the different SED models (Welch's T-test: $p\text{-value}= 3.07 \times 10^{-35}$ compared with the primary SED model and $p\text{-value}= 2.06 \times 10^{-24}$ in comparison with the model with constant SFH and fixed metallicity).

Comparing the inferred ages of the two SED models with constant SFH, Figure \ref{fig:hsc_phys_props_gradient_age_SFH} shows that the fixed metallicity model generally leads to younger estimated ages than if the metallicity is fitted alongside the stellar age. However, the age gradients as a function of the radial distance are not significantly different (Table \ref{tab:hsc_phys_props_gradient_age_variations}, Welch's T-test: $p\text{-value}= 0.104$). In contrast, the model using a delayed $\tau$ SFH tends to infer older ages for the clumps and the age gradient resulting from that model are significantly flatter. The gradient is still negative but the 95\% confidence interval overlaps with the confidence interval of the gradient determined for the simulated clumps (Table \ref{tab:hsc_phys_props_gradient_age_variations}).

\begin{table*}
	\centering
	\caption[Values and uncertainties of the radial clump stellar age gradient $m$ calculated for different SED model assumptions and galaxy parameters.]{Similar to Table \ref{tab:hsc_phys_props_gradient_mass_variations} but showing the gradient values $m$ of the linear models $\log(t_{\mathrm{age}}/\mathrm{yr}) = m (\mathrm{d}_{\mathrm{cl}}/r_{\mathrm{eff}})+c$ fitted to the distributions of the clump stellar age as a function of the normalised radial distance calculated for different SED model assumptions and galaxy parameters. The values for the baseline samples are repeated from Table \ref{tab:hsc_phys_props_gradient_age}.}
    \label{tab:hsc_phys_props_gradient_age_variations}
	\footnotesize
        \begin{tabular}{lrrrrr}
		\hline
		Sample & \multicolumn{1}{l}{Count} & \multicolumn{1}{l}{Gradient $m$} & \multicolumn{1}{l}{Constant $c$} & \multicolumn{1}{l}{95\% conf. interval}  & \multicolumn{1}{l}{$p$-value} \\
		\hline
        \multicolumn{5}{l}{\textbf{\textit{ugrizy} clumps}} \\
        \hline
        Baseline & 14,249 & $-0.235 \pm 0.010$ & $8.652 \pm 0.012$ & $[-0.255, -0.216]$ & $2.1 \times 10^{-124}$ \\
        Spectroscopic redshift & 3,933 & $-0.229 \pm 0.020$ & $8.548 \pm 0.026$ & $[-0.269, -0.190]$ & $4.23 \times 10^{-30}$ \\
        Photometric redshift & 10,316 & $-0.213 \pm 0.011$ & $8.665 \pm 0.013$ & $[-0.235, -0.191]$ & $2.06 \times 10^{-80}$ \\
        Half-light radius (HSC) & 11,559 & $-0.139 \pm 0.009$ & $8.643 \pm 0.014$ & $[-0.157, -0.122]$ & $8.05 \times 10^{-55}$ \\
        Half-light radius (SDSS) & 14,115 & $-0.192 \pm 0.010$ & $8.607 \pm 0.012$ & $[-0.211, -0.172]$ & $1.19 \times 10^{-83}$ \\
        delayed $\tau$ SFH & 14,907 & $-0.080 \pm 0.008$ & $8.574 \pm 0.009$ & $[-0.095, -0.064]$ & $5.8 \times 10^{-24}$ \\
        const. SFH, $\log(Z/Z_\odot)=0.0$ & 14,550 & $-0.217 \pm 0.011$ & $8.363 \pm 0.013$ & $[-0.238, -0.195]$ & $1.26 \times 10^{-85}$ \\
		\hline
        \multicolumn{5}{l}{\textbf{\textit{grizy} clumps}} \\
        \hline
        Baseline & 655,992 & $-0.187 \pm 0.002$ & $8.445 \pm 0.002$ & $[-0.190, -0.184]$ & $0.000$ \\
        Spectroscopic redshift & 189,772 & $-0.197 \pm 0.003$ & $8.385 \pm 0.004$ & $[-0.203, -0.191]$ & $0.000$ \\
        Photometric redshift & 466,220 & $-0.167 \pm 0.002$ & $8.452 \pm 0.002$ & $[-0.170, -0.163]$ & $0.000$ \\
        Half-light radius (HSC) & 538,070 & $-0.099 \pm 0.001$ & $8.421 \pm 0.002$ & $[-0.102, -0.096]$ & $0.000$ \\
        Half-light radius (SDSS) & 649,616 & $-0.145 \pm 0.002$ & $8.405 \pm 0.002$ & $[-0.149, -0.142]$ & $0.000$ \\
		\hline
        \multicolumn{5}{l}{\textbf{Simulated clumps}} \\
        \hline
        Baseline & 1,586 & $-0.017 \pm 0.025$ & $8.293 \pm 0.039$ & $[-0.066, 0.032]$ & $0.497$ \\
        Spectroscopic redshift & 572 & $0.018 \pm 0.041$ & $8.108 \pm 0.068$ & $[-0.063, 0.099]$ & $0.666$ \\
        Photometric redshift & 1,014 & $-0.006 \pm 0.031$ & $8.352 \pm 0.047$ & $[-0.067, 0.056]$ & $0.851$ \\
        Half-light radius (HSC) & 915 & $-0.033 \pm 0.034$ & $8.368 \pm 0.061$ & $[-0.101, 0.034]$ & $0.331$ \\
        Half-light radius (SDSS) & 1,374 & $-0.009 \pm 0.027$ & $8.282 \pm 0.042$ & $[-0.061, 0.044]$ & $0.746$ \\
        \hline
	\end{tabular}
\end{table*}

%%%%%%%%%%%%%%%%%%%%%%%%%%%%%%%%%%%%%%%%%%%%%%%%%%

% Don't change these lines
\bsp	% typesetting comment
\label{lastpage}
\end{document}